\documentclass{actorbased}

\begin{document}

\maketitle

\begin{abstract}
To date, most work regarding the formal analysis of access control schemes has
focused on quantifying and comparing the expressive power of a set of schemes.
Although expressive power is important, it is a property that exists in an
\emph{absolute} sense, detached from the application-specific context within
which an access control scheme will ultimately be deployed. In this paper, by
contrast, we formalize the access control \emph{suitability analysis problem},
which seeks to evaluate the degree to which a set of candidate access control
schemes can meet the needs of an application-specific workload. This process
involves both reductions to assess whether a scheme is \emph{capable} of
implementing a workload, as well as cost analysis using ordered measures to
quantify the \emph{overheads} of using each candidate scheme to service the
workload. We develop a mathematical framework for analyzing instances of the
suitability analysis problem, and evaluate this framework both formally (by
quantifying its efficiency and accuracy properties) and practically (by
exploring a group-based messaging workload from the literature). An ancillary
contribution of our work is the identification of \emph{\aums}, which are a
useful class of modifications that can be made to enhance the expressive power
of an access control scheme without negatively impacting the safety properties
of the scheme.
\end{abstract}

\section{Introduction}
\label{sec:intro}

Access control is one of the most fundamental aspects of computer security, and
has been the subject of much formal study. However, existing work on the formal
analysis of access control schemes has focused largely on comparing the
\emph{relative expressive power} of two or more access control schemes
(e.g.,~\cite{hru, lipton, early-sim, statetransition, tripli, osborn00tissec,
sandhu92jcs, atam}). Although expressive power is an interesting and meaningful
basis for comparing access control schemes, it exists only as a comparison made
in absolute terms. That is, the knowledge that a scheme \(\scm{S}\) is more
expressive than another scheme \(\scm{S}^\prime\) provides no assurance that
\(\scm{S}\) is the best access control scheme for use within a particular
real-world application context. It could be the case, for instance, that
\(\scm{S}^\prime\) is \emph{expressive enough} for a particular application and
also has lower administrative overheads than \(\scm{S}\) would in the same
situation. As was noted in a recent NIST report, access control is not an area
with ``one size fits all'' solutions and, as such, systems should be evaluated
and compared relative to application-specific metrics~\cite{nist}. This report
notes a variety of possible access control quality metrics, but provides little
guidance for actually applying these metrics and carrying out \emph{practical}
analyses of access control schemes.

Considering the wide availability of many diverse access control schemes and the
relative difficulty of designing and building new secure systems from the ground
up, an interesting topic for exploration is that of \emph{suitability analysis}.
Informally, this problem can be stated as follows: \emph{Given a description of
a system's access control needs and a collection of access control schemes,
which scheme best meets the needs of the system?} Instances of this question can
arise in many different scenarios, encompassing both the deployment of new
applications and the reexamination of existing applications as assumptions and
requirements evolve. Modern software applications are complex entities that may
control access to both digital (e.g., files) and physical (e.g., doors)
resources. Given that organizations are typically afforded little guidance in
choosing appropriate security solutions, suitability analysis could help
software developers sort through the myriad available security frameworks and
the multiple access control schemes embedded in each.

In this paper, we identify and formalize the access control suitability problem,
and develop a mathematical framework and techniques to facilitate suitability
analysis. We first formalize the notion of an access control workload to
abstract the application's access control needs and the expected uses of these
functionalities. Analysis then consists of two orthogonal tasks: (i)
demonstrating that each candidate access control scheme is capable of
\emph{safely} implementing the workload, and (ii) quantifying the costs
associated with the use of each candidate scheme. Within this context, we
develop techniques for safely extending the functionality of candidate schemes
that require additional expressive power, develop guidelines for formally
specifying a wide range of access control cost metrics, and present a simulation
framework for carrying out Monte Carlo-based cost analysis within our
mathematical model. In doing so, we make the following contributions:
\begin{itemize}

\item We formalize the \emph{access control suitability analysis problem}, and
articulate a set of requirements that should be satisfied by suitability
analysis frameworks.

\item We present the first formal definition of an \emph{access control
workload}. This enables system administrators to clearly and concisely specify
the functionalities that must be provided by access control schemes that are to
be used within a given context, as well as identify the ways in which these
schemes are envisioned to be exercised.

\item We develop a \emph{two-phase analysis framework} for assessing the
suitability of an access control scheme with respect to a particular workload.
We first establish whether the candidate system is expressive enough to safely
implement the functionality of the workload via reduction. We then utilize a
constrained, actor-based workload invocation structure to drive a cost analysis
simulation that explores the expected costs of deployment.

\item To address issues of fragility that arise when constructing reductions
between a workload and a candidate scheme, we introduce the notion of
\emph{access control \aums} (\AMs). From a practical perspective, \aums
represent ``tweaks'' that can be made to an existing scheme to increase the
range of questions that it can answer. From a theoretical perspective, \AMs
describe a class of enhancements to a scheme's expressive power that \emph{do
not} alter its safety properties (i.e., those that \emph{strictly} expand the
set of policies that can be represented). We prove the safety guarantees of
\AMs, and demonstrate their use during suitability analysis.

\item We present a detailed case study demonstrating how our framework can be
used to gain insight into a realistic scenario. Namely, we investigate a
workload derived from a group messaging
scenario~\cite{krishnan2009sacmat,krishnan2009asiaccs,krishnan2011}. We confirm
the intuition that such a scenario can be implemented in commonly-used
general-purpose access control schemes (though extensions are required to do so
safely). We also found that such implementations differ widely in their costs,
confirming the belief that addressing group-based sharing using general-purpose
access control (even with a scheme that is expressive enough) can lead to
inferior results. This emphasizes the importance of suitability analysis when
making access control decisions.

\end{itemize}

The remainder of this paper will be structured as follows. In
\cref{sec:related}, we describe related work. In \cref{sec:approach}, we present
a formal problem statement, solution requirements, and an overview of our
proposed analysis framework. In \cref{sec:expressiveness,sec:cost}, we describe
the two phases of our framework in detail (expressiveness evaluation and cost
analysis, respectively). In \cref{sec:extensions}, we discuss techniques for
extending an under-expressive scheme so that it may implement a workload that it
otherwise could not. We describe our case study and present its results in
\cref{sec:casestudy}. In \cref{sec:discussion}, we evaluate the degree to which
our analysis framework meets the requirements in articulated in
\cref{sec:approach}, and discuss a number of interesting open problems related
to the suitability analysis problem. We conclude in \cref{sec:conclusion}.

\section{Related Work}
\label{sec:related}

The formal study of access control schemes began with the seminal paper by
Harrison, Ruzzo, and Ullman that investigated the rights leakage
problem~\cite{hru}. This paper formalized a general access control model and
proved that determining whether a particular access right could ever be granted
to a specific individual---the so-called ``safety problem''---was undecidable.
Shortly thereafter, Lipton and Snyder showed that in a more restricted access
control system, safety was not only decidable, but decidable in linear
time~\cite{lipton}. These two results introduced the notion that the most
capable system is not always the right choice---that restricting our system can
yield higher efficiency and greater ease in solving relevant security problems.
This led to many results investigating the relative expressive power of various
access control schemes, often leveraging some notion of (bi)simulation
(e.g.,~\cite{early-sim,statetransition,osborn00tissec,sandhu92jcs,atam}).

Further work by Ammann et al.~\cite{early-sim}, Chander et
al.~\cite{statetransition}, and Li et al.~\cite{safety} developed
simulation-based frameworks for comparing the expressive power of various access
control schemes. These simulation frameworks proved to be too relaxed, allowing
almost any reasonable scheme to be shown equivalent to all others. To address
this, Tripunitara and Li~\cite{tripli} developed a more restrictive notion of
expressive power. Their framework supersedes the more informal notions of
simulation developed in prior works by requiring the use of specific types of
mappings between systems that guarantee relevant security properties are
preserved under simulation; this provides a greater level of precision when
ranking access control schemes in terms of their expressiveness. Unfortunately,
none of these frameworks supports the comparison of access control schemes with
regards to their ability to perform \emph{well} within a particular environment.

The need for application-aware evaluation of access control systems was
reinforced by a recent NIST report, which states that ``when it comes to access
control mechanisms, one size does not fit all''~\cite{nist}. The report bemoans
the lack of established quality metrics for access control systems, going so far
as to list numerous possibilities, but stopping short of explaining how one
might choose between them or evaluate systems with respect to one's specific
requirements. In this paper, we develop a formal framework for exploring exactly
this problem.

Wang et al.~\cite{delegation} described methods to safely extend role-based
access control schemes with delegation primitives. However, role-based access
control is only one particular scheme, and delegation is only one particular
access control feature. Thus, this work provides no guideline for extending
other access control schemes, or using extensions to allow different classes of
abilities. In our work, we discuss the general problem of extending access
control schemes, and present a particular class of safe and useful extensions,
called \aums.

As a result of the lack of tools for evaluating suitability in access control,
there is little work in the field for generating synthetic traces that are
representative of an access control application. Thus, for inspiration in
designing the access control workload's invocation component (see
\cref{sec:cost:invocations}), we turn to work in other domains. In the field of
disk benchmarking, Ganger~\cite{disk} observed that interleaved workloads
provided the most accurate approximation of recorded traces. Thus, mechanisms
for representing access control workloads must be capable of simulating the
interleaved actions of multiple actors. This view is reinforced by the design of
IBM's SWORD workload generator for stream processing systems~\cite{sword}. This
work also points out that synthetic workloads need to replicate both volumetric
and contextual properties of an execution environment in order to provide an
accurate indication of a system's performance within that environment. Thus, we
conjecture that access control workloads as well may need to be capable of
expressing not only volumetric statistics such as number of documents created,
but also contextual statistics such as the type of content in created documents.

Recent work in workflow systems has analyzed the complexity of the workflow
satisfiability problem (\WSP), which determines whether a workflow can be
completed by the participants in the
system~\cite{wang2007esorics,crampton2012ccs}. This problem turns out to be
important for our approach, since our analysis framework includes a simulation
procedure that utilizes workflow systems for describing behavior. Without an
efficient method of solving \WSP, our simulation would suffer either
intractability or incorrect behavior.

\section{A New Approach}
\label{sec:approach}

Historically, evaluating the expressive power of access control schemes has
allowed researchers to separate schemes into equivalence classes and answer
important policy analysis questions. However, absolute assessments tell us very
little about the performance and suitability of a particular access control
scheme for a given application. In this section, we identify the access control
suitability analysis problem, develop a set of requirements that solutions to
this problem must satisfy, and overview our solution approach.

\subsection{Problem Definition}
\label{sec:approach:problem}

Given a formalization of an application's access control requirements, we
postulate that assessing the \emph{suitability} of an access control scheme for
that application will involve two classes of suitability measures:
\emph{expressiveness} and \emph{cost}. As such, suitability analysis is
necessarily a two-phased process.

In the first phase, one must ensure that candidate schemes for use within an
application are expressive enough to safely meet the needs of the application;
that is, whether the candidate schemes can admit \emph{at least} the policies
required by the application. In this expressiveness phase, the analyst
formalizes the candidate access control schemes, the operations required by the
application, and the set of properties that a safe implementation must satisfy.
Examples of potential implementation requirements range from simply enforcing
the same accesses to ensuring a strict bisimulation over state transitions. Upon
completion of this phase, the analyst should be able to narrow down the list of
schemes to those that are expressive enough to operate within the application
while satisfying all required properties.

The notion of costs, on the other hand, requires examining ordered measures of
suitability such as administrative overheads, workflow throughput, or degree of
reliance on system extensions (e.g., to increase expressiveness) that result
from the choice of a particular candidate access control scheme. In the cost
analysis phase, the analyst formalizes the cost measures of interest, the
expected usage of the access control system, and the expected costs of
individual actions within each scheme. This information can be used to conduct a
cost analysis that determines a partial order over the candidate schemes that
expresses their relative suitability to the application with respect to the cost
measures of interest.

More formally, we address the following problem:
\begin{problem}[Suitability Analysis]
Given an access control workload \(W\), a set of candidate access
control schemes \(\mathcal{S} = \set{\scm{S}_1, \ldots, \scm{S}_n}\), a notion
of safe implementation \(\mathcal{I}\), and a set of ordered cost measures
\(\mathcal{C} = \set{C_1, \ldots, C_m}\), determine:
\begin{enumerate}
\item[(i)] the subset \(\mathcal{S}^\prime \subseteq \mathcal{S}\) of schemes
that admit implementations of \(W\) preserving \(\mathcal{I}\)
\item[(ii)] the schemes within \(\mathcal{S}^\prime\) whose cost assessments
are optimal within the lattice \(C_1 \times \cdots \times C_m\)
\end{enumerate}
\end{problem}

\subsection{Solution Requirements}
\label{sec:approach:requirements}

We now explore requirements for suitability analysis frameworks. First, we
consider requirements in how an access control workload (\(W\)) is represented.
These requirements ensure that a suitability analysis framework is capable of
modeling the tasks carried out within an organization, and the interactions
required to support and process these tasks. Considering both facets of a
workload is critical, as neither one alone can fully parameterize the behavior
of an organization.\footnote{For instance, the well-documented shortcomings of
the U.S. military's access control scheme result not from some core inability to
process data, but instead from overheads associated with scaling these processes
to support high volumes of data and dynamic sharing
patterns~\cite{airforce,jason}.}

\begin{itemize}

\item \hypertarget{req:DE}{\emph{Domain exploration:}} Large organizations are
complex systems with subtle interactions. The emergent behaviors of such systems
may not be captured during the static process of workload specification. It must
be possible to efficiently explore many initial conditions (e.g., types of
actors, operations supported, organization size, and operation distributions) to
examine the effects of various levels of concurrency and resource limitation.

\item \hypertarget{req:CI}{\emph{Cooperative interaction:}} Tasks within large
organizations typically require the interaction of many individuals. To model
these interactions, a suitability analysis framework should support the use of
operational workflows, as well as constraints on their execution (e.g., to model
separation or binding of duty).

\end{itemize}

Next, we ensure that the analyst is able to tune the suitability analysis
framework to meet the specific needs of her application. 
believe that For maximum flexibility, it must be possible to choose the metrics
used to assess the suitability of an access control scheme for a given workload.
This should include both the binary metrics used in expressiveness evaluation
(\(\mathcal{I}\)) and the ordered metrics used in cost evaluation
(\(\mathcal{C}\)).

\begin{itemize}

\item \hypertarget{req:TS}{\emph{Tunable safety:}} Given a particular workload
and scheme, there may be many different ways for the scheme to implement the
workload. Without enforcing structure on the mapping encoding this
implementation, even the most under-expressive schemes can appear to implement a
workload~\cite{tripli}. However, as mentioned in \cref{sec:approach:problem},
the particular properties that any given implementation is required to satisfy
will depend on the application in which the access control system will be
utilized.

\item \hypertarget{req:TC}{\emph{Tunable cost:}} There is no single notion of
cost that is sensible for use in every analysis instance. As evidenced by a
recent NIST report~\cite{nist}, the costs that are relevant in evaluating access
control schemes are very application-dependent. Any suitability analysis
framework should be flexible enough to represent many types of costs, including
computational, communication, and administrative costs. It must also be possible
to examine multiple notions of cost simultaneously during an analysis.

\end{itemize}

Finally, we consider requirements that ensure that the suitability analysis
framework remains practical to use---in terms of runtime efficiency and
accuracy---even for large-scale application workloads.

\begin{itemize}

\item \hypertarget{req:EF}{\emph{Tractability:}} Steps of the analysis process
that can be automated should be done so using tractable (e.g., polynomial time
or fixed-parameter tractable) algorithms that remain feasible to use even for
very large systems.

\item \hypertarget{req:AC}{\emph{Accuracy:}} In many cases, full exploration of
all possible system traces for the purposes of cost analysis (e.g., via model
checking) will be impractical. As such, it must be possible to approximate the
expected error of costs obtained by exploring only a subset of these traces.

\end{itemize}

We have allowed these requirements to drive the development of our suitability
analysis framework, and will thus refer to them when justifying various design
decisions throughout the following sections. We discuss our framework's success
in achieving each of these requirements in \cref{sec:discussion:requirements}.

\subsection{Framework Overview}
\label{sec:approach:overview}

\begin{figure*}
\centering
\includegraphics[width=.75\textwidth]{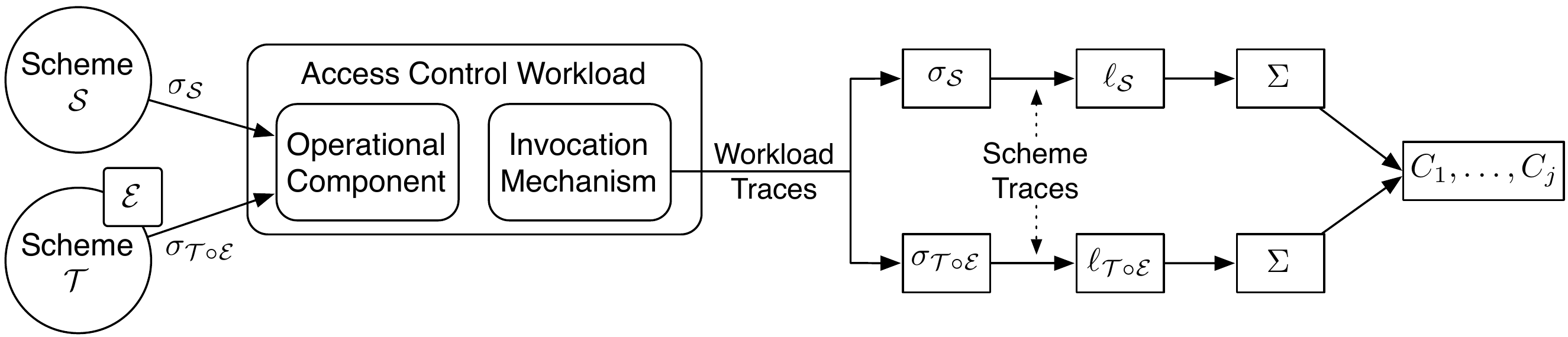}
\caption{Overview of an application-aware analysis framework for access
control\label{fig:overview}}
\end{figure*}

\Cref{fig:overview} presents a overview of the technical approach that we
propose for analyzing instances of the access control suitability analysis
problem. The first phase of this process is largely manual, and begins by
capturing the requirements of the application in what we call an \emph{access
control workload}. The workload includes a state machine that formalizes the
application's required protection state and supported commands and queries
(\cref{sec:expressiveness:workloads}). In addition, the workload contains a
specification of the expected utilization patterns of this functionality,
encoding individual behaviors using actor-based probabilistic models, and
collaborative tasks via constrained workflows (\cref{sec:cost:invocations}).
Candidate access control schemes are then specified as state machines, using the
same formalism as the operational aspects of the access control workload
(\cref{sec:expressiveness:schemes}).

The representational similarity between the workload's operational description
and the candidate schemes grants us the ability to construct
\emph{implementations} of the workload. This is done by mapping states,
commands, and queries in the workload to states, (sequences of) commands, and
queries in the candidate schemes (e.g., \(\sigma_\mathcal{S}\) and
\(\sigma_\mathcal{T}\) in \cref{fig:overview}). Security properties that must be
upheld by a workload implementation can be expressed as constraints on these
mappings, and proofs are manually constructed by the analyst indicating that
these properties are upheld (\cref{sec:expressiveness:implementation}). In this
way, the process of constructing implementations is a conceptual extension of
prior work in expressive power analysis
(e.g.,~\cite{early-sim,statetransition,osborn00tissec,sandhu92jcs,atam,safety}).
It may be necessary for a candidate scheme to be augmented to support such a
safe implementation (\cref{sec:extensions}). The result of the first phase of
analysis determines goal (i) of the Suitability Analysis problem: the subset of
schemes that admit implementations of the workload while preserving the
requisite security properties.

After the initial specification of cost measures to quantify the costs of
interest to the analyst (\cref{sec:cost:measures}) and cost functions to assign
cost distributions to actions taken within each candidate access control scheme
(\cref{sec:cost:functions}), Phase 2 of the suitability analysis process is
largely automated. Specifically, our approach makes use of Monte Carlo
simulation to carry out cost analysis: input parameters (e.g., number of users,
frequency of execution for various types of processes, etc.) are sampled from
appropriate distributions, the actor-based probabilistic model of workload
utilization is walked to generate concurrent traces of workload activities,
these activities are mapped to (sequences of) actions in each candidate scheme
being analyzed, actions are carried out, and the resulting costs are aggregated.
This process is repeated until either (i) adequate coverage of the input space
is obtained or (ii) adequate confidence intervals can be placed on the costs for
specific points within this input space (\cref{sec:cost:simulation}). The result
of this phase of analysis determines goal (ii) of the Suitability Analysis
Problem: the set of schemes whose cost assessments are optimal within the
lattice formed by the collection of all costs measures.

\section{Phase 1: Expressiveness Evaluation}
\label{sec:expressiveness}

In this section, we discuss the first phase of suitability analysis,
expressiveness evaluation. In this phase, the analyst formalizes the workload
and the candidate access control schemes, then constructs expressiveness
mappings to ensure that each scheme has the expressiveness necessary to properly
implement the workload.

\subsection{Formalizing Access Control Schemes}
\label{sec:expressiveness:schemes}

At the heart of an access control system is the access control \emph{model}, the
collection of data structures used to store the information needed to make
access control decisions. An access control model is formalized as a set of
\emph{states}, the possible configurations of these data structures. An access
control \emph{scheme}, then, defines the set of commands and queries that can be
used to interact with the model's states. Lastly, an access control
\emph{system} is an instantiation of a scheme, defining the subset of the
scheme's commands that are immediately available, as well as an initial state.
Previous work has shown distinctions in expressiveness between schemes with
identical models but different commands~\cite{statetransition,tripli} or
queries~\cite{atam,tripli}. However, there seems to be little benefit in
including a system's initial state in an analysis, since generalizing over
\emph{all} states allows us to make stronger claims about its properties. Thus,
in this paper, our analysis considers access control \emph{schemes}.

Our particular formalism is adapted from prior work~\cite{tripli}, and
represents an access control scheme as a state transition system operating over
a set of protection states \(\Gamma\). States in \(\Gamma\) contain all
information necessary for the operation of the access control scheme (e.g., sets
of principals, objects, roles, etc.) \emph{Queries} provided by the scheme
enable inspection of this state, while \emph{commands} enable transitions
between states. We now more formally define these concepts.

\begin{definition}[Query]
\label{def:query}
Given a set of access control states, \(\Gamma\), an \emph{access control query}
over \(\Gamma\) is a question that can be asked of an access control system,
defined as \(q = \tup{n, P, {\entails}}\), where:
\begin{itemize}
\item \(n\) names the query
\item \(P = \tup{P_1, \ldots, P_j}\) is the set of parameter spaces from which
the query's \(j\) parameters are drawn (e.g., the set of subjects, objects,
roles), where \(p_1 \in P_1\) represents the entity executing the
query.\footnote{The first parameter of a query or command represents the
executing entity. For queries, this allows the access control scheme to respond
differently to different queriers (e.g., a user may not be allowed to find out
the existence of another user's documents). For commands, this allows the scheme
to determine whether the requested execution is allowed.} We denote \(P_1 \times
\cdots \times P_j\) as \(P^*\).
\item \({\entails} : \Gamma \times P^* \to \set{\true, \false}\) is the
entailment relation that maps each state and parameterization to a truth value,
asserting that truth value of the query in the given state with the given
parameters
\end{itemize}
\end{definition}

Given access control query \(q = \tup{n_q, P_q, \entails_q}\) over \(\Gamma\),
state \(\gamma \in \Gamma\), and parameterization \(p \in P_q^*\), we say that
\(\gamma \entails q\prm{p_1, \ldots, p_j}\) to indicate that
\({\entails_q}\prm{\gamma, p_1, \ldots, p_j} = \true\). To better explain the
intuition behind queries, we present the following example.

\begin{example}
\label{ex:query}
Consider the access control state \(\gamma\), in which Alice has no
access to user Bob's document, \textstt{foo}. Bob may choose to
verify whether Alice has access to \textstt{foo} by asking a query
where \(n =\) \textstt{can\_read}, which takes parameters from spaces
\(\tup{U, U, D}\), the sets of users, users, and documents,
respectively, and whose entailment \(\entails\) maps \(\gamma\) and
the parameterization \textstt{(Bob, Alice, foo)} to \false,
indicating that Alice does not have the access in question.
\end{example}

While queries are used to inspect access control states, commands are used to
modify these states.

\begin{definition}[Command]
\label{def:command}
Given a set of access control states, \(\Gamma\), an \emph{access control
command} over \(\Gamma\) is the mechanism for state transformations, defined as
\(c = \tup{n, P, e}\), where:
\begin{itemize}
\item \(n\) names the command
\item \(P = \tup{P_1, \ldots, P_j}\) is the set of parameter spaces from which
the command's \(j\) parameters are drawn (e.g., the set of subjects, objects,
roles), where \(p_1 \in P_1\) represents the entity executing the command. We
denote \(P_1 \times \cdots \times P_j\) as \(P^*\).
\item \(e : \Gamma \times P^* \to \Gamma\), the effect mapping, which maps each
state and parameterization to the state that results from the execution of the
command with the given parameters in the given state.
\end{itemize}
\end{definition}

We now give an example command to clarify \cref{def:command}.

\begin{example}
\label{ex:command}
Consider the state \(\gamma\) from \cref{ex:query}. Bob may choose to grant
Alice read access to \textstt{foo} by executing a command where \(n =\)
\textstt{grant\_read}, which takes parameters from spaces \(\tup{U, U, D}\), the
sets of users, users, and documents, respectively, and whose effect mapping
\(e\) maps \(\gamma\) and the parameterization \textstt{(Bob, Alice, foo)} to
\(\gamma\Prime\), an identical state with the exception of Alice being granted
read access to \textstt{foo}.
\end{example}

Given a set of access control commands \(\Psi\) over \(\Gamma\), and two states
\(\gamma, \gamma^\prime \in \Gamma\), we say that \(\gamma \mapsto_\Psi
\gamma^\prime\) if there exists a command \(\psi = \tup{n, P, e} \in \Psi\) and
a parameterization \(p \in P^*\) such that \(e\prm{\gamma, p} = \gamma^\prime\).
We use \(\gamma \Mapsto_\Psi \gamma^\prime\) to denote the transitive closure of
\(\mapsto_\Psi\): i.e., there exists a sequence of commands \(\tup{\psi_1 =
\tup{n_1, P_1, e_1}, \ldots, \psi_k = \tup{n_k, P_k, e_k}}\) and a sequence of
parameterizations \(\tup{p_1 \in P_1^*, \ldots, p_k \in P_k^*}\) of these
commands such that \(e_k\prm{\ldots e_1\prm{\gamma, p_1}, \ldots, p_k} =
\gamma^\prime\). We can now precisely formalize an access control scheme.

\begin{definition}[Scheme]
\label{def:scheme}
An \emph{access control scheme} is a state transition system \(\scm{S} =
\tup{\Gamma, \Psi, Q}\), where \(\Gamma\) is the set of access control states,
\(\Psi\) is the set of commands over \(\Gamma\), and \(Q\) is the set of queries
over \(\Gamma\).
\end{definition}

We now give an example to demonstrate the structure.

\begin{example}
\label{ex:scheme}
DAC is the discretionary access control scheme, defined by \(\scm{D} =
\tup{\Gamma^\scm{D}, \Psi^\scm{D}, Q^\scm{D}}\). Its states, \(\Gamma^\scm{D}\),
are defined by the sets \(\tup{S, O, R, M}\), where:
\begin{itemize}
\item \(S\) is the set of subjects
\item \(O\) is the set of objects
\item \(R\) is the set of rights
\item \(M: S \times O \to 2^R\) is the access matrix
\end{itemize}

DAC's commands, \(\Psi^\scm{D}\), include the following.
\begin{itemize}
\item \textstt{CreateObject(\(S\), \(O\))}, which adds an object
\item \textstt{DestroyObject(\(S\), \(O\))}, which deletes an object
\item \textstt{CreateSubject(\(S\), \(S\))}, which adds an subject
\item \textstt{DestroySubject(\(S\), \(S\))}, which deletes an subject
\item \textstt{Grant(\(S\), \(S\), \(O\), \(R\))}, which grants a right over an
object to a subject
\item \textstt{Revoke(\(S\), \(S\), \(O\), \(R\))}, which revokes a right over
an object from a subject
\end{itemize}

Finally, DAC's queries, \(Q^\scm{D}\), include the following.
\begin{itemize}
\item \textstt{Access(\(S\), \(S\), \(O\), \(R\))}, which asks whether a user
has a right over an object
\item \textstt{SubjectExist(\(S\), \(S\))}, which asks whether a subject exists
\end{itemize}
\end{example}

\subsection{Formalizing Workloads}
\label{sec:expressiveness:workloads}

An access control workload describes an abstraction of the access control needs
of an environment. A workload specifies both an \emph{operational} component
describing the relevant operations that must be supported, as well as an
\emph{invocation} component that describes how those operations are expected to
be used. The operational component can be viewed as the collection of high-level
commands and queries that the application would like to execute, and hence can
be formalized as an (abstract) access control state machine using
\cref{def:scheme}. We note that, while formalized in the same way, workloads and
schemes differ in their intention. While a scheme represents a functioning piece
of software, a workload is built by the analyst to represent the higher-level
\emph{desired} functionality of a system, without necessarily being appropriate
for direct implementation. We discuss possible ways to more formally express
this difference in intention in \cref{sec:discussion}.

The invocation component describes the ways in which the system is typically
used; i.e., the order in which the high-level commands and queries are executed.
At a minimum, the invocation component should be able to dictate the
probabilities with which various commands are executed and queries are asked
during paths of execution. Our framework allows the invocation component to
remain flexible. We discuss the invocation mechanism and present a particular
instantiation of this concept in \cref{sec:cost:invocations}.

\begin{definition}[Workload]
\label{def:workload}
An \emph{access control workload} is defined by \(\tup{\scm{S}, I^\scm{S}}\),
where:
\begin{itemize}
\item \(\scm{S} = \tup{\Gamma, \Psi, Q}\) is an abstract access control scheme
and acts as the operational component
\item \(I^\scm{S}\) is an invocation mechanism over \(\scm{S}\), e.g., an
instance of \cref{def:invocation} (cf. \cref{sec:cost:invocations})
\end{itemize}
\end{definition}

Note that it is not always obvious how to transform an abstract description of a
desired access control policy into a machine-level specification for use as the
scheme component of a workload. We discuss this problem in
\cref{sec:discussion}. We now give an example of a workload operational
component.

\begin{example}
\label{ex:workload}
Consider an environment that grants users discretionary control over their own
resources, but allows administrators to have full access to any object. This
workload, \(W_\scm{A} = \tup{\scm{A}, I^\scm{A}}\), utilizes as its operational
component the administrative DAC scheme (ADAC), \(\scm{A}\). The ADAC scheme is
similar to the DAC scheme from \cref{ex:scheme}, but must also maintain the set
of administrators, who have full access to each object in the system. It is
defined as \(\scm{A} = \tup{\Gamma^\scm{A}, \Psi^\scm{A}, Q^\scm{A}}\). Its
states, \(\Gamma^\scm{A}\), are defined by the sets \(\tup{S, A, O, R, M}\),
where:
\begin{itemize}
\item \(S\) is the set of subjects
\item \(A \subseteq S\) is the set of administrators
\item \(O\) is the set of objects
\item \(R\) is the set of rights
\item \(M: S \times O \to 2^R\) is the access matrix
\end{itemize}

ADAC's commands, \(\Psi^\scm{A}\), include the following.
\begin{itemize}
\item \textstt{CreateObject(\(S\), \(O\))}, which adds an object
\item \textstt{DestroyObject(\(S\), \(O\))}, which deletes an object
\item \textstt{CreateSubject(\(S\), \(S\))}, which adds an subject
\item \textstt{DestroySubject(\(S\), \(S\))}, which deletes an subject
\item \textstt{Grant(\(S\), \(S\), \(O\), \(R\))}, which grants a right over an
object to a subject
\item \textstt{Revoke(\(S\), \(S\), \(O\), \(R\))}, which revokes a right over
an object from a subject
\item \textstt{GrantAdmin(\(S\), \(S\))}, which grants administrative status
\item \textstt{RevokeAdmin(\(S\), \(S\))}, which revokes administrative status
\end{itemize}

Finally, ADAC's queries, \(Q^\scm{A}\), include the following.
\begin{itemize}
\item \textstt{Access(\(S\), \(S\), \(O\), \(R\))}, which asks whether a user
has a right over an object
\item \textstt{SubjectAdmin(\(S\), \(S\))}, which asks whether a subject is an
administrator
\item \textstt{SubjectExist(\(S\), \(S\))}, which asks whether a subject exists
message
\end{itemize}
\end{example}

\subsection{Implementing a Workload in a Scheme}
\label{sec:expressiveness:implementation}

Once the analyst selects an appropriate set of candidate access control schemes,
she must verify each scheme's ability to safely execute the operations required
by the workload. To do so, the analyst demonstrates the existence of mappings
from the workload's operational component to each of the candidate access
control schemes. These mappings provide a translation from the workload's state
representation and actions to those of each candidate scheme. Moreover, these
mappings are used to guarantee that the safety properties of the workload are
preserved in each candidate scheme.

\begin{definition}[Implementation]
\label{def:implementation}
Given an access control workload \(W = \tup{\scm{W}, I^\scm{W}}\) in which
\(\scm{W} = \tup{\Gamma^\scm{W}, \Psi^\scm{W}, Q^\scm{W}}\), and an access
control scheme \(\scm{S} = \tup{\Gamma^\scm{S}, \Psi^\scm{S}, Q^\scm{S}}\), an
\emph{implementation} of \(\scm{W}\) in \(\scm{S}\) is a set of mappings
\(\sigma = \tup{\sigma_\Gamma, \sigma_\Psi, \sigma_Q}\), where:
\begin{itemize}
\item \(\sigma_\Gamma: \Gamma^\scm{W} \to \Gamma^\scm{S}\) is the state mapping
\item \(\sigma_\Psi: \Psi^\scm{W} \to \prm{\Psi^\scm{S}}^+\) is the command
mapping (each \(\psi \in \Psi^\scm{W}\) is mapped to a sequence \(\tup{\psi_1,
\ldots, \psi_k}\), where each \(\psi_i\) is a command in \(\Psi^{S}\))
\item \(\sigma_Q: Q^\scm{W} \to Q^\scm{S}\) is the query mapping
\end{itemize}
\end{definition}

While \cref{def:implementation} describes the structure of an implementation,
the properties that such an implementation must satisfy are defined by the
application in question. One particularly natural set of properties that an
implementation might be required to preserve is the set of \emph{compositional
security analysis instances}~\cite{tripli}. The compositional security analysis
instance is a generalization of simple safety analysis~\cite{hru} to arbitrary
quantified boolean formulas over queries.

\begin{definition}[Compositional Security Analysis]
\label{def:csai}
Given an access control scheme \(\scm{S} = \tup{\Gamma, \Psi, Q}\), a
\emph{compositional security analysis instance} has the form \(\tup{\gamma,
\varphi, \Pi}\), where \(\gamma \in \Gamma\) is a state, \(\varphi\) is a
propositional formula over \(Q\), and \(\Pi \in \set{{\exists}, {\forall}}\) is
a quantifier. If \(\Pi = {\exists}\), the instance asks whether there exists
\(\gamma\Prime \in \Gamma\) such that \(\gamma \Mapsto_\Psi \gamma\Prime\) and
\(\gamma\Prime \entails \varphi\) (whether \(\varphi\) is \emph{possible}). If
\(\Pi = {\forall}\), the instance asks whether for every \(\gamma\Prime \in
\Gamma\) such that \(\gamma \Mapsto_\Psi \gamma\Prime\), \(\gamma\Prime \entails
\varphi\) (whether \(\varphi\) is \emph{necessary}).
\end{definition}

The compositional security analysis instance is a natural language for
expressing many types of practical policies (e.g., ``Bob cannot edit payroll
data while his wife, Alice, is also an employee.''~\cite{tripli}). An
implementation that preserves all compositional security analysis instances is
said to be \emph{strongly security-preserving}. Unfortunately, directly proving
that a mapping is strongly security-preserving can be quite expensive, as it
requires the analysis of all possible compositional security analysis instances.
For this reason, Tripunitara and Li presented the \emph{state-matching
reduction}~\cite{tripli}, a type of mapping that is defined by a set of
structural properties that are necessary and sufficient for being strongly
security-preserving. Using the state-matching reduction is advantageous, as it
is easier to prove that a mapping satisfies these structural requirements than
it is to directly prove that it preserves all compositional security analysis
instances. We now present the state-matching implementation, a type of
implementation based on (and maintaining the security properties of) Tripunitara
and Li's state-matching reduction.

\begin{definition}[State-Matching Implementation]
\label{def:smimplementation}
Given an access control workload \(W = \tup{\scm{W}, I^\scm{W}}\) in which
\(\scm{W} = \tup{\Gamma^\scm{W}, \Psi^\scm{W}, Q^\scm{W}}\), an access control
scheme, \(\scm{S} = \tup{\Gamma^\scm{S}, \Psi^\scm{S}, Q^\scm{S}}\), and an
implementation \(\sigma = \tup{\sigma_\Gamma, \sigma_\Psi, \sigma_Q}\) of
\(\scm{W}\) in \(\scm{S}\), we say that two states \(\gamma^\scm{W}\) and
\(\sigma_\Gamma\prm{\gamma^\scm{W}} = \gamma^\scm{S}\) are \emph{equivalent}
with respect to the implementation \(\sigma\) (and denote this equivalence as
\(\gamma^\scm{W} \eqv{\sigma} \gamma^\scm{S}\)) when for every \(q^\scm{W} =
\tup{n, P, {\entails}} \in Q^\scm{W}\) (with \(q^\scm{S} =
\sigma_Q\prm{q^\scm{W}}\)) and every \(p^\scm{W} \in P^*\) (with \(p^\scm{S} =
\sigma_\Gamma\prm{p^\scm{W}}\)), \(\gamma^\scm{W} \entails
q^\scm{W}\prm{p^\scm{W}}\) if and only if \(\gamma^\scm{S} \entails
q^\scm{S}\prm{p^\scm{S}}\).

An implementation \(\sigma\) of \(\scm{W}\) in \(\scm{S}\) is said to be a
\emph{state-matching implementation} if for every \(\gamma^\scm{W} \in
\Gamma^\scm{W}\), with \(\gamma^\scm{S} = \sigma_\Gamma\prm{\gamma^\scm{W}}\),
the following two properties hold:
\begin{enumerate}
\item For every state \(\gamma^\scm{W}_1 \in \Gamma^\scm{W}\) such that
\(\gamma^\scm{W} \Mapsto_{\Psi^\scm{W}} \gamma^\scm{W}_1\), there exists a state
\(\gamma^\scm{S}_1 \in \Gamma^\scm{S}\) such that \(\gamma^\scm{S}
\Mapsto_{\Psi^\scm{S}} \gamma^\scm{S}_1\) and \(\gamma^\scm{W}_1 \eqv{\sigma}
\gamma^\scm{S}_1\).
\item For every state \(\gamma^\scm{S}_1 \in \Gamma^\scm{S}\) such that
\(\gamma^\scm{S} \Mapsto_{\Psi^\scm{S}} \gamma^\scm{S}_1\), there exists a state
\(\gamma^\scm{W}_1 \in \Gamma^\scm{W}\) such that \(\gamma^\scm{W}
\Mapsto_{\Psi^\scm{W}} \gamma^\scm{W}_1\) and \(\gamma^\scm{W}_1 \eqv{\sigma}
\gamma^\scm{S}_1\).
\end{enumerate}
\end{definition}

The following \lcnamecref{thm:smimplementation} demonstrates the power of this
notion of implementation. The proof of \cref{thm:smimplementation} can be found
in \cref{appx:proofs:smimplementation}.

\begin{proposition}
\label{thm:smimplementation}
Given an access control workload \(W = \tup{\scm{W}, I^\scm{W}}\) in which
\(\scm{W} = \tup{\Gamma^\scm{W}, \Psi^\scm{W}, Q^\scm{W}}\), an access control
scheme, \(\scm{S} = \tup{\Gamma^\scm{S}, \Psi^\scm{S}, Q^\scm{S}}\), and an
implementation \(\sigma = \tup{\sigma_\Gamma, \sigma_\Psi, \sigma_Q}\) of
\(\scm{W}\) in \(\scm{S}\), \(\sigma\) is a state-matching implementation if and
only if it is strongly security-preserving; that is, every compositional
security analysis instance in \(\scm{W}\) is true if and only if the image of
the instance under \(\sigma\) is true in \(\scm{S}\).
\end{proposition}

\begin{proofsketch}
In proving \cref{thm:smimplementation}, we utilize a previous result from
Tripunitara and Li~\cite{tripli}, which states that a mapping is a
state-matching \emph{reduction} if and only if it is strongly
security-preserving. We show that, if an implementation \(\sigma\) is a
state-matching implementation of \(\scm{W}\) using \(\scm{S}\), there exist
\(\scm{W}\Prime\) and \(\scm{S}\Prime\), schemes under Tripunitara and Li's
definition that are equivalent to \(\scm{W}\) and \(\scm{S}\), respectively. We
prove that this equivalence is strongly security-preserving, and then than the
implementation corresponds to a state-matching reduction, \(\sigma\Prime\), from
\(\scm{W}\Prime\) to \(\scm{S}\Prime\). This proves that \(\sigma\Prime\) is
strongly security-preserving, and finally that \(\sigma\) is. (This ends the
``if'' direction.)

For the ``only if'' direction, we consider an implementation that is strongly
security-preserving. Again, we show that this implementation corresponds to a
Tripunitara-Li mapping, this time deducing a state-matching reduction from the
strongly security-preserving mapping. We show that this state-matching reduction
is equivalent to our implementation, and thus that the implementation is a
state-matching implementation.
\end{proofsketch}

As a result of these security guarantees, we will require that, in order for an
implementation of workload \(W\) in scheme \(\scm{S}\) to be considered a
\emph{safe implementation}, that implementation be state-matching. In addition,
in this work we restrict implementations to preserve the semantics of the
respective model with respect to accesses. We accomplish this by requiring the
query mapping to map the access queries in the workload to the access queries in
the scheme, thus forcing the implementation of the workload to use the same
procedure for deciding accesses as the scheme uses. These restrictions on
implementations are examples of how an application can dictate how a scheme can
be used for the purpose of ``simulating'' a workload. Previous work on various
expressiveness properties (e.g.,~\cite{logicalfoundations}) can provide
guidelines for choosing the type of implementation that best suits
expressiveness evaluation in the context of the relevant application. Exploring
the full range of possible implementation properties and their corresponding
implementation structure is a subject of future work (see
\cref{sec:discussion:fw}).

Consider ADAC from \cref{ex:workload} and DAC from \cref{ex:scheme}. Despite the
similarities between them, DAC does not seem to admit a state-matching
implementation of ADAC using DAC, since DAC is unable to maintain information
about the set of administrators. For scenarios such as this, we explore the
ability to \emph{extend} schemes in \cref{sec:extensions}.

\section{Phase 2: Cost Analysis}
\label{sec:cost}

As discussed in \cref{sec:approach}, our approach to suitability analysis is
two-phased. In the previous section, we discussed the first phase,
expressiveness evaluation, which allows the analyst to ensure that all candidate
schemes are expressive enough to safely meet the needs of the application. In
this section, we present the details of the second phase, cost analysis, which
explores more quantitative suitability measures.

\subsection{Actor-based Invocation Mechanism}
\label{sec:cost:invocations}

Recall from \cref{sec:expressiveness:workloads} that an access control workload,
\(W = \tup{\scm{W}, I^\scm{W}}\), consists of an operational component,
\(\scm{W}\), and an invocation mechanism over \(\scm{W}\). The invocation
mechanism describes the expected usage of the access control system within the
application being described. Most simply, this mechanism could be a recorded
trace of operations that will be ``played back'' while its operating costs are
recorded. However, this violates several of the requirements from
\cref{sec:approach:requirements}. For example, \reqDE{} requires that we are
able to alter input parameters. While this type of static invocation mechanism
does not preclude the varying of the initial access control state, it does not
allow the trace to react to these changes (e.g., more users typically means more
frequent execution of commands and queries).

To overcome these types of issues, we define an invocation mechanism utilizing
the concepts of \emph{actors} carrying out \emph{actions} within the system.
Actors are human users, daemons, and other entities that act on the access
control system. We determine the set of actors by extracting the active entities
from an access control state. We express the various ways in which actors
cooperate to complete a task using \emph{constrained workflows}. Within this
structure, \emph{workflows} express the dependency between related actions, and
\emph{constraints} express the restrictions placed on which user can execute
each action in a task. Finally, \emph{actor machines} express the behavior
models for the actors within the constraints imposed upon them by the
constrained workflow. Together, these structures enable the modeling and
simulation of complex and concurrent behaviors of the entities that are active
within a given workload.

We now formalize the notion of an \emph{action}, which is the basic component of
work executed by an actor in the system. An action is a partially parameterized
command or query. The free parameters are assigned statically by the executing
actor's behavior machine or dynamically during execution.

\begin{definition}[Action]
\label{def:action}
Let \(\scm{S} = \tup{\Gamma, \Psi, Q}\) be an access control scheme and
\(\mathfrak{V}\) a set of variable symbols. An \emph{access control action} from
scheme \(\scm{S}\) is defined as \(\alpha = \tup{n, a, C}\), where:
\begin{itemize}
\item \(n\) names the action
\item \(a \in \Psi \cup Q \cup \set{\varnothing}\) is the command or query
(whose set of parameter spaces is \(P = \tup{P_1, \ldots, P_j}\)) that the
action executes. A value of \(\varnothing\) indicates that the action does not
execute a command or query in the access control system.
\item \(C \in \prm{P_1 \cup \mathfrak{V}} \times \ldots \times \prm{P_j \cup
\mathfrak{V}} \cup \set{\varnothing}\) is the partial parameterization. For each
parameter in \(P\), \(C\) specifies a parameter value or a variable from
\(\mathfrak{V}\). For actions that do not execute a command or query, \(C\),
like \(a\), is \(\varnothing\).
\end{itemize}
\end{definition}

Although actions that do not execute commands or queries within a scheme seem
counter-intuitive, they become important in the context of workflows that link
together multi-user tasks within a workload. To describe various dependencies
between actions (executed by a single actor or a set of actors in coordination),
we present the notion of a \emph{constrained access control workflow}, which
organizes the execution of actions. Formally, this structure specifies the
partial order describing action dependence as well as a set of constraints that
restrict the set of users that can execute various actions.

\begin{definition}[Constrained Workflow]
\label{def:constrainedworkflow}
Let \(\scm{S} = \tup{\Gamma, \Psi, Q}\) be an access control scheme and
\(\mathfrak{A}\) a set of access control actors within \(\scm{S}\). We say that
\(W = \tup{A, {\prec}, C}\) is an \emph{constrained access control workflow}
over the scheme \(\scm{S}\), where:
\begin{itemize}
\item \(A\) is the set of actions from scheme \(\scm{S}\)
\item \({\prec} \subset A \times A\) is the partial order describing the
dependency relation between actions. If \(\alpha_1 \prec \alpha_2\), then
\(\alpha_2\) depends on \(\alpha_1\). That is, \(\alpha_2\) cannot be executed
unless a corresponding execution of \(\alpha_1\) has occurred.
\item \(C\) is the set of constraints, where each constraint is of the form
\(\tup{\rho, \alpha_1, \alpha_2}\). Here, \(\rho\) is a binary operator of the
form \(\mathfrak{A} \times \mathfrak{A} \to \set{\true, \false}\). A constraint
restricts execution of actions \(\alpha_1\) and \(\alpha_2\) to actors who
satisfy the binary operator \(\rho\). For example, \(\tup{{\neq}, \alpha_1,
\alpha_2}\) says that \(\alpha_1\) and \(\alpha_2\) must be executed by
different actors.
\end{itemize}
\end{definition}

Within a workflow \(\tup{A, {\prec}, C}\), subsets of \(A\) that are pairwise
disjoint with respect to \(\prec\) are referred to as \emph{tasks}, and each
action within a task is referred to as a \emph{step} in that task.

\begin{figure}
\centering
\includegraphics[width=.8\columnwidth]{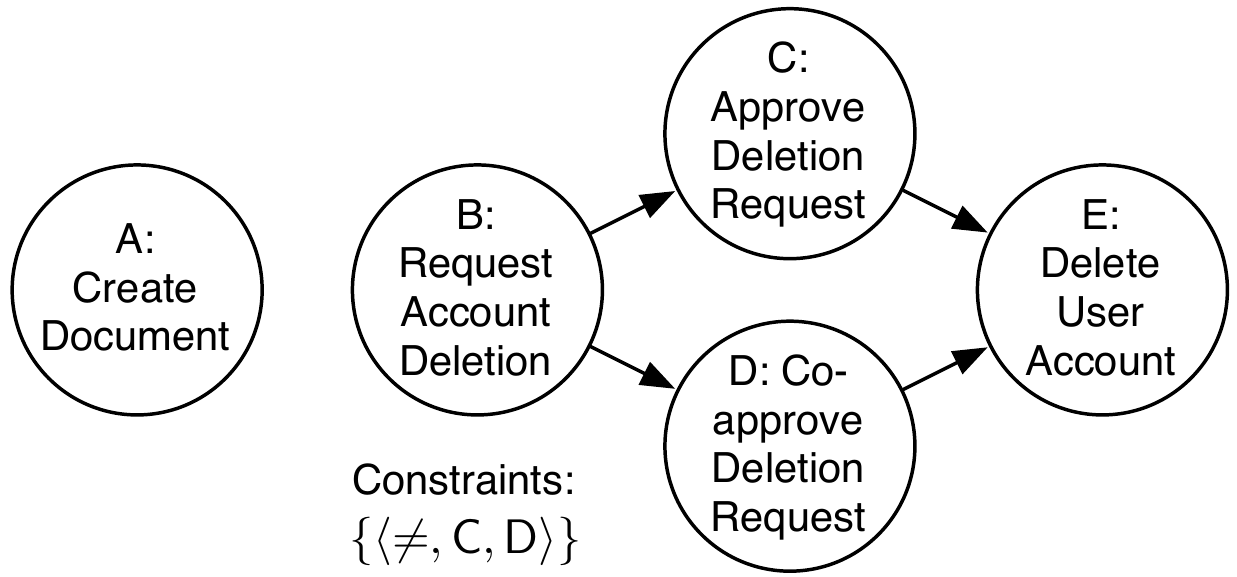}
\caption{An example of a constrained workflow\label{fig:workflow}}
\end{figure}

\begin{example}
\label{ex:workflow}
\Cref{fig:workflow} displays a constrained workflow that includes two tasks,
corresponding to document creation and account deletion. The former is a
degenerate task containing a single action. Execution of this action is thus
effectively unconstrained by the workflow. However, the task of deleting an
account requires the approval of two different administrators. The workflow
allows administrators to approve deletion of accounts only after the deletion
request, and the deletion can only happen after it has been approved twice.
Furthermore, the example constraint requires that the two approval actions be
executed by two different administrators.
\end{example}

We draw a distinction between the use of workflows here and their use in, e.g.,
\textsf{\smaller R^{2}BAC}~\cite{wang2007esorics}. While there exist access
control schemes with the native ability to enforce workflow semantics, our goal
is to represent workflow properties at the access control workload level, and
utilize implementations of these workloads to ensure tasks execute according to
these higher-level constraints. This allows us to utilize even simple access
control schemes while still constraining actors to work within such
organizational policies as separation of duty.

To describe the patterns with which actors execute actions, we employ
\emph{actor machines}, which are state machines that describe each actor's
behavior. Each state in the machine is labeled with an action name and a
refining parameterization (which assigns values to parameters that were left as
variables in the action specification). Transitions in this state machine are
labeled with \emph{rates} akin to those used in continuous-time Markov processes
(e.g.,~\cite{liggett2010continuous}). We then generate representative traces of
actor behavior by probabilistically walking this machine, following transitions
with probabilities proportional to their rates.

\begin{definition}[Actor Machine]
\label{def:actormachine}
Let \(\scm{S} = \tup{\Gamma, \Psi, Q}\) be an access control scheme, \(W =
\tup{A, {\prec}, C}\) a constrained workflow over \(\scm{S}\), and
\(\mathfrak{V}\) a set of variable symbols. An \emph{actor
machine} for \(\scm{S}\) and \(G\) is the state machine \(\tup{S, \Phi, R}\),
where:
\begin{itemize}
\item \(S\) is the set of states
\item \(\Phi: S \to A \times \prm{P_1 \cup \mathfrak{V}} \times \ldots \times
\prm{P_j \cup \mathfrak{V}}\) labels each state with an action and a
\emph{refinement} of the action's parameterization (i.e., parameters assigned by
the action remain the same, while parameters not assigned by the action may be
assigned to values)
\item \(R: S \times S \to \mathbb{R}\) is the set of rates of transitioning from
state to state
\end{itemize}
\end{definition}

The semantics of the execution of an actor machine are as follows. \(R\)
describes the rates of transitioning from one state to another. In order to
achieve the Markov property, the time spent waiting to exit a state is
exponentially distributed, with rate parameter proportional to the sum of the
rates of all exiting transitions. When executing, an actor carries out a state's
action upon \emph{entering} the state. We distinguish between \emph{entering} a
state and \emph{remaining in} a state. Transitioning from a state back to itself
will result in a re-execution of the state's action. Remaining in a state while
waiting for the next transition to trigger will \emph{not} result in a
re-execution.

\begin{figure}
\centering
\includegraphics[width=.9\columnwidth]{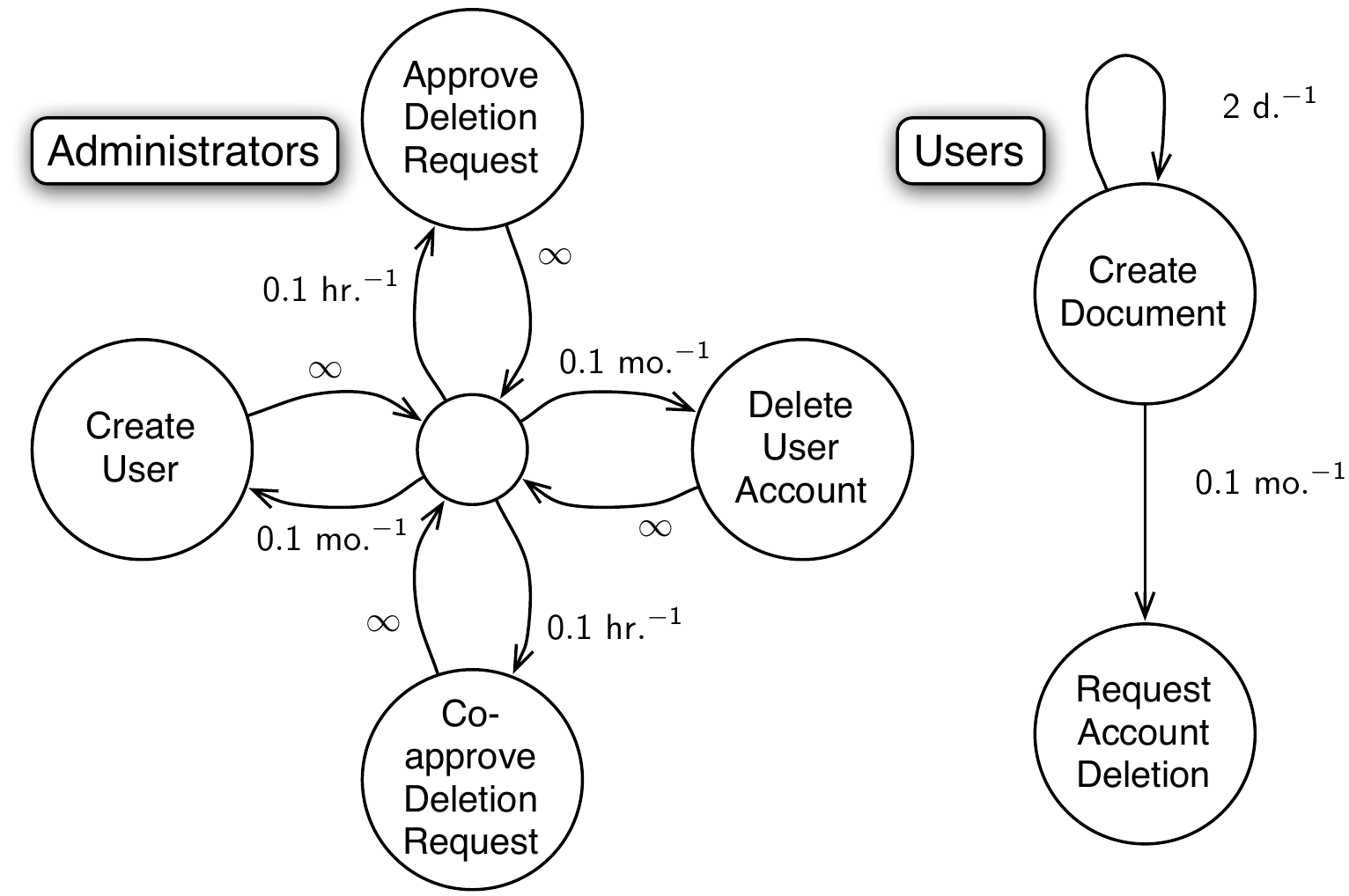}
\caption{Example actor machines\label{ex:actormachine}}
\end{figure}

Example actor machines are demonstrated in \cref{ex:actormachine}. In this
example, we classify users into two categories of actors: administrators and
non-administrators. The former add users and approve and execute user deletions,
while the latter generate documents and occasionally request to be deleted. Due
to the labeled rates on this machine, each administrator creates users at the
expected rate of one per month, and roughly 10\% of non-administrative users
request deletion each month. High rates on transitions leading to, e.g.,
approving deletions indicate the rate at which these actions will be executed
\emph{when enabled}. Transitions labeled with \(\infty\) occur immediately after
completing the preceding action.

Our actor-based invocation mechanism that will complete \cref{def:workload},
then, consists of a constrained workflow, a set of actor machines, and a method
for extracting the current actors and their assigned machines from an access
control state.

\begin{definition}[Actor-Based Invocation]
\label{def:invocation}
Let \(\scm{S} = \tup{\Gamma, \Psi, Q}\) be an access control scheme. We say that
\(I^\scm{S} = \tup{W, \mathfrak{A}, A, G_A, g}\) is an \emph{constrained, actor-based
access control invocation mechanism} over the scheme \(\scm{S}\), where:
\begin{itemize}
\item \(W\) is a constrained workflow over \(\scm{S}\)
\item \(\mathfrak{A}\) is the set of all actors
\item \(A : \Gamma \to \powerset\prm{\mathfrak{A}}\) is the actor relation,
mapping each access control state to the set of actors active in that state
\item \(G_A\) is the set of actor machines
\item \(g: \mathfrak{A} \to G_A\) is the actor machine assignment, mapping each
actor to its actor machine
\end{itemize}
\end{definition}

\subsection{Cost Measures}
\label{sec:cost:measures}

An important part of cost analysis is choosing relevant cost measures. These
measures should be representative of the ``problem'' (i.e., what types of cost
the analyst cares about), while also enabling the definition of a cost function
for each candidate scheme (see \cref{sec:cost:functions}). For example, while
``operational cost per day'' may be representative of access control evaluation
goals in industry, it is hard to assign costs in this measure to each fully
parameterized access control action. A measure such as ``average administrative
personnel-hours spent per access control operation,'' on the other hand, is more
easily quantified and enables the same types of analyses.

In this paper, we make no commitment to any particular cost measures but rather
develop an analysis framework that operates on any measure satisfying a number
of simple properties. A cost measure must include a set of elements representing
the costs, an associative and commutative operator that combines two costs to
produce another cost (e.g., addition), and a partial order for comparing costs.
Finally, we enforce that there are no ``negative'' costs.

\begin{definition}[Cost Measure]
\label{def:measure}
A \emph{cost measure} is defined by the ordered abelian monoid \(\mathbf{G} =
\tup{G, {\op}, {\cmp}}\), where \(G\) is the set of costs, \(\op\) is the
closed, associative, commutative accrual operator over \(G\) with identity
\(0_G\), and \(\cmp\) is a partial order over \(G\) such that \(\forall a, b \in
G: a \cmp a \op b \land b \cmp a \op b\).
\end{definition}

\Cref{def:measure} can be used to encode a variety of interesting access control
measures, including several of those noted in a recent NIST report on the
assessment of access control schemes~\cite{nist}. For example, costs like
``steps required for assigning and dis-assigning user capabilities'' and
``number of relationships required to create an access control policy'' can be
represented using the cost measure \(\tup{\mathbb{N}, {+}, {\leq}}\). Our notion
of measure is general enough to represent many other types of costs as well.
Measures for human work such as ``personnel-hours per operation'' and
``proportion of administrative work to data-entry work'' can be represented
using the cost measures \(\tup{\mathbb{Z}^{+}, {+}, {\leq}}\) and
\(\tup{\mathbb{Z}^{+} \times \mathbb{Z}^{+}, {+}, {\leq}}\), respectively.
Maximum memory usage can be represented using \(\tup{\mathbb{N}, {\max},
{\leq}}\).

A common desire is for an analyst to evaluate an access control scheme using
several different cost measures in parallel. Thus, we define a \emph{vector} of
cost measures.

\begin{definition}[Vector of Measures]
\label{def:vector}
Given cost measures
\(\mathbf{N}_1 = \tup{N_1, {\op[1]}, {\cmp[1]}}\),
\(\mathbf{N}_2 = \tup{N_2, {\op[2]}, {\cmp[2]}}\),
\ldots,
\(\mathbf{N}_i = \tup{N_i, {\op[i]}, {\cmp[i]}}\),
let \(\mathbf{M} = \tup{M, {\op[*]}, {\cmp[*]}}\) be the \emph{vector} of
cost measures \(\mathbf{N}_1, \mathbf{N}_2, \ldots, \mathbf{N}_i\), where:
\begin{itemize}
\item
  \(M = N_1 \times N_2 \times \cdots \times N_i\).
\item
  Given
  \(a_1, b_1 \in \mathbf{N}_1\),
  \(a_2, b_2 \in \mathbf{N}_2\),
  \ldots,
  \(a_i, b_i \in \mathbf{N}_i\),
  \(\tup{a_1, a_2, \ldots, a_i} \op[*]
    \tup{b_1, b_2, \ldots, b_i} =
    \tup{a_1 \op[1] b_1,
         a_2 \op[2] b_2,
         \ldots,
         a_i \op[i] b_i}
  \).
\item
  Given
  \(a_1, b_1 \in \mathbf{N}_1\),
  \(a_2, b_2 \in \mathbf{N}_2\),
  \ldots,
  \(a_i, b_i \in \mathbf{N}_i\),
  \(\tup{a_1, a_2, \ldots, a_i} \cmp[*]
    \tup{b_1, b_2, \ldots, b_i}
  \) if and only if
  \(a_1 \cmp[1] b_1 \land
    a_2 \cmp[2] b_2 \land
    \ldots \land
    a_i \cmp[i] b_i
  \).
\end{itemize}
\end{definition}

\Cref{def:vector} gives a simple way of combining several measures. As the
following \lcnamecref{thm:vector} states, a vector of cost measures is also a
cost measure, enabling the analyst to use a combination of measures within our
analysis framework. We prove \cref{thm:vector} in \cref{appx:proofs:vector}.

\begin{proposition}
\label{thm:vector}
Given cost measures \(\mathbf{N}_1, \mathbf{N}_2, \ldots, \mathbf{N}_i\) and
their associated cost vector, \(\mathbf{M}\), \(\mathbf{M}\) is a cost measure.
\end{proposition}

\begin{proofsketch}
Given the definition of measure, we know that all of \(\mathbf{N}_i\) satisfy
closure, associativity, identity, and non-negativity. By algebra we show that,
given these properties and \cref{def:vector}, we can derive closure,
associativity, identity, and non-negativity for \(\mathbf{M} = N_1 \times N_2
\times \cdots \times N_i\).
\end{proofsketch}

Once a measure is chosen, the analyst must next model how each candidate access
control scheme accrues costs using that measure. This requires assigning costs
associated with each fully parameterized access control action (command or query
execution). Such an assignment is a \emph{cost function}.

\subsection{Cost Functions}
\label{sec:cost:functions}

In order to calculate the total cost of a particular implementation, costs of
executing the various actions within the implementing schemes must be
determined. Sometimes, the cost of any execution of a particular command or
query is constant (e.g., creating a document requires a constant amount of
\io{}). In other cases, the parameters of the command or query affect the cost
(e.g., adding a user to the system is more expensive for classes of users with
greater capabilities). In addition, some costs depend on the current state
(e.g., granting access to all documents with a certain property may require
inspecting each document, a procedure that grows in cost with the number of
documents in the system). Thus, in general, the cost function is required to map
each (command, parameterization, state) or (query, parameterization, state) to
an element of the relevant cost measure.

\begin{definition}[Cost Function]
\label{def:function}
Let \(\scm{S} = \tup{\Gamma, \Psi, Q}\) be an access control scheme, \(A\) a set
of actions from scheme \(\scm{S}\), and \(\mathbf{G} = \tup{G, {\op}, {\cmp}}\)
a cost measure. A \emph{cost function} for \(\scm{S}\) in \(\mathbf{G}\) is a
function \(\cost^\scm{S}_\mathbf{G}: A \times \Gamma \to G\), which which maps
each access control action and state to the member of the cost measure that best
represents the costs associated with executing the given action in the given
state.
\end{definition}

Although most cost functions are infinite (since the number of states and
parameterizations are usually infinite), we can often generalize (as mentioned
above) for actions whose costs do not depend on the state and\slash{}or
parameterization. In addition, when state or parameters do affect the cost, the
correlation is generally formulaic (e.g., proportional to the size of certain
state elements) and is thus simple to describe in a compact way. Finally, in
cases where the relation between the state or parameterization and the resulting
cost is more complex, we can often take advantage of the simulation-based nature
of the cost analysis process and the law of large numbers by abstracting out
parameters or state (or both) and reproducing their effect via a probability
distribution.

In addition to the cost functions that are of specific interest to the analyst,
our simulation process (\cref{sec:cost:simulation}) also requires the
specification of each scheme's \emph{time function}. The time function is
formalized as a cost function, describing the duration of time required to
complete an access control action. The cost measure of this specialized cost
function is \(\tup{\mathbb{R} \times \text{time}, {+}, {\leq}}\).

\subsection{Cost Analysis via Monte Carlo Simulation}
\label{sec:cost:simulation}

\begin{algorithm}
\begin{algorithmic}
\smaller
\Input \(\mathfrak{S}\), set of candidate schemes
\Input \(\Sigma\), set of implementations (\(\forall \scm{S} \in \mathfrak{S}: \sigma\sb{\scm{S}} \in \Sigma\))
\Input \(\mathbf{C}\), set of cost measures (\(\tau = \tup{\mathbb{R} \times \text{time}, {+}, {\leq}} \in \mathbf{C}\))
\Input \(L\), set of cost functions (\(\forall \scm{S} \in \mathfrak{S}, C \in \mathbf{C}: \cost\sp{\scm{S}}\sb{C} \in L\))
\Input \(I = \tup{W, \mathfrak{A}, A, G_A, g}\), invocation mechanism
\Input \(\gamma_0 \in \Gamma_\scm{W}\), start state
\Input \(T_f\), goal time
\Input \(t\), time step
\Statex

\Procedure {\proc{ACCostEvalSim}} {\(\mathfrak{S}, \Sigma, \mathbf{C}, L, I, \gamma\sb{0}, T\sb{f}, t\)\relax}
  \State \(\mathbf{S} \is \set{}\) \Comment {Initialize set of running AC systems}
  \State \(T \is 0\) \Comment {Initialize master clock}
  \ForAll {\(\scm{S} = \tup{\Gamma, \Psi, Q} \in \mathfrak{S}\)} \Comment {Initialize state}
    \State \(\mathbf{S} \is \mathbf{S} \cup \set{\scm{S}}\)
    \State \(\mathbf{A}_\scm{S} \is \set{}\) \Comment {Set of running actor machines}
    \State \(\gamma_\scm{S} \is \sigma_\scm{S}\prm{\gamma_0}\) \Comment {Current state of scheme \(\scm{S}\)}
    \ForAll {\(C \in \mathbf{C}\)}
      \State \(c^\scm{S}_C \is 0_C\) \Comment {Total cost of scheme \(\scm{S}\) in \(C\)}
    \EndFor
    \ForAll {\(\alpha \in A\prm{\gamma_\scm{S}}\)}
      \State \(\mathbf{A}_\scm{S} \is \mathbf{A}_\scm{S} \cup \set{g\prm{\alpha}}\)
      \State \(T_\alpha \is 0\) \Comment {Per-actor clock}
    \EndFor
  \EndFor
  \While {\(T \leq T_f\)} \Comment {Main loop}
    \State \(T \is T + t\) \Comment {Increment clock}
    \ForAll {\(\scm{S} \in \mathbf{S}\)} \Comment {Each AC system}
      \State \(K = \set{}\) \Comment {Clear action list}
      \ForAll {\(\alpha \in \mathbf{A}_\scm{S}\)} \Comment {Choose next actions}
        \If {\(T_a < T\)} \Comment {Check actor busy state}
          \State \(\tup{k, P_k} \is \call{nextAction}{g\prm{\alpha}}\)
          \If {\(k \neq \varnothing \land \call{WSat}{k, \alpha, P_k} \neq \varnothing\)}
            \State {\(T_a \is T + \cost^{\scm{S}}_{\tau}\prm{k}\)} \Comment{Busy state}
            \State {\(K \is K \cup \set{\tup{k, \alpha, P_k}}\)} \Comment {Save action}
          \EndIf
        \EndIf
      \EndFor
      \ForAll {\(\tup{k, \alpha, P_k} \in K\)} \Comment {Compile costs}
        \ForAll {\(C \in \mathbf{C}\)}
          \State {\(c^\scm{S}_C \is c^\scm{S}_C \op[C] \cost^\scm{S}_C\prm{\sigma_\scm{S}\prm{\tup{k, \alpha, P_k}}}\)}
        \EndFor
        \If {\(k\) is a command}
          \State \(\gamma_\scm{S} \is \sigma_\scm{S}\prm{e_k\prm{\gamma_\scm{S}, P_k}}\) \Comment {Update state}
        \EndIf
      \EndFor
    \EndFor
  \EndWhile
  \ForAll {\(\scm{S} \in \mathfrak{S}\)}
    \State Log \(\tup{\scm{S}, c^\scm{S}_{C_1}, \ldots, c^\scm{S}_{C_m}}\)
  \EndFor
\EndProcedure
\end{algorithmic}
\caption{Cost analysis simulation algorithm \label{alg:simulation}}
\end{algorithm}

In \cref{sec:expressiveness:implementation}, we discussed the construction of
implementations, which (in addition to their role in expressiveness evaluation)
provide a recipe for using each candidate scheme to execute the access control
actions needed by the application of interest. In \cref{sec:cost:invocations},
we discussed an actor-based invocation mechanism, which serves as the second
component of the access control workload and provides us with a mechanism for
generating traces of access control actions that are characteristic of usage
within the desired application. Finally, in \cref{sec:cost:functions}, we
discussed cost functions, including the time function, which allow us to
quantify the costs of individual access control actions as well as track the
passage of time during the execution of generated traces. Given these inputs, we
can utilize an automated cost analysis procedure that generates traces of
workload actions, translates these into traces of scheme actions, then
calculates the costs of these scheme actions.

\Cref{alg:simulation} describes such a simulation procedure. First, each
candidate scheme is instantiated as a system. An actor machine is then launched
for each actor in the state of each system. During the main loop, the clock is
incremented and each actor machine is inspected for the correct action to
execute next, as per the execution semantics of the actor machine described in
\cref{sec:cost:invocations}. If an action is to be executed by the actor during
this time step, a reference monitor for the workflow satisfiability problem
(procedure \proc{WSat}) is consulted to ensure that---with respect to the
workflow and constraints---the actor can execute the action without rendering
the workflow instance \emph{unsatisfiable}. For independent actions (i.e., those
in \(\set{a : \nexists a\Prime, a\Prime \prec a}\)), the workflow instance in
question is a new, blank instance added to the pool of partially executed
instances. For dependent actions (i.e., those in \(\set{a : \exists a\Prime,
a\Prime \prec a}\)), the instance is chosen from the existing instances which
belong to the same task as the current action in question.

After all action executions for a time step are collected (and verified by the
reference monitor), they are simulated within the access control state and their
costs are accrued into a running total for each scheme\slash{}cost measure
combination. (We note that costs may also be accrued per user, per workflow,
etc., by trivially extending \cref{alg:simulation}.) The final step in the loop
adjusts the set of actors according to changes in the state. Once a specified
amount of time has passed in the simulated system (denoted the \emph{goal
time}), the main loop breaks and the total costs are output.

To address the requirement of \reqEF{}, we present the following \lcnamecref{thm:complexity}
regarding the runtime of \cref{alg:simulation}. The proof of this \lcnamecref{thm:complexity}
utilizes previous work by Wang and Li~\cite{wang2007esorics} on the complexity
of deciding workflow satisfiability.
\begin{theorem}
\label{thm:complexity}
Assuming that workflow constraints are restricted to the binary operators
\(\set{{=}, {\neq}}\) (i.e., constraints expressing binding of duty and
separation of duty)\footnote{A recent result by Crampton et
al.~\cite{crampton2012ccs} allows the use of a wider range of constraints
(including those over organizational hierarchies) while preserving the
complexity result. For brevity and simplicity, we consider only \(\set{{=},
{\neq}}\) as constraint operators in this work.}, the simulation procedure
described in \cref{alg:simulation} is pseudo-polynomial in the number of
simulated steps and \FPT with parameter \(\alpha\), the number of actions in
the largest task (i.e., the size of the largest disjoint subgraph of the
workflow graph).
\end{theorem}

\begin{proofsketch}
By far, the step of \cref{alg:simulation} that dominates its complexity is the
call to \proc{WSat}, as the workflow satisfiability problem (\WSP) is
\NP-complete. The call to \proc{WSat} is nested within loops which will cause it
to be called \(S \cdot T \cdot A\) times. By~\cite{wang2007esorics}, \WSP is
solvable in \(\bigO{C \cdot A^\alpha}\), yielding a total complexity of
\(\bigO{S \cdot C \cdot T \cdot A^{\alpha+1}}\), which is in \FPT with fixed
parameter \(\alpha\) (maximum number of actions in a task).
\end{proofsketch}

\begin{algorithm}
\begin{algorithmic}
\smaller
\Input \(\mathfrak{S}\), set of candidate schemes
\Input \(\Sigma\), set of implementations (\(\forall \scm{S} \in \mathfrak{S}: \sigma\sb{\scm{S}} \in \Sigma\))
\Input \(\mathbf{C}\), set of cost measures (\(\tau = \tup{\mathbb{R} \times \text{time}, {+}, {\leq}} \in \mathbf{C}\))
\Input \(L\), set of cost functions (\(\forall \scm{S} \in \mathfrak{S}, C \in \mathbf{C}: \cost\sp{\scm{S}}\sb{C} \in L\))
\Input \(I = \tup{W, \mathfrak{A}, A, G_A, g}\), invocation mechanism
\Input \(\Pr\prm{\gamma}\), probability distribution over start states
\Input \(\chi\), number of Monte Carlo runs
\Input \(T_f\), goal time
\Input \(t\), time step
\Statex

\Procedure {\proc{ACCostEvalMC}} {\(\mathfrak{S}, \Sigma, \mathbf{C}, L, I, \Pr\prm{\gamma}, \chi, T\sb{f}, t\)\relax}
  \ForAll {\(\iirange{1, \chi}\)} \Comment {Monte Carlo loop}
    \State {\(\gamma_0 \is\) random sample from \(\Pr\prm{\gamma}\)}
    \State {\(\call{ACCostEvalSim}{\mathfrak{S}, \Sigma, \mathbf{C}, \mathfrak{L}, I, \gamma_0, T_f, t}\)}
  \EndFor
\EndProcedure
\end{algorithmic}
\caption{Monte Carlo application of \cref{alg:simulation} \label{alg:montecarlo}}
\end{algorithm}

\Cref{alg:simulation} executes a single run of the system. We next discuss two
approaches to utilizing this algorithm: using the Monte Carlo technique to
generate large numbers of data points for trend analysis using scatter plots,
and using fixed-sample-size point estimates for calculating cost assessments
with a particular confidence interval for a small set of important input
configurations. \Cref{alg:montecarlo} demonstrates the former. This algorithm
repeatedly calls \cref{alg:simulation} using randomly sampled start states in an
attempt to exploit the potentially large variance between executions. An
advantage of this approach is the detection of trends across a variety of start
states. Furthermore, the repeated execution contributes to the complexity of the
full analysis by only a multiplicative factor. As such, Monte Carlo
analysis---like single run analysis---is in \FPT.

\begin{algorithm}
\begin{algorithmic}
\smaller
\Input \(\mathfrak{S}\), set of candidate schemes
\Input \(\Sigma\), set of implementations (\(\forall \scm{S} \in \mathfrak{S}: \sigma\sb{\scm{S}} \in \Sigma\))
\Input \(\mathbf{C}\), set of cost measures (\(\tau = \tup{\mathbb{R} \times \text{time}, {+}, {\leq}} \in \mathbf{C}\))
\Input \(L\), set of cost functions (\(\forall \scm{S} \in \mathfrak{S}, C \in \mathbf{C}: \cost\sp{\scm{S}}\sb{C} \in L\))
\Input \(I = \tup{W, \mathfrak{A}, A, G_A, g}\), invocation mechanism
\Input \(\gamma_0\), start state
\Input \(T_f\), goal time
\Input \(t\), time step
\Input \(u \in \eerange{0, 1}\), desired confidence level
\Input \(v \in \eerange{0, 1}\), desired tolerance
\Statex

\Procedure {\proc{ACCostEvalCI}} {\(\mathfrak{S}, \Sigma, \mathbf{C}, L, I, \gamma\sb{0}, T\sb{f}, t, u, v\)\relax}
  \State \(n \is \emptyset\)
  \While {\(t_{\card{n}-1,1-u/2} \sqrt{\frac{S^2\prm{n}}{\card{n}}} > v \cdot \bar{X}\prm{n}\)}
    \State {\(n \is n \cup \call{ACCostEvalSim}{\mathfrak{S}, \Sigma, \mathbf{C}, L, I, \gamma_0, T\sb{f}, t}\)}
  \EndWhile
\EndProcedure
\end{algorithmic}
\caption{Confidence-bounding application of \cref{alg:simulation} \label{alg:confidence}}
\end{algorithm}

In the interest of the \reqAC{} requirement, we consider a second approach,
which allows the analyst to achieve an intended confidence in the cost value
generated for a particular start state. With this approach, we decide the number
of simulation runs to conduct based on a desired confidence and the assumption
of a normal distribution of costs across runs. We use the fixed-sample-size
procedure for point estimate of a mean, which says that the confidence interval
for a mean is:
\[\bar{X}\prm{n} \pm t_{\card{n}-1,1-\frac{\alpha}{2}} \sqrt{\frac{S^2\prm{n}}{\card{n}}}\]
where \(\bar{X}\prm{n}\) is the sample mean, \(\frac{S^2\prm{n}}{\card{n}}\) is
the sample variance, and \(t_{\nu,\gamma}\) is the critical point for the
\(t\)-distribution with \(\nu\) degrees of freedom. The resulting range is an
approximate \(100 (1-\alpha)\)-percent confidence interval for the expected
average cost of the scheme. During simulation, we repeatedly calculate the
confidence interval for incrementing \(n\), terminating when a satisfactory
confidence is reached. For example, assuming we desire a 90-percent confidence
interval of no more than 0.1 of the mean, we run the simulation repeatedly
until:
\[t_{\card{n}-1,0.95} \sqrt{\frac{S^2\prm{n}}{\card{n}}} \leq 0.1 \bar{X}\prm{n}\]
\Cref{alg:confidence} demonstrates the use of this approach to execute
\cref{alg:simulation} until a desired confidence is reached, rather than
executing for a fixed number of runs.

We note that our cost analysis procedure evaluates particular implementations of
the workload within candidate schemes, and thus cannot make formal claims about
schemes in general. However, in practice, an analyst will be concerned primarily
with the costs associated with the specific implementation she has designed; the
existence of more efficient, though unknown, implementations is not particularly
helpful in choosing an access control scheme. Finding optimal implementations is
an orthogonal problem that we discuss in \cref{sec:discussion}.

\section{Access Control Extensions}
\label{sec:extensions}

In the event that a scheme does not admit a safe implementation of the workload,
the analyst may attempt to enable the construction of such an implementation by
augmenting the scheme with additional functionality.\footnote{Note that it is
not always possible to extend a scheme in a way that enables a particular
implementation.} Intuitively, extending an access control scheme expands its
protected state, commands, and\slash{}or queries. One must use care, however,
when constructing such extensions. Although virtually any changes to an access
control scheme will yield another valid scheme, not all changes will yield a
scheme that preserves the security properties of the original. As an extreme
example, almost any scheme will be ``broken'' if we add a \textstt{grant-all}
command that grants all permissions to all subjects (similar to McLean's System
Z~\cite{mclean1987oakland}).

To maintain the intuition behind the concept of an extension, we require that
the changes made to the scheme at most enable \emph{additional} implementations
(i.e., do not preclude the use of any implementations possible in the original).
Specifically, in order to \emph{safely} extend a scheme, one must prove that the
extended scheme does not violate any of the security properties of the original.
One can prove safety by viewing the original scheme as a workload operational
description, and demonstrating a state-matching implementation of the original
scheme within the extended scheme. This proves that the extended scheme can be
used transparently in place of the original, and is therefore a safe extension.
The violation of even simple safety resulting from extending a scheme with the
above \textstt{grant-all} command can be detected by attempting (and failing) to
construct such an implementation of the original scheme within this extended
version while preserving simple safety.

In this paper, we explore a particular class of extensions that we call
\emph{\aums} (\AMs).

\begin{definition}[\AuM]
\label{def:auxm}
An \emph{access control \aum} for augmenting an access control scheme over the
set of access control states \(\Gamma_0\) is a state-transition system
\(\tup{\Gamma, \Psi, Q}\), where:
\begin{itemize}
\item \(\Gamma\) is the set of auxiliary states.
\item \(\Psi\) is the set of commands over \(\Gamma_0 \times \Gamma\) where we
enforce that \(\forall \tup{n, P, e} \in \Psi, p \in P^{*}, \gamma_0 \in
\Gamma_0, \gamma \in \Gamma, \exists \gamma^\prime \in \Gamma:
e\prm{\tup{\gamma_0, \gamma}, p} = \tup{\gamma_0, \gamma^\prime}\) (i.e.,
commands can reference the original scheme's state, but cannot alter it).
\item \(Q\) is the set of queries over \(\Gamma_0 \times \Gamma\) (i.e., that
can reference the original scheme's state).
\end{itemize}
\end{definition}

\begin{figure}
\centering
\includegraphics[width=.8\columnwidth]{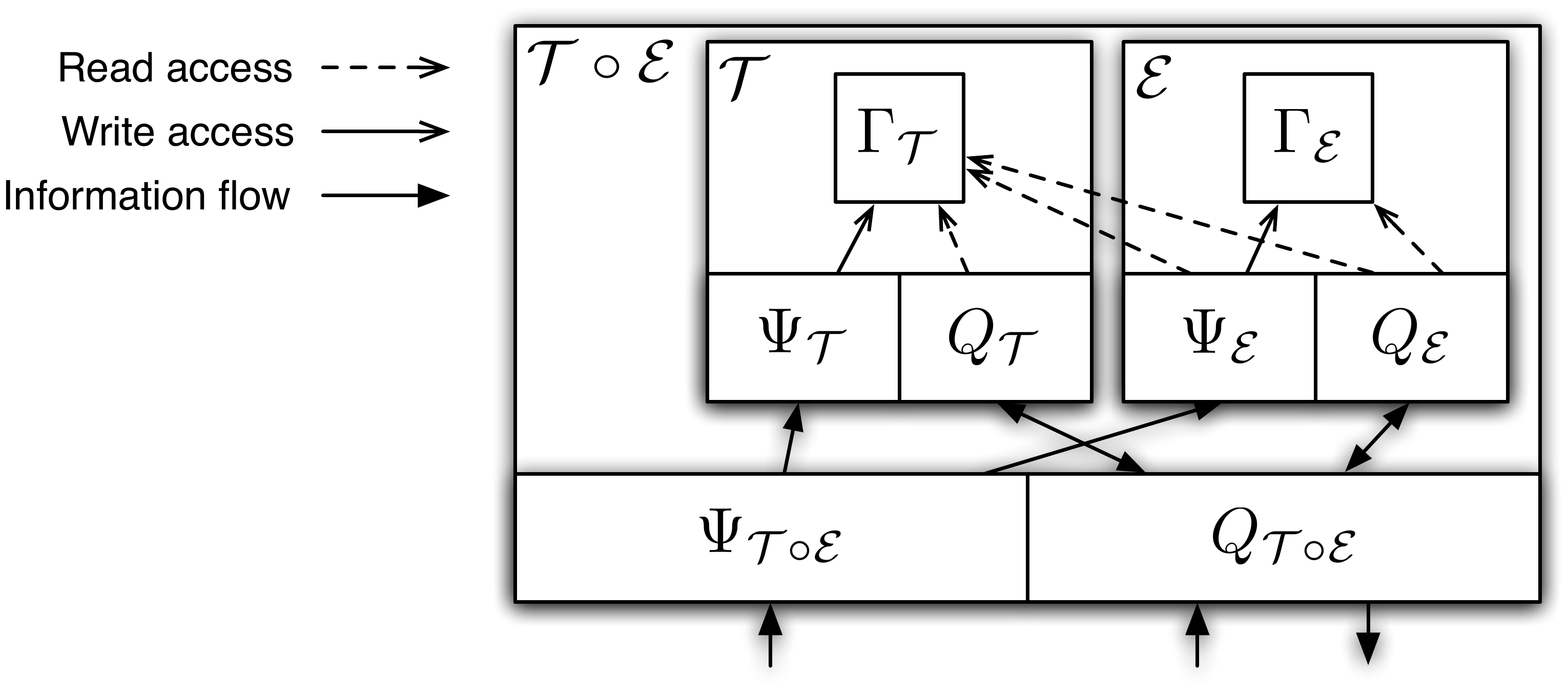}
\caption{A graphical representation of an access control scheme \(\scm{T}\)
augmented with an \aum \(\scm{E}\)\label{fig:augment}}
\end{figure}

Augmenting an access control scheme with an \aum is achieved by computing the
cross product of the states of the two machines and the union of the commands
and queries, as follows, and is represented graphically in \cref{fig:augment}.

\begin{definition}[Augmented Scheme]
\label{def:augmented}
Let \(\scm{S} = \tup{\Gamma^\scm{S}, \Psi^\scm{S}, Q^\scm{S}}\) be an access
control scheme, \(\scm{U} = \tup{\Gamma^\scm{U}, \Psi^\scm{U}, Q^\scm{U}}\) be
an access control \aum. The \emph{augmented access control scheme} formed by
augmenting scheme \(\scm{S}\) with \AM \(\scm{U}\), is the scheme \(\scm{S}
\circ \scm{U} = \tup{\Gamma^{\scm{S} \circ \scm{U}}, \Psi^{\scm{S} \circ
\scm{U}}, Q^{\scm{S} \circ \scm{U}}}\) where:
\begin{itemize}
\item \(\Gamma^{\scm{S} \circ \scm{U}} = \Gamma^\scm{S} \times \Gamma^\scm{U}\)
\item \(\Psi^{\scm{S} \circ \scm{U}} = \Psi^\scm{S} \cup \Psi^\scm{U}\)
\item \(Q^{\scm{S} \circ \scm{U}} = Q^\scm{S} \cup Q^\scm{U}\)
\end{itemize}
\end{definition}

\Cref{def:auxm,def:augmented} give us the following \lcnamecref{thm:auxm}, which
proves that the class of extensions that can be represented as \aums encode safe
extensions to \emph{any} access control scheme with respect to the
state-matching implementation.

\begin{theorem}
\label{thm:auxm}
Given access control scheme \(\scm{S} = \tup{\Gamma^\scm{S}, \Psi^\scm{S},
Q^\scm{S}}\) and access control \aum \(\scm{U} = \tup{\Gamma^\scm{U},
\Psi^\scm{U}, Q^\scm{U}}\), there exists a state-matching implementation of
\(\scm{S}\) in \(\scm{S} \circ \scm{U}\).
\end{theorem}

\begin{proofsketch}
Intuitively, a scheme extended with an \aum can behave exactly as it would
without the \AM---it must answer the original queries in the same way as the
original scheme, and it is forbidden from modifying elements of the original
scheme's state in ways the original could not. Thus, to satisfy property~(1) of
the state-matching implementation, we map each state and action in the original
to the same state or action in the augmented scheme, and the \AM state is not
utilized.

When considering only the original queries, the unmodified scheme can also
easily mimic the augmented scheme, since these queries are guaranteed only to
reference state that both schemes change in the same way (via the original
commands). This satisfies property~(2) of the state-matching implementation.
\end{proofsketch}

While these security properties of \aums and augmented schemes enable the
analyst to use the constructs without fear of contaminating the original
schemes, they do not imply that the use of \AMs (or extensions in general) is
without penalty. Since \AMs would be implemented as additional trusted code that
communicates in a secure way with the original access control software, one may
be concerned if a high proportion of the total state is stored within the \AM,
or if a large amount of communication needs to occur between the original state
and the \AM state. These types of concerns can be addressed by choosing
appropriate cost measures for cost analysis.

Having presented a notion of scheme extensions and proven that it is safe, we
now revisit the implementation of ADAC (\cref{ex:workload}) using DAC
(\cref{ex:scheme}).

\begin{example}
\label{ex:auxm}
Recall that the workload \(W_\scm{A}\) (\cref{ex:workload}) differs from the DAC
scheme \(\scm{D}\) (\cref{ex:scheme}) mainly in that \(W_\scm{A}\) has
administrators with full rights to the system. In particular, the query
\textstt{SubjectAdmin} is problematic, as the DAC scheme \(\scm{D}\) has no way
of maintaining the list of administrative users. One natural attempt at fixing
this problem is to create a special object within \(\scm{D}\), rights over which
indicate administrator status. Another possibility is to create a special right
that administrators have over all objects. Such approaches fail to allow a safe
implementation, because they invalidate security analysis instances. In
particular, these approaches alter the value of query \textstt{Access} for
certain parameterizations.

Instead, we construct an \aum that stores the additional information and answers
the additional query regarding administrative status of subjects. We extend DAC
with \aum \(\scm{M} = \tup{\Gamma^\scm{M}, \Psi^\scm{M}, Q^\scm{M}}\). The \AM's
states, \(\Gamma^\scm{M}\), are defined by the sets \(\tup{A, N}\), where:
\begin{itemize}
\item \(A \subseteq S\)  is the set of administrators
\item \(N: A \times O \to 2^R\) is the ``hidden'' access matrix that keeps track
of the access rights each administrative subject would revert to upon losing
administrator status
\end{itemize}

The extension's commands, \(\Psi^\scm{M}\), include the following.
\begin{itemize}
\item \textstt{GrantAdmin(\(S\), \(S\))}, which grants administrative privilege
to a subject
\item \textstt{RevokeAdmin(\(S\), \(S\))}, which revokes administrative
privilege from a subject
\item \textstt{SoftGrant(\(S\), \(S\), \(O\), \(R\))}, which grants a right over
an object to a subject in the hidden access matrix
\item \textstt{SoftRevoke(\(S\), \(S\), \(O\), \(R\))}, which revokes a right
over an object from a subject in the hidden access matrix
\end{itemize}

Finally, \(Q^\scm{M}\) includes the following.
\begin{itemize}
\item \textstt{SubjectAdmin(\(S\), \(S\))}, which asks whether a subject is an
administrator
\item \textstt{HiddenAccess(\(S\), \(S\), \(O\), \(R\))}, which asks whether a
user has a right over an object in the hidden access matrix
\end{itemize}
\end{example}

\begin{example}
\label{ex:implementation}
The \AM described in \cref{ex:auxm} can augment the DAC scheme with the ability
to keep track of which subjects are administrators, as well as which rights each
would have if they lost such status. The implementation of \(W_\scm{A}\) using
\(\text{DAC} \circ \scm{M}\) then has several non-trivial tasks. When a subject
is added to \(A\), the system must copy all current access for that subject from
\(M\) to \(N\) and then grant that subject all accesses in \(M\). This procedure
is reversed when removing a user from \(A\), and any rights granted to or
revoked from a user in \(A\) are recorded in \(N\) and do not affect \(M\).
\end{example}

\section{Case Study}
\label{sec:casestudy}

In this section, we discuss an example scenario which we will use to demonstrate
a full analysis using our framework. This case study explores a workload based
on a group messaging scenario with conflicts of interest.

\subsection{Workload description}
\label{sec:casestudy:description}

Group-centric Secure Information Sharing (g-SIS)~\cite{krishnan2009sacmat} has
been proposed as a new approach to access control that differs from the
dissemination-centric approach that has inspired the development of schemes such
as RBAC and DAC. Dissemination-centric models focus on bestowing policies upon
objects as they are produced, sometimes refining these policies at later times.
These policies are then referenced as consumers access these objects. The g-SIS
approach, on the other hand, addresses collaboration- and subscription-based
systems. In the g-SIS models, groups can be brought together to share
information as they work toward a common goal. Accesses are decided not by
attaching policies to objects, but in a time-variant way by inspecting the
users' historical membership in groups. For example, an online periodical may
offer a base subscription in which users have access to issues published during
their subscription, and only while they remain subscribed. They might also offer
(for an additional fee) current subscribers access to back issues, or former
subscribers the ability to access issues published during their subscription.

The current state-of-the-art in implementations based on g-SIS is a formal
specification in linear temporal logic for formal analysis~\cite{krishnan2011}.
The creators of g-SIS speculate that, with respect to expressiveness, these
models may be equivalent to more traditional, dissemination-centric sharing
models. However, they believe that the g-SIS approach will better enable the
type of information sharing common in collaborative settings. Schemes inspired
by g-SIS, then, would aim to provide an application-specific solution to access
control. These schemes would aim to satisfy a category of applications that
current models fail to capture, despite (possibly) possessing the expressive
power necessary to express the applications' policies. Our framework is designed
to investigate and quantify exactly this type of scenario, and as such this
problem is a natural application of our framework. Thus, to verify and quantify
the claims about g-SIS, we have modeled a group messaging workload after a
particular use case within g-SIS, and evaluated within this workload the
expressiveness and costs of common dissemination-centric schemes as well as a
particular instantiation of the g-SIS approach within trust management.

\subsection{Our g-SIS Workload}
\label{sec:casestudy:workload}

In our group messaging scenario, the main objects of interest are
\emph{messages} posted to \emph{groups}. Current members of a group have access
to the messages posted to it. When joining a group, a user can choose to request
a \emph{strict join} (in which access to previously posted messages is not
granted) or a \emph{liberal join} (in which access to all previous messages is
granted). A similar decision is made when leaving a group. In the spirit of
discussions such as those that take place in program committee meetings, we
model workflows that accommodate users who must temporarily take leave from a
group due to conflicts of interest, appointing temporary group administrators
(if necessary) during this time.

\begin{figure}
\centering
\includegraphics[width=0.85\columnwidth]{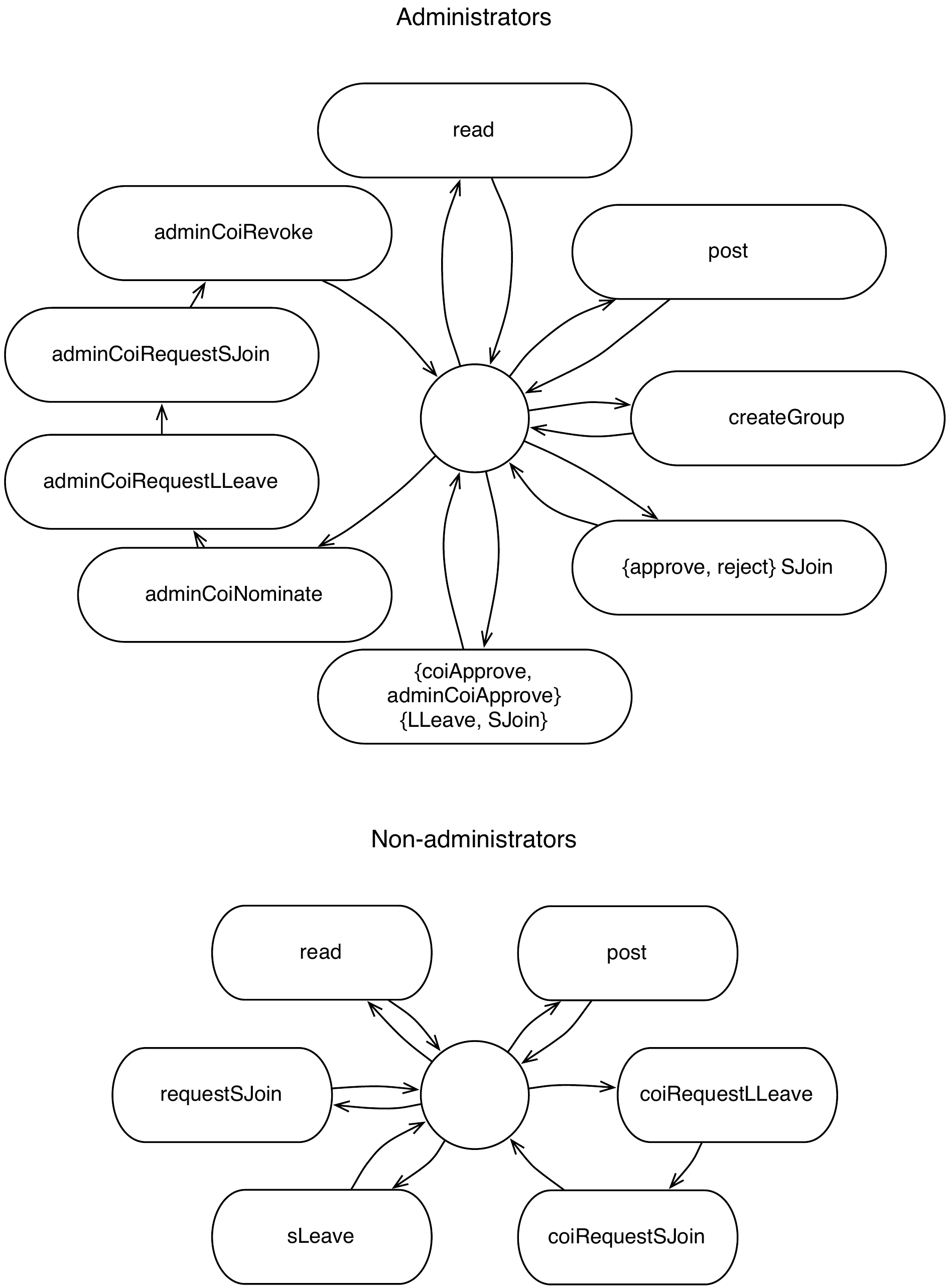}
\caption{Actor machines for the group messaging
workload\label{fig:gmactormachines}}
\end{figure}

The group messaging workload, \(\scm{W}^\scm{G} = \tup{\scm{G}, I^\scm{G}}\),
utilizes as its operational component the abstract group messaging scheme
(GMS), \(\scm{G}\). GMS is defined as \(\scm{G} =
\tup{\Gamma^\scm{G}, \Psi^\scm{G}, Q^\scm{G}}\). Its states, \(\Gamma^\scm{G}\),
are defined by the sets \(\tup{U, G, M, T, T_c, O, A, R, TX}\), where:
\begin{itemize}
\item \(U\) is the set of users
\item \(G\) is the set of groups
\item \(M\) is the set of messages
\item \(T\) is the ordered set of timestamps, including special timestamp \(\infty\)
\item \(T_c\) is the current timestamp
\item \(O \subseteq U \times G\) is the group ownership relation
\item \(A \subseteq U \times G\) is the group administration relation
\item \(R \subseteq U \times G \times T \times T\) is the group membership record
\item \(TX \subseteq G \times M \times T\) is the messaging transcript
\end{itemize}

GMS's commands, \(\Psi^\scm{G}\), include the following.
\begin{itemize}
\item \textstt{CreateGroup(\(U\), \(G\))}, which creates a group
\item \textstt{GrantAdmin(\(U\), \(U\), \(G\))}, which grants a user
administrative permission for a group
\item \textstt{RevokeAdmin(\(U\), \(U\), \(G\))}, which revokes from a user
administrative permission for a group
\item \textstt{SAddMember(\(U\), \(U\), \(G\))}, which strict-adds a user to a
group (i.e., adds the user without granting permission to view existing
messages)
\item \textstt{LAddMember(\(U\), \(U\), \(G\))}, which liberal-adds a user to a
group (i.e., adds the user and grants permission to view existing messages)
\item \textstt{SRemoveMember(\(U\), \(U\), \(G\))}, which strict-removes a user
from a group (i.e., removes the user and revokes permission to view currently
existing messages)
\item \textstt{LRemoveMember(\(U\), \(U\), \(G\))}, which liberal-removes a user
from a group (i.e., removes the user without revoking permission to view
currently existing messages)
\item \textstt{Post(\(U\), \(G\), \(M\))}, which posts a message to a group
\end{itemize}

Finally, GMS's queries, \(Q^\scm{G}\), include the following.
\begin{itemize}
\item \textstt{Access(\(U\), \(M\))}, which asks whether a user can view a
message
\end{itemize}

\begin{figure}
\centering
\includegraphics[width=0.88\columnwidth]{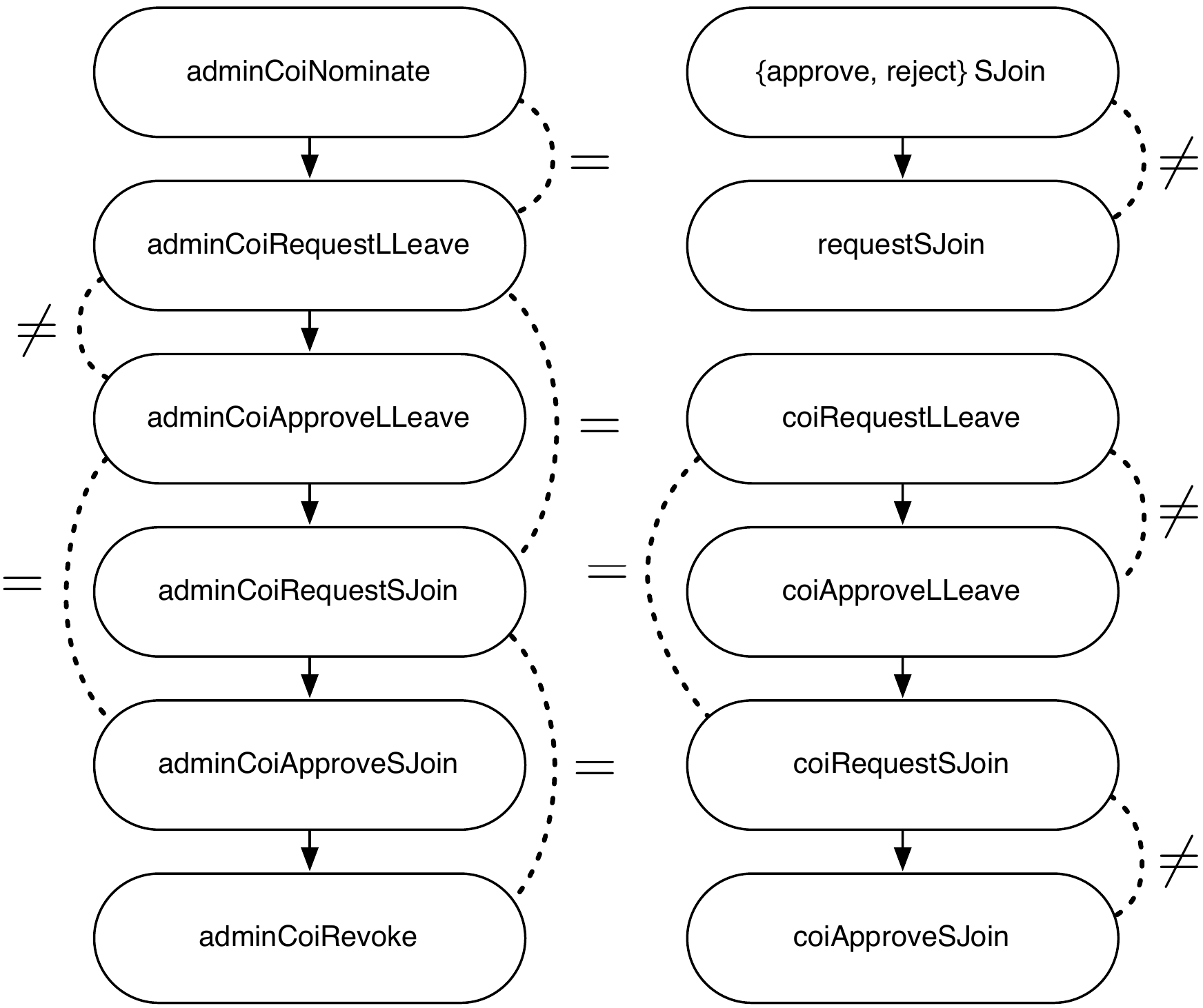}
\caption{Constrained workflow for the group messaging
workload\label{fig:gmworkflow}}
\end{figure}

We fully define GMS in \cref{appx:expressiveness:gms}. The invocation mechanism
for the group messaging workload, \(I^\scm{G}\), is described by the actor
graphs depicted in \cref{fig:gmactormachines}\footnote{We omit the rates in
\cref{fig:gmactormachines}, as these are varied during our cost analysis.} and
the constrained workflow depicted in \cref{fig:gmworkflow}.

\subsection{Expressiveness Evaluation}
\label{sec:casestudy:expressiveness}

In the first phase of suitability analysis, we examine the expressive power of
our candidate schemes to determine which are capable of safely implementing the
workload. In \cref{appx:expressiveness}, we formally describe these
implementations and prove that they are state-matching implementations. In this
section, we omit these details in favor of an intuitive discussion. In
particular, we explored the use of the following access control schemes to
implement the GMS workload:

\begin{itemize}

\item \emph{SD3-GM} is a specially-parameterized instantiation of the trust
management language SD3~\cite{sd3}. Given the flexibility offered by a logical
policy language, SD3-GM easily implements the group messaging workload.

\item \emph{DAC} is a discretionary access control scheme based on the
Graham-Denning scheme~\cite{grahamdenning}. DAC does not admit an obvious
state-matching implementation. Thus, we extended DAC with an \aum to manage the
group-based metadata (e.g., the group membership relation). DAC's access matrix
is updated after changes are made to the \AM data, allowing the \textstt{Access}
query to be answered as in the original DAC scheme.

\item \emph{RBAC} is a role-based access control scheme based on NIST
RBAC~\cite{nist-rbac}. While SD3-GM is a near perfect fit for the workload, and
DAC is reduced to having its native internal state used only as a projection of
an \aum, RBAC's role relation can be used to maintain more relevant state
natively. We still utilize an \AM for RBAC, mainly to maintain the message-group
relation which cannot be maintained in any obvious way within the RBAC state.

\item \emph{GTRBAC} (Generalized Temporal RBAC) is an extended version of RBAC
that adds temporal features such as the time-constrained activation of
roles~\cite{joshi05tkde}. However, it does not include the ability to make
access decisions based on the time at which a user joined a role or the time an
object was created. Thus, the features of GTRBAC beyond those of RBAC do not
contribute to a more efficient implementation of GMS, and we thus dropped GTRBAC
from consideration.

\end{itemize}

\begin{figure}[t]
\centering
\includegraphics[width=\columnwidth]{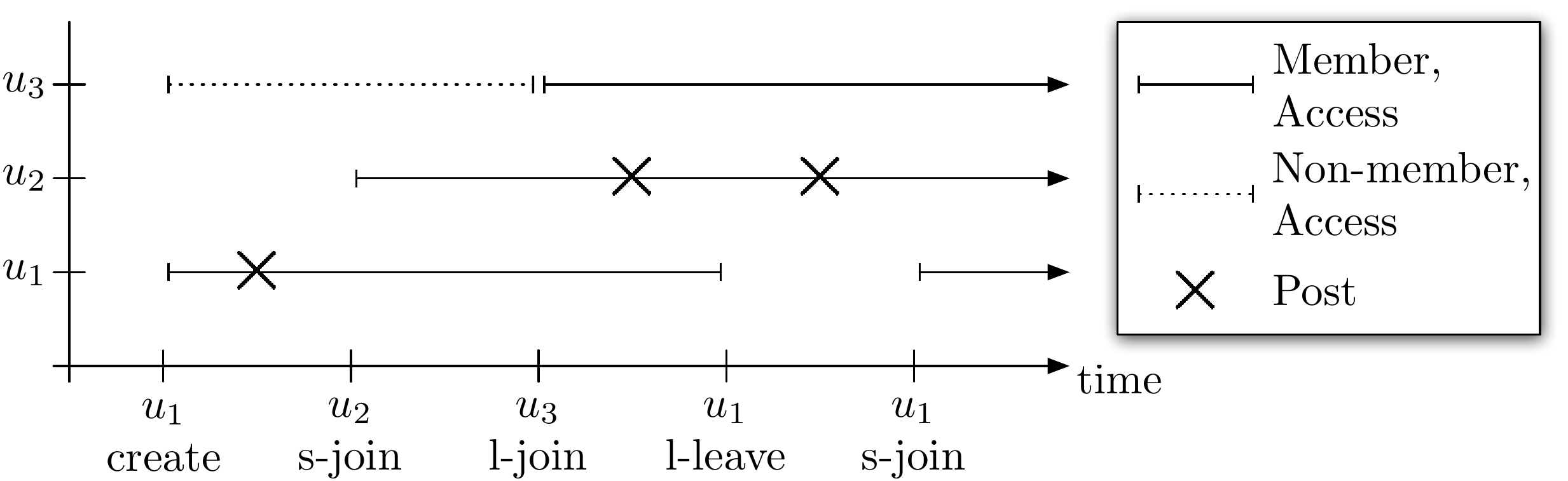}
\caption{An example scenario of accesses in a single group in the
group-messaging scenario\label{fig:gmexample}}
\end{figure}

At first blush, it may seem counter-intuitive that DAC and RBAC require
extensions to correctly support the GMS workload. However, as demonstrated by
\cref{fig:gmexample}, the group messaging scenario can be unexpectedly difficult
to represent in dissemination-centric models. Although a group may seem to
conceptually resemble a role in role-based access control, roles grant the same
accesses to all members, while \cref{fig:gmexample} shows that even a simple
series of events within a single group containing a small number of users can
lead to multiple disjoint sets of accesses in GMS. In this particular example,
all three users have a different ``view'' of the objects in the group, despite
all being members. This single-group scenario is impossible to represent in a
role-based scheme with fewer than three roles, indicating that any
implementation of GMS in a role-based scheme is very likely to exceed a role per
user, reducing the administrative value of utilizing roles at
all~\cite{LuHongLiu2007,RoleVAT}.

The following \lcnamecref{thm:implementations} asserts that each of our three
remaining candidate schemes satisfies our requirements for a safe implementation
of the group messaging workload. This \lcnamecref{thm:implementations} is proved
(individually for each scheme) in \cref{appx:expressiveness}.

\begin{theorem}
\label{thm:implementations}
There exists a state-matching implementation of GMS in SD3-GM, and in
each of our extended versions of RBAC and DAC.
\end{theorem}

\subsection{Cost Analysis}
\label{sec:casestudy:cost}

\begin{figure*}[t]
\centering
\subfloat[]{\includegraphics[width=0.33\textwidth]{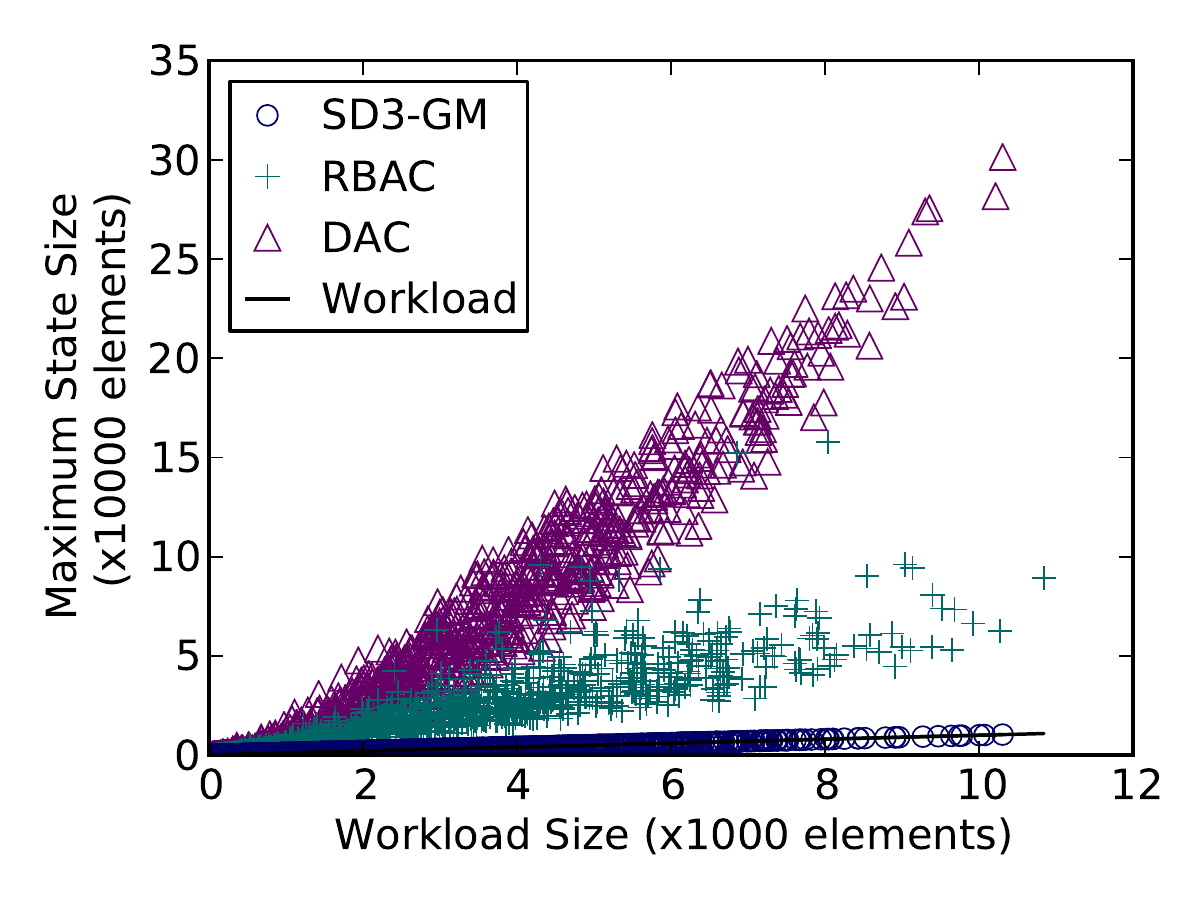}\label{chart:workloadstate}}
\subfloat[]{\includegraphics[width=0.33\textwidth]{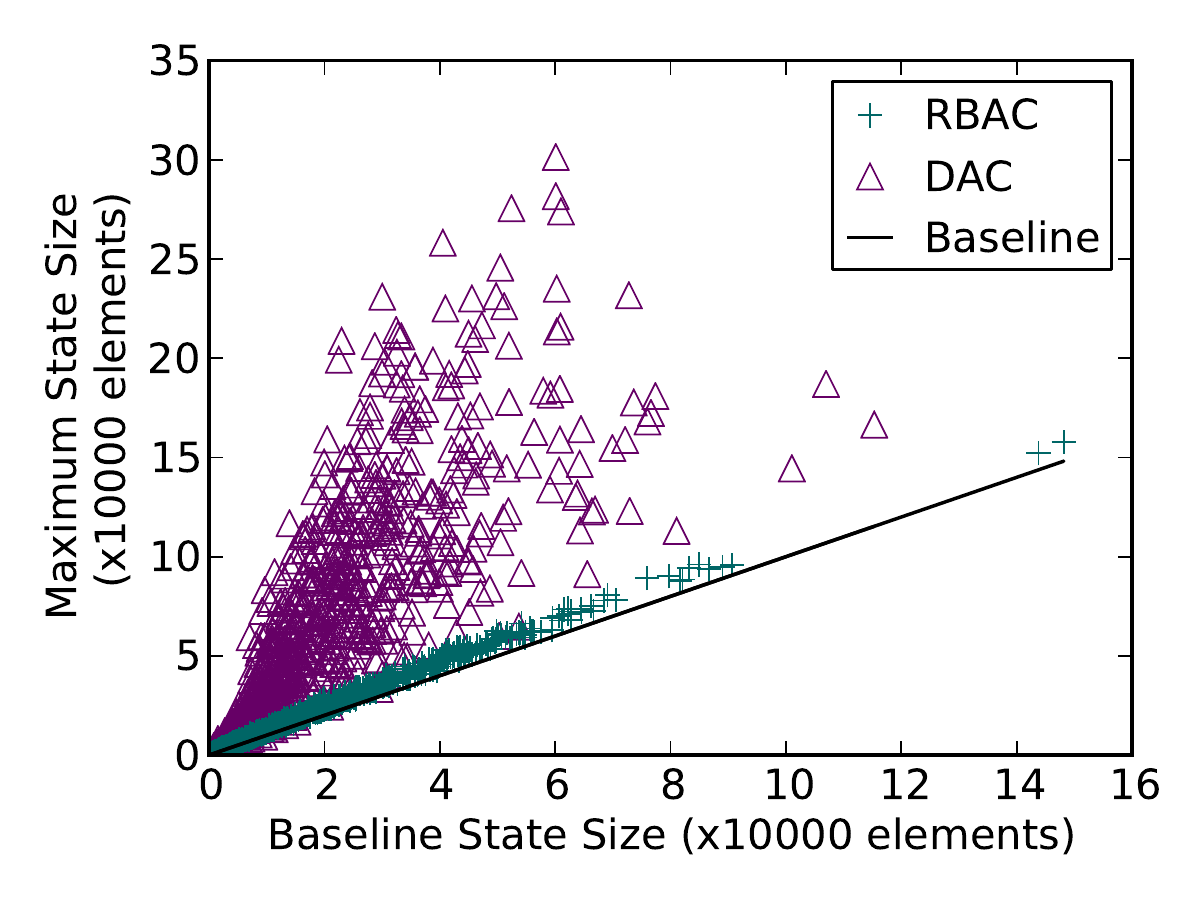}\label{chart:baselinestate}}
\subfloat[]{\includegraphics[width=0.33\textwidth]{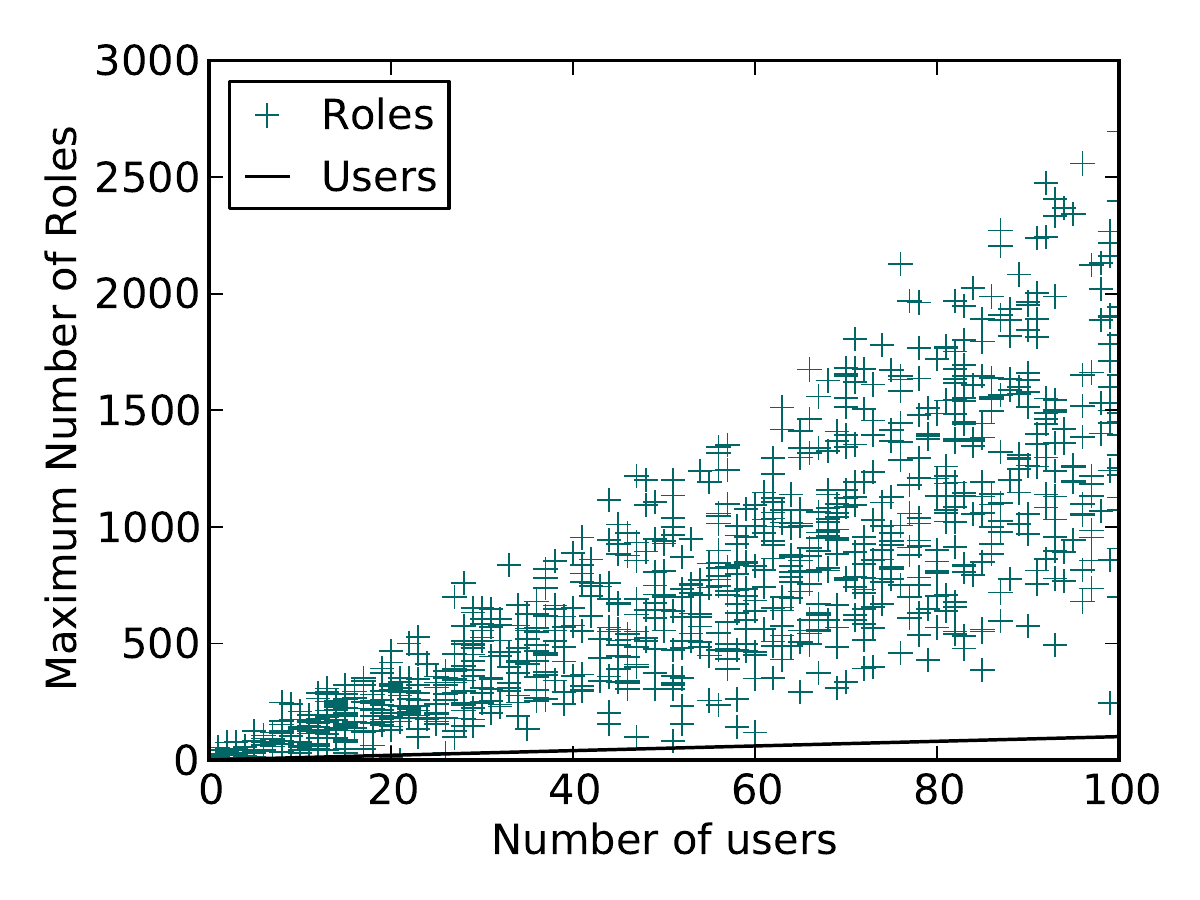}\label{chart:roles}}
\\
\subfloat[]{\includegraphics[width=0.33\textwidth]{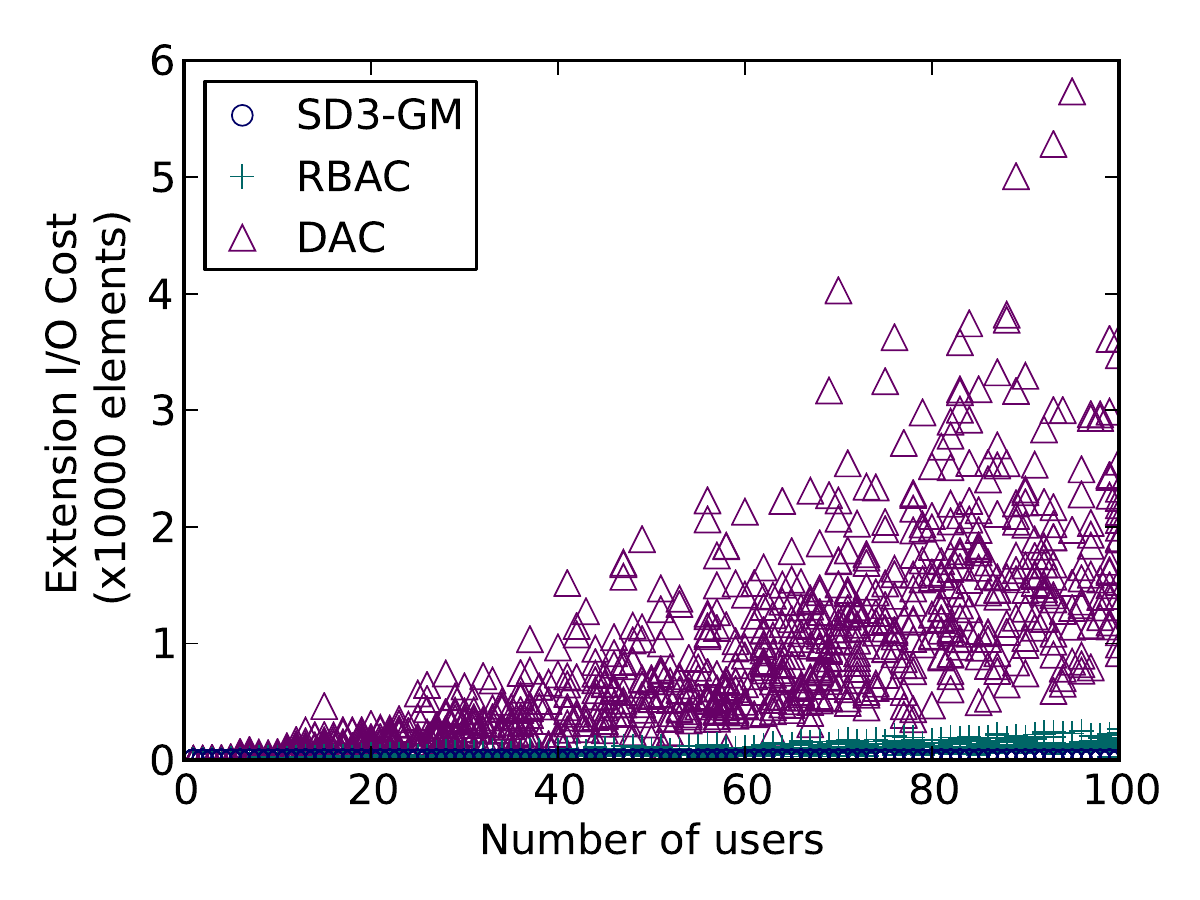}\label{chart:extension}}
\subfloat[]{\includegraphics[width=0.33\textwidth]{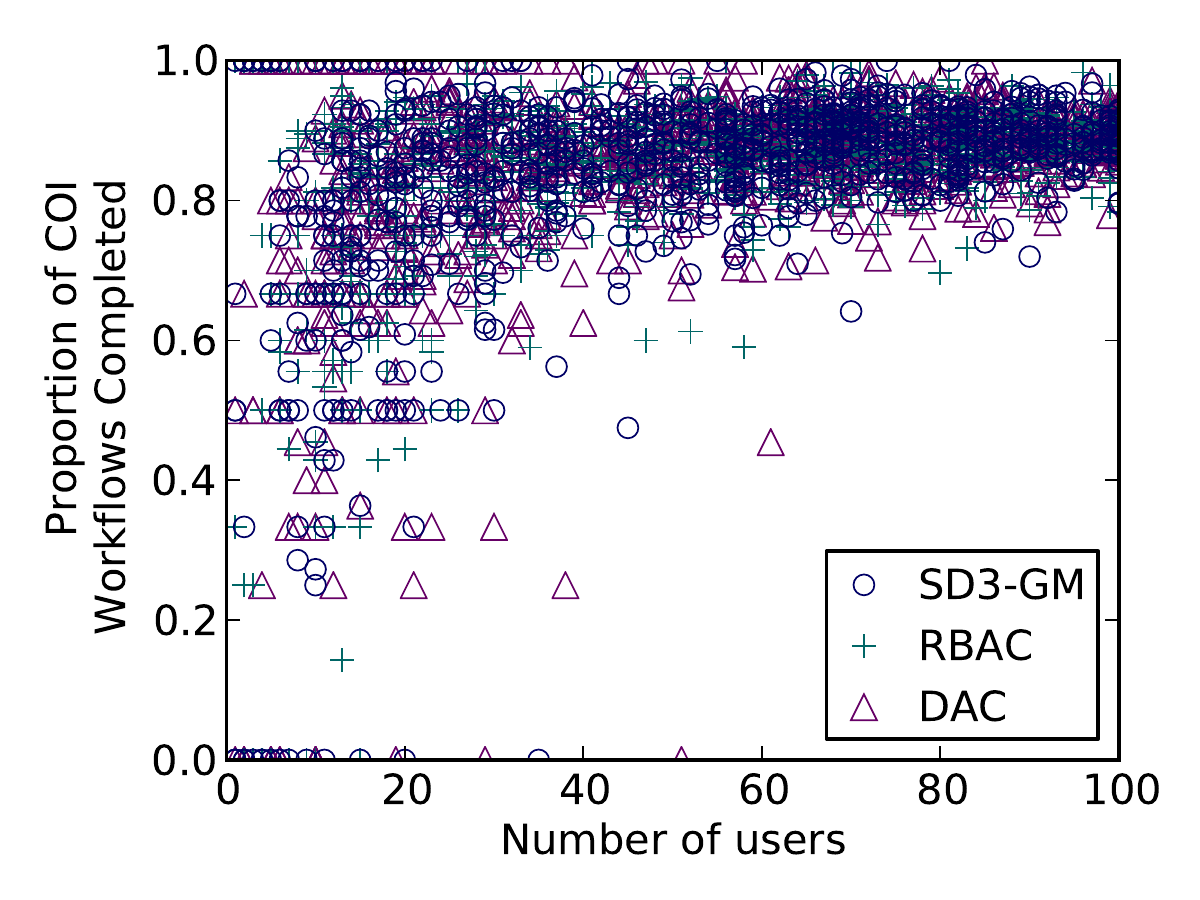}\label{chart:coi}}
\subfloat[]{\includegraphics[width=0.33\textwidth]{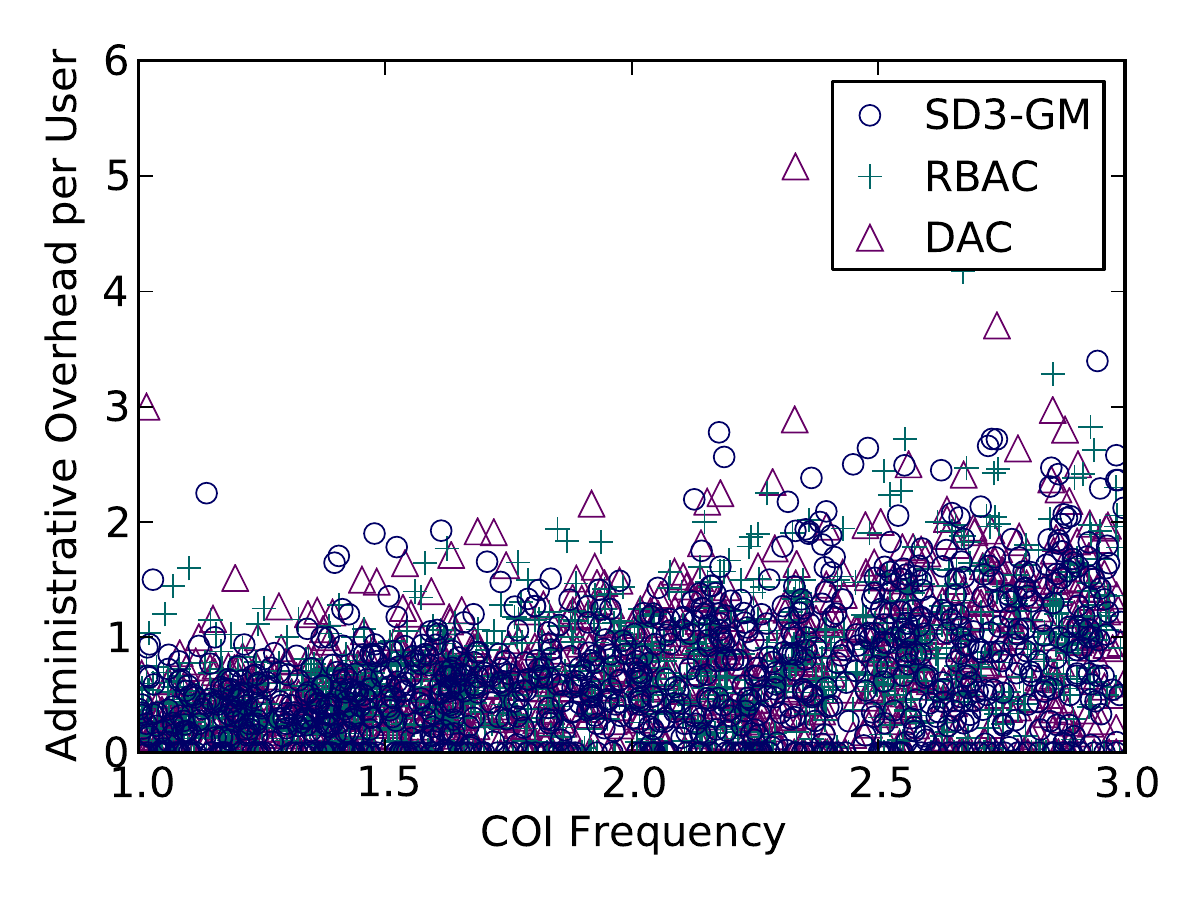}\label{chart:useradmin}}
\caption{Group messaging case study results\label{chart:results}}
\end{figure*}

To perform cost analysis for the group messaging case study described above, we
consider cost measures representing communication with the \aum (where
applicable) and maximum state size. We then defined cost functions over these
cost measures for RBAC, DAC, and SD3-GM. We used these cost functions as inputs
to an implementation we built of (extended versions of)
\cref{alg:simulation,alg:montecarlo}. The implementation of our simulator
consists of about 2000 lines of Java code. We take the Monte Carlo approach in
order to gain insight into the trends in the implementations' costs across the
variety of start states, altering the number of users, number of administrators,
global rate of conflict-of-interest scenarios, and global rate of message
posting. We simulated the messaging environment for 8-hour periods of
interleaved action traces as described by the group messaging workload's actor
machines (\cref{fig:gmactormachines}) and constrained workflow
(\cref{fig:gmworkflow}). We repeated this simulation for 1,000 Monte Carlo runs.

\Cref{chart:results} shows the results of the our evaluation of the
implementations of GMS. In \cref{chart:workloadstate}, we compare maximum state
size to the state size occupied by the equivalent GMS state, demonstrating the
additional storage needed to utilize each candidate access control scheme. While
SD3-GM utilized a small constant amount of additional storage, both RBAC and DAC
required many times the storage of GMS.

In \cref{chart:baselinestate}, we look at maximum state size in a different
way---compared to the ``baseline state,'' which describes the amount of storage
needed to use the scheme naively to reproduce the same accesses as the workload.
For DAC, this is the access matrix, including the appropriate accesses. For
RBAC, this includes the user-role and role-permission relations, assigning each
user to her own role with the permissions the user has access to. Although the
baseline state does not maintain enough information to enable a state-matching
(i.e., safe) implementation, it allows a comparison to the storage of using the
scheme naively. As \cref{chart:baselinestate} shows, storage in both RBAC and
DAC exceeds this baseline, with DAC being particularly excessive.

In \cref{chart:roles}, we compare the number of users in the system to the
number of roles needed to represent the GMS state in RBAC. The administrative
value of the RBAC model diminishes when the number of roles exceeds the number
of users~\cite{LuHongLiu2007,RoleVAT}, and thus we assume that this scenario is
evidence of the RBAC system being used outside of the use cases it was designed
for. Thus, \cref{chart:roles} is particularly strong evidence of the
ill-suitedness of RBAC to the group messaging workload, since systems with less
than 100 users can exceed 2,000 roles, and on average there were over 14 times
as many roles as users.

Finally, as another proxy for implementation complexity, \cref{chart:extension}
shows the amount of communication with an \aum that occurred during a run,
compared with the number of users in the system. DAC's much larger extension
cost was the result of this scheme having no appropriate state elements that
could store most of the information needed in the group messaging workload,
while RBAC performed better due to its ability to store group membership,
ownership, and administration relations within its role relation.

In addition, we present in \cref{chart:results} several findings that, although
they do not support the selection of one scheme over another, nonetheless
provided insight into the group messaging workload. In \cref{chart:coi}, we show
the relationship between the number of users in the system and the proportion of
attempted conflict-of-interest workflows that are successfully completed within
the simulation run. We found that the main bottleneck for completing COIs was
the number of users. Thus, runs with fewer users had both fewer COIs complete
and a longer duration of time between the initialization and completion of those
that did. In \cref{chart:useradmin}, we present the results of our exploration
of a related facet of this relationship, in which we found a clear positive
trend between the frequency with which COIs are initiated and administrative
work completed by non-administrative users (i.e., those that had been nominated
to fulfill administrative duties temporarily).

\subsection{Summary of Findings}
\label{sec:casestudy:summary}

The creators of g-SIS believed that dissemination-centric sharing models could
represent group-centric sharing~\cite{krishnan2009sacmat}. However, we found
that, without extensions, commonly-used examples DAC and RBAC were not able to
\emph{safely} implement a workload inspired by a particular scenario for which
g-SIS is well-suited. Although SD3 can also represent dissemination-centric
sharing, we have used a particular parameterization of SD3 (which we call
SD3-GM) to represent an instantiation of the g-SIS model and implement group
messaging workload in an efficient way. Thus, it seems that at least some access
control schemes are capable of performing well in both dissemination-centric and
group-centric scenarios. By evaluating the relative suitability of
dissemination-centric schemes (extended versions of DAC and RBAC) to the group
messaging workload, we have confirmed the suspicions
in~\cite{krishnan2009sacmat} that, although these schemes can represent the
workload, they cannot address it as naturally, and suffer from inefficiencies.
This highlights the importance of conducting suitability analysis, especially
for novel applications, and confirms that expressiveness alone is not enough to
make decisions about access control schemes.

\section{Discussion}
\label{sec:discussion}

In this section, we first revisit each of the requirement for suitability
analysis frameworks outlined in \cref{sec:approach:requirements}, and then
discuss a number of areas of future work related to the suitability analysis
problem.

\subsection{Requirements Revisited}
\label{sec:discussion:requirements}

In \cref{sec:approach:requirements}, we outlined six requirements to guide the
development of our suitability analysis framework. We now discuss the degree to
which each requirement was met.

\begin{figure}
\centering
\includegraphics[width=.33\textwidth]{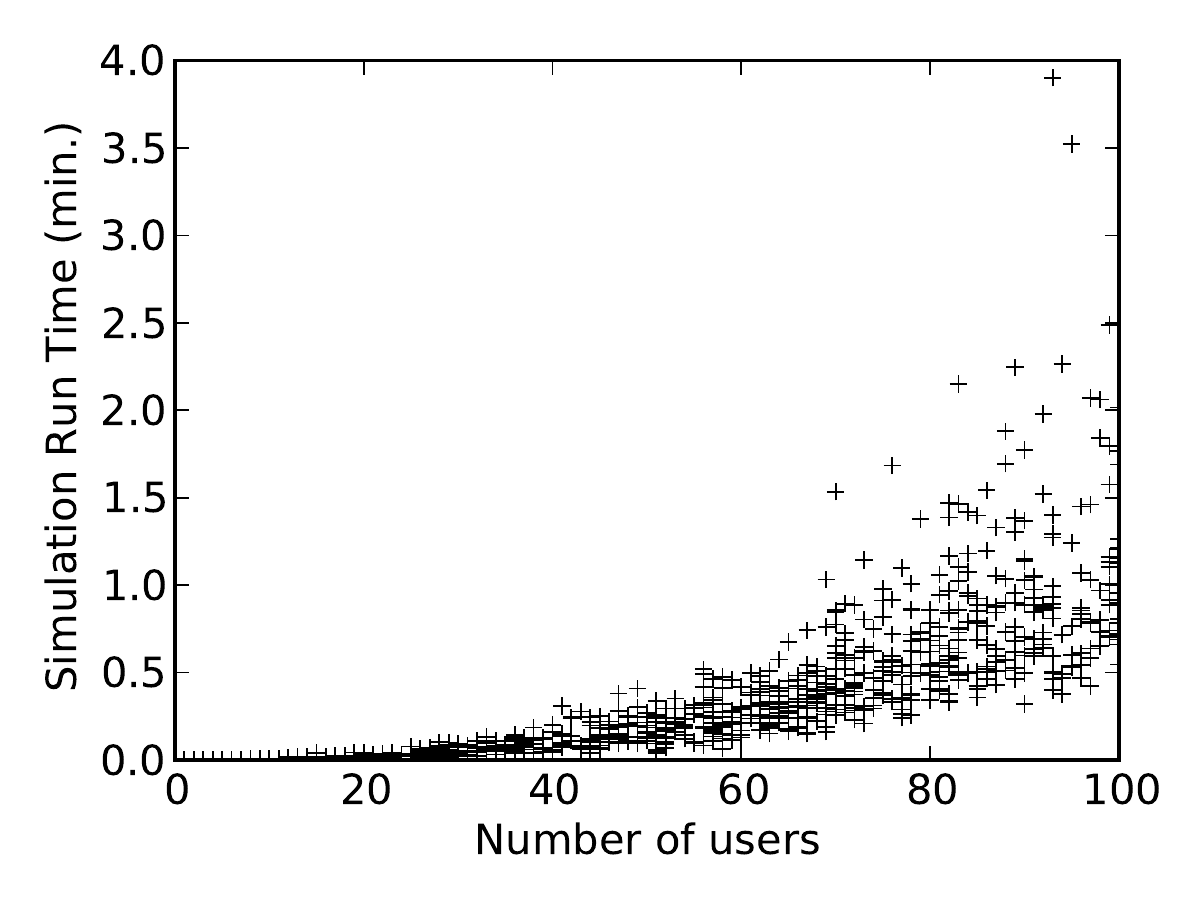}
\caption{Run time for simulating 8-hour periods in the group messaging
workload\label{chart:time}}
\end{figure}

\begin{itemize}

\item The \reqDE{} requirement is addressed equally by the workload formalism
developed in \cref{sec:expressiveness:workloads} and the Monte Carlo simulation
procedure described in \cref{sec:cost:simulation}: the former leaves the state
defining the workload and the mechanisms that can alter it completely in the
hands of the analyst, while the latter facilitates cost analysis over many such
instances of the workload.

\item \reqCI{} is met by combining the workflow and actor graph formalisms
developed in \cref{sec:cost:invocations} with the \WSP solver leveraged by
\cref{alg:simulation} in \cref{sec:cost:simulation}. Specifically, constrained
workflows articulate the ways in which cooperation must be carried out, while
the use of actor graphs and the \WSP solver ensures that all traces generated
during cost analysis are compliant with these workflows.

\item With respect to safety, we focused in this paper on a particular notion of
safe implementation---i.e., the state-matching implementation (cf.
\cref{sec:expressiveness:implementation})---and its use in extending access
control schemes and implementing workloads. However, the use of this particular
notion of safe implementation is not required by our framework: proofs of safety
are carried out manually, and thus any other notion of safe implementation could
be used. As such, our framework provides \reqTS{}.

\item In contrast to safety analysis, cost analysis is a largely automated
procedure that is constrained by our framework. However, as was demonstrated in
\cref{sec:cost:measures,sec:casestudy}, the notion of cost measure developed in
this paper is capable of representing a wide range of system- and human-centric
costs. Further, \cref{thm:vector} shows that any vector of measures is itself a
cost measure, so many costs can be considered in parallel. As such, our
framework meets the \reqTC{} requirement.

\item Supporting multi-user workflows is seemingly at odds with the \reqEF{}
requirement, as the workflow satisfiability problem has been shown to be
NP-complete~\cite{wang2007esorics}. However, the proof of \cref{thm:complexity}
makes use of recent results~\cite{wang2007esorics, crampton2012ccs} to show that
our Monte Carlo analysis process (via \cref{alg:simulation,alg:montecarlo}) is
fixed-parameter tractable if the length of workflows within the system is
treated as a small constant, as is typically the case in practice. In addition,
\cref{chart:time} shows the time required for 8-hour simulation runs using our
Java-based simulator on a 3.06 GHz Core 2 Duo with respect to the number of
users in the system (the most significant variable in the run time). This
\lcnamecref{chart:time} shows that even with many users, simulating 8-hour runs
takes less than four minutes on commodity hardware. In addition, since we
utilize a Monte Carlo approach, the multiple simulation runs are inherently
parallelizable.

\item In terms of the \reqAC{} requirement, \cref{sec:cost:simulation} discusses
how to calculate confidence intervals for point estimates of cost. Further,
\cref{alg:simulation,alg:confidence} demonstrate how the cost analysis process
can be guided by a desired confidence interval for specific configurations of
interest within the workload's parameter space.

\end{itemize}

The analysis framework developed in this paper meets each of the desiderata
outlined in \cref{sec:approach:requirements}, and provides a flexible,
efficient, and precise mechanism for analyzing instances of the access control
suitability analysis problem.

\subsection{Open Problems and Future Work}\label{sec:discussion:fw}

We now discuss the future of application-aware suitability analysis, including
refinements to our existing framework and ways in which this approach can be
extended to the formalization of more general security workloads.

\paragraph*{Implementation Non-Existence} A proof that a particular
implementation does not exist is typically harder to produce than a constructive
existence proof. Thus, in our work so far, when discussing a lack of an
implementation, we often resort to informal arguments for justification.
Ideally, it would be possible to more easily prove the non-existence of of an
implementation, since such proofs give higher confidence in the necessity of
extending access control schemes.

\paragraph*{Implementation Optimality} The constructive nature of an
implementation of a workload in a scheme leads quite naturally to the cost
analysis of this scheme, as workload actions can be translated into scheme
actions by the implementation. Given an access control scheme \(\scm{S}\) and a
workload \(W\), we therefore carry out the cost analysis of a \emph{particular}
implementation of \(W\) in \(\scm{S}\), rather than the \emph{best}
implementation of \(W\) in \(\scm{S}\). It would be useful to develop techniques
for proving the optimality of an implementation. This would enable analysts to
make strong claims about the (sub-)optimality of an access control scheme for a
given workload without needing to justify or defend the implementations used
during their analysis.

\paragraph*{Alternate Notions of Implementation} Recall that we consider a type
of safe implementation based on the state-matching reduction, the strongest type
of mapping studied in previous work~\cite{tripli}. However, other notions of
implementation certainly exist in the literature (e.g.,
see~\cite{early-sim,statetransition,osborn00tissec,sandhu92jcs,atam}), and are
likely applicable within certain classes of workloads. Understanding the
benefits and limitations of using relaxed notions of implementation is an
important area of future work. It is also important to explore relationships
(e.g., implication, equivalence) between known access control implementations,
as well as between implementations and mappings from other domains. For example,
the state-matching reduction shares certain structural properties with weak
simulations in model checking. Alternate formalizations of the access control
problem could enable the application of analysis techniques from other domains
toward access control.

\paragraph*{Quantifying Human Costs} Although the cost measures and cost
functions formalized in this paper are capable of representing a wide-range of
interesting costs, capturing human-centric costs---such as, e.g., cognitive
overheads for various tasks, or error rates in policy formulation---is a
difficult task. Our focus in this paper lies in the utilization of these types
of costs measures, rather than in their capture. However, we are inspired by
recent work within the usable security community on measuring exactly these
types of phenomena
(e.g.,~\cite{reeder2011chi,bauer2008chi,reeder2008chi,clark2007soups,mazzia2012soups}).
These types of studies provide a roadmap for suitability analysts that wish to
incorporate human costs into their analyses, and signal a shift in security
analysis: quantitative analysis of these systems cannot be done in a strictly
pencil-and-paper fashion, but must also include studies of the humans who
manipulate and administer these systems.

\paragraph*{Beyond Access Control} This paper focuses on one particular instance
of the suitability analysis problem that is specific to access control schemes.
However, we believe that suitability analysis can be cast in a more general
manner and applied to broader security workloads, as solutions to many security
problems need to balance formal requirements to be upheld by a system with the
real-world impacts and costs of these solutions. As an example, consider the
public key infrastructure (PKI) upon which many web authentication and
authorization frameworks are built. Recently, there have been high profile
compromises of CAs in the web PKI domain (e.g.,~\cite{diginotar,comodo}). These
failures have made clear the fragility of the trust model and revocation
mechanisms in the web space, and have inspired the community to examine methods
both for reinforcing the system's mechanisms to prevent fraudulent certificate
issuances and improving the robustness of the revocation infrastructure (e.g.,
Perspectives~\cite{perspectives}, Sovereign Keys~\cite{sovereign-keys}, CA
Transparency~\cite{ca-transparency}, etc.). However, there is considerable
debate in the community regarding what the appropriate metrics for judging
replacement systems should be, and how the different proposals compare under
realistic conditions. A more general formulation of the suitability analysis
problem could enable better understanding of the trade-offs between the formal
guarantees and the real-world costs incurred by such candidate infrastructures.

\section{Conclusion}
\label{sec:conclusion}

Historically, most work regarding the formal analysis of access control schemes
has focused on evaluating expressive power in absolute terms. By contrast, our
goal in this paper was to formalize the \emph{suitability analysis problem} and
to develop a methodology for application-specific evaluation of access control
schemes. To this end, we have developed a formal framework for specifying access
control workloads, reasoning about the abilities of candidate access control
schemes to safely service these workloads, safely augmenting schemes that are
incapable of implementing a given workload, and carrying out cost-based analysis
of the suitability of each candidate scheme for servicing the workload. Formal
proofs demonstrate the soundness of our approach, and a detailed case study
drawn from the literature illustrates the applicability of our framework for
conducting real world suitability analyses. The framework that we have developed
is a first step toward understanding the application-specific strengths of
access control systems. However, the basic techniques used in this framework
appear to be applicable to broader security problems, in which several systems
may be capable of meeting a set of security goals, but the costs of using each
candidate system vary.

\textbf{Acknowledgements}\, This work was supported in part by the
National Science Foundation under awards CNS-0964295, CNS-1228697, and
CNS-1228947.


\appendices
\section{Proofs}
\label[appendix]{appx:proofs}

\subsection{Proof of \Cref{thm:smimplementation}}
\label[appendix]{appx:proofs:smimplementation}

First, we restate our definition of access control scheme from
\cref{sec:expressiveness:schemes}. For the purposes of this proof, we refer to
this notion of scheme as the GLH scheme.

\renewcommand{\freeheader}{\Cref{def:scheme}}
\begin{freedefinition}
A \emph{GLH scheme} is a state transition system \(\scm{S} = \tup{\Gamma, \Psi,
Q}\), where \(\Gamma\) is the set of access control states, \(\Psi\) is the set
of commands over \(\Gamma\), and \(Q\) is the set of queries over \(\Gamma\).
\end{freedefinition}

Next, we state the definition of scheme used by Tripunitara and
Li~\cite{tripli}, which we refer to in this proof as the TL scheme.

\begin{definition}
A \emph{TL scheme} is a state-transition system \(\scm{S} = \tup{\Gamma, Q,
{\entails}, \Psi}\), in which \(\Gamma\) is a set of states, \(Q\) is a set of
queries, \({\entails}: \Gamma \times Q \to \set{\true, \false}\) is called the
entailment relation, and \(\Psi\) is a set of state-transition rules.
\end{definition}

Recall that GLH schemes formalize transitions and state inspection using
commands and queries that accept parameters. For transitions, a TL scheme
specifies a set of state transition rules, each a binary relation on the set of
states. A running system, then, must specify which transition rule is active.
For queries, a TL scheme specifies a set of (non-parameterized) queries, and the
entailment relation (rather than being individually specified as a component of
the query structure) is specified for all queries as a separate component of the
scheme.

A main result of Tripunitara and Li's framework closely mirrors
\cref{thm:smimplementation}, except for TL schemes and the state-matching
reduction rather than GLH schemes and the state-matching implementation. We now
present the definition of state-matching reduction and the related
\lcnamecref{thm:smreduction} to \cref{thm:smimplementation}.

\begin{definition}[State-Matching Reduction~\cite{tripli}]
\label{def:smreduction}
Given two access control schemes \(\scm{A} = \tup{\Gamma^\scm{A}, \Psi^\scm{A},
Q^\scm{A}, \entails^\scm{A}}\) and \(\scm{B} = \tup{\Gamma^\scm{B},
\Psi^\scm{B}, Q^\scm{B}, \entails^\scm{B}}\), and a mapping from \(\scm{A}\) to
\(\scm{B}\), \(\sigma : \prm{\Gamma^\scm{A} \times \Psi^\scm{A}} \cup Q^\scm{A}
\to \prm{\Gamma^\scm{B} \times \Psi^\scm{B}} \cup Q^\scm{B}\), we say that two
states \(\gamma^\scm{A}\) and \(\gamma^\scm{B}\) are \emph{equivalent} under
the mapping \(\sigma\) when for every \(q^\scm{A} \in Q^\scm{A}\),
\(\gamma^\scm{A} \entails^\scm{A} q^\scm{A}\) if and only if \(\gamma^\scm{B}
\entails^\scm{B} \sigma\prm{q^\scm{A}}\). A mapping \(\sigma\) from \(\scm{A}\)
to \(\scm{B}\) is said to be a \emph{state-matching reduction} if for every
\(\gamma^\scm{A} \in \Gamma^\scm{A}\) and every \(\psi^\scm{A} \in
\Psi^\scm{A}\), \(\tup{\gamma^\scm{B}, \psi^\scm{B}} =
\sigma\prm{\tup{\gamma^\scm{A}, \psi^\scm{A}}}\) has the following two
properties:
\begin{enumerate}
\item For every state \(\gamma^\scm{A}_1\) in scheme \(\scm{A}\) such that
\(\displaystyle \gamma^\scm{A} {\mathop{\mapsto}^*}_{\psi^\scm{A}}
\gamma^\scm{A}_1\), there exists a state \(\gamma^\scm{B}_1\) such that
\(\displaystyle \gamma^\scm{B} {\mathop{\mapsto}^*}_{\psi^\scm{B}}
\gamma^\scm{B}_1\) and \(\gamma^\scm{A}_1\) and \(\gamma^\scm{B}_1\) are
equivalent under \(\sigma\).
\item For every state \(\gamma^\scm{B}_1\) in scheme \(\scm{B}\) such that
\(\displaystyle \gamma^\scm{B} {\mathop{\mapsto}^*}_{\psi^\scm{B}}
\gamma^\scm{B}_1\), there exists a state \(\gamma^\scm{A}_1\) such that
\(\displaystyle \gamma^\scm{A} {\mathop{\mapsto}^*}_{\psi^\scm{A}}
\gamma^\scm{A}_1\) and \(\gamma^\scm{A}_1\) and \(\gamma^\scm{B}_1\) are
equivalent under \(\sigma\).
\end{enumerate}
\end{definition}

\begin{theorem}[Rephrased, from~\cite{tripli}]
\label{thm:smreduction}
Given two schemes \(\scm{A}\) and \(\scm{B}\), and a mapping, \(\sigma\), from
\(\scm{A}\) to \(\scm{B}\), \(\sigma\) is a state-matching reduction if and
only if it is strongly security-preserving; that is, every compositional
security analysis instance in \(\scm{A}\) is true if and only if the image of
the instance under \(\sigma\) is true in \(\scm{B}\).
\end{theorem}

Next, we restate the definition of state-matching implementation from
\cref{sec:expressiveness:implementation}.

\renewcommand{\freeheader}{\Cref{def:smimplementation}}
\begin{freedefinition}
Given an access control workload \(W = \tup{\scm{W}, I^\scm{W}}\) in which
\(\scm{W} = \tup{\Gamma^\scm{W}, \Psi^\scm{W}, Q^\scm{W}}\), an access control
scheme, \(\scm{S} = \tup{\Gamma^\scm{S}, \Psi^\scm{S}, Q^\scm{S}}\), and an
implementation \(\sigma = \tup{\sigma_\Gamma, \sigma_\Psi, \sigma_Q}\) of
\(\scm{W}\) in \(\scm{S}\), we say that two states \(\gamma^\scm{W}\) and
\(\sigma_\Gamma\prm{\gamma^\scm{W}} = \gamma^\scm{S}\) are \emph{equivalent}
with respect to the implementation \(\sigma\) (and denote this equivalence as
\(\gamma^\scm{W} \eqv{\sigma} \gamma^\scm{S}\)) when for every \(q^\scm{W} =
\tup{n, P, {\entails}} \in Q^\scm{W}\) (with \(q^\scm{S} =
\sigma_Q\prm{q^\scm{W}}\)) and every \(p^\scm{W} \in P^*\) (with \(p^\scm{S} =
\sigma_\Gamma\prm{p^\scm{W}}\)), \(\gamma^\scm{W} \entails
q^\scm{W}\prm{p^\scm{W}}\) if and only if \(\gamma^\scm{S} \entails
q^\scm{S}\prm{p^\scm{S}}\).

An implementation \(\sigma\) of \(\scm{W}\) in \(\scm{S}\) is said to be a
\emph{state-matching implementation} if for every \(\gamma^\scm{W} \in
\Gamma^\scm{W}\), with \(\gamma^\scm{S} = \sigma_\Gamma\prm{\gamma^\scm{W}}\),
the following two properties hold:
\begin{enumerate}
\item For every state \(\gamma^\scm{W}_1 \in \Gamma^\scm{W}\) such that
\(\gamma^\scm{W} \Mapsto_{\Psi^\scm{W}} \gamma^\scm{W}_1\), there exists a state
\(\gamma^\scm{S}_1 \in \Gamma^\scm{S}\) such that \(\gamma^\scm{S}
\Mapsto_{\Psi^\scm{S}} \gamma^\scm{S}_1\) and \(\gamma^\scm{W}_1 \eqv{\sigma}
\gamma^\scm{S}_1\).
\item For every state \(\gamma^\scm{S}_1 \in \Gamma^\scm{S}\) such that
\(\gamma^\scm{S} \Mapsto_{\Psi^\scm{S}} \gamma^\scm{S}_1\), there exists a state
\(\gamma^\scm{W}_1 \in \Gamma^\scm{W}\) such that \(\gamma^\scm{W}
\Mapsto_{\Psi^\scm{W}} \gamma^\scm{W}_1\) and \(\gamma^\scm{W}_1 \eqv{\sigma}
\gamma^\scm{S}_1\).
\end{enumerate}
\end{freedefinition}

Finally, we restate and prove \cref{thm:smimplementation}.

\renewcommand{\freeheader}{\Cref{thm:smimplementation}}
\begin{freetheorem}
Given an access control workload \(W = \tup{\scm{W}, I^\scm{W}}\) in which
\(\scm{W} = \tup{\Gamma^\scm{W}, \Psi^\scm{W}, Q^\scm{W}}\), an access control
scheme, \(\scm{S} = \tup{\Gamma^\scm{S}, \Psi^\scm{S}, Q^\scm{S}}\), and an
implementation \(\sigma = \tup{\sigma_\Gamma, \sigma_\Psi, \sigma_Q}\) of
\(\scm{W}\) in \(\scm{S}\), \(\sigma\) is a state-matching implementation if
and only if it is strongly security-preserving; that is, every compositional
security analysis instance in \(\scm{W}\) is true if and only if the image of
the instance under \(\sigma\) is true in \(\scm{S}\).
\end{freetheorem}

\begin{nproof}
Consider workload operational component \(\scm{W}\), scheme \(\scm{S}\), and
implementation \(\sigma^\scm{W}\) of \(\scm{W}\) using \(\scm{S}\). We assume
that \(\sigma^\scm{W}\) is a state-matching implementation and show that it must
be strongly security-preserving.

Construct TL scheme \(\scm{A}\) from \(\scm{W}\) (and, similarly, \(\scm{B}\)
from \(\scm{S}\)) as follows. Make \(\Gamma^\scm{A}\) equal to
\(\Gamma^\scm{W}\) (preserve state information exactly). Map each
query-parameter pair \(q, P\) in \(\scm{W}\) to the single query \(q_P\) in
\(Q^\scm{A}\) in scheme \(\scm{A}\). Encode these queries' individual entailment
relations (over parameters) in scheme \(\scm{A}\)'s entailment relation (over
queries). Collapse all commands in \(\scm{W}\) into a binary relation over
states. Encode this binary relation as a single state transition rule in
\(\Psi^\scm{A}\), where \(\tup{\gamma_1, \gamma_2} \in \psi^\scm{A}\) if and
only if \(\exists \psi \in \Psi^\scm{W}, P \in P^* : e\prm{\gamma_1, P} =
\gamma_2\).

Construct the reduction \(\sigma^\scm{A} : (\Gamma^\scm{A} \times \Psi^\scm{A}
\to \Gamma^\scm{B} \times \Psi^\scm{B}) \cup (Q^\scm{A} \to Q^\scm{B})\) from
the implementation \(\sigma^\scm{W} : (\Gamma^\scm{W} \to \Gamma^\scm{S}) \cup
(\Psi^\scm{W} \to \Psi^\scm{S}) \cup (Q^\scm{W} \to Q^\scm{S})\) as follows.
Encode the (state, transition rule) mapping to be equivalent to the state
mapping of \(\sigma^\scm{W}\) (trivial since there is only one transition rule).
Copy the query mapping in the obvious way.

Since \(\scm{A}\) and \(\scm{B}\) are crafted to encode the states, queries, and
reachability properties of \(\scm{W}\) and \(\scm{S}\), and \(\sigma^\scm{A}\)
encodes \(\sigma^\scm{W}\), it is clear that \(\sigma^\scm{A}\) is strongly
security-preserving if and only if \(\sigma^\scm{W}\) is. Thus, if we show that
\(\sigma^\scm{A}\) is a state-matching reduction, then by
\cref{thm:smreduction}, it is strongly security-preserving, and thus
\(\sigma^\scm{W}\) is as well. These equivalences in encoding also make it clear
that \(\sigma^\scm{A}\) is, indeed, a state-matching reduction
(\cref{def:smreduction}), by observation of the properties of \(\sigma^\scm{W}\)
as defined in \cref{def:smimplementation}.

Thus, we have shown that an implementation is a state-matching implementation
\emph{only if} it is strongly security-preserving.

Next we assume that \(\sigma^\scm{W}\) is strongly security-preserving and show
that it is a state-matching implementation. Construct \(\scm{A}\), \(\scm{B}\),
and \(\sigma^\scm{A}\) as above. Since \(\sigma^\scm{W}\) is strongly
security-preserving, so is \(\sigma^\scm{A}\). Thus, \(\sigma^\scm{A}\) is a
state-matching reduction by \cref{thm:smreduction}. Finally, using an argument
as above, \(\sigma^\scm{W}\) is a state-matching implementation.

Thus, we have shown that an implementation is a state-matching implementation
\emph{if} it is strongly security-preserving, and thus an implementation is
state-matching \emph{if and only if} it is strongly security-preserving.
\end{nproof}

\subsection{Proof of \Cref{thm:vector}}
\label[appendix]{appx:proofs:vector}

First, we present several requisite definitions.

\begin{definition}[Abelian Monoid]
An \emph{abelian monoid}, \(\mathbf{S} = \tup{S, {\op}}\), is a set, \(S\),
together with a binary operation, \(\op\), that satisfies the following
properties.
\begin{enumerate}
\item \textbf{Closure} \;
  \(\forall a, b \in S, a \op b \in S\)
\item \textbf{Associativity} \;
  \(\forall a, b, c \in S, \prm{a \op b} \op c = a \op \prm{b \op c}\)
\item \textbf{Commutativity} \;
  \(\forall a, b \in S, a \op b = b \op a\)
\item \textbf{Identity} \;
  \(\exists 0 \in S, \forall a \in S, a \op 0 = a\)
\end{enumerate}
\end{definition}

\begin{definition}[Partially Ordered Set]
A \emph{partially ordered set}, \(\mathbf{S} = \tup{S, {\cmp}}\), is a set,
\(S\), together with a binary relation, \(\cmp\), that satisfies the following
properties.
\begin{enumerate}
\item \textbf{Reflexivity} \;
  \(\forall a \in S, a \cmp a\)
\item \textbf{Antisymmetry} \;
  \(\forall a, b \in S, a \cmp b \land b \cmp a \Rightarrow a = b\)
\item \textbf{Transitivity} \;
  \(\forall a, b, c \in S, a \cmp b \land b \cmp c \Rightarrow a \cmp c\)
\end{enumerate}
\end{definition}

\begin{definition}[Ordered Abelian Monoid]
An \emph{ordered abelian monoid}, \(\mathbf{S} = \tup{S, {\op}, {\cmp}}\), is a
set, \(S\), together with a binary operator, \(\op\), and binary relation,
\(\cmp\), that satisfies the following properties.
\begin{enumerate}
\item \(\tup{S, {\op}}\) is an abelian monoid
\item \(\tup{S, {\cmp}}\) is a partially ordered set
\end{enumerate}
\end{definition}

Now, we restate the definition of a vector of cost measures from
\cref{sec:cost:measures}.

\renewcommand{\freeheader}{\Cref{def:vector}}
\begin{freedefinition}[Vector of Cost Measures]
Given cost measures
\(\mathbf{N}_1 = \tup{N_1, {\op[1]}, {\cmp[1]}}\),
\(\mathbf{N}_2 = \tup{N_2, {\op[2]}, {\cmp[2]}}\),
\ldots,
\(\mathbf{N}_i = \tup{N_i, {\op[i]}, {\cmp[i]}}\),
let \(\mathbf{M} = \tup{M, {\op[*]}, {\cmp[*]}}\) be the \emph{vector} of
cost measures \(\mathbf{N}_1, \mathbf{N}_2, \ldots, \mathbf{N}_i\), where:
\begin{itemize}
\item
  \(M = N_1 \times N_2 \times \cdots \times N_i\).
\item
  Given
  \(a_1, b_1 \in \mathbf{N}_1\),
  \(a_2, b_2 \in \mathbf{N}_2\),
  \ldots,
  \(a_i, b_i \in \mathbf{N}_i\),
  \(\tup{a_1, a_2, \ldots, a_i} \op[*]
    \tup{b_1, b_2, \ldots, b_i} =
    \tup{a_1 \op[1] b_1,
         a_2 \op[2] b_2,
         \ldots,
         a_i \op[i] b_i}
  \).
\item
  Given
  \(a_1, b_1 \in \mathbf{N}_1\),
  \(a_2, b_2 \in \mathbf{N}_2\),
  \ldots,
  \(a_i, b_i \in \mathbf{N}_i\),
  \(\tup{a_1, a_2, \ldots, a_i} \cmp[*]
    \tup{b_1, b_2, \ldots, b_i}
  \) if and only if
  \(a_1 \cmp[1] b_1 \land
    a_2 \cmp[2] b_2 \land
    \ldots \land
    a_i \cmp[i] b_i
  \).
\end{itemize}
\end{freedefinition}

\renewcommand{\freeheader}{\Cref{thm:vector}}
\begin{freetheorem}
Given cost measures
\(\mathbf{N}_1 = \tup{N_1, {\op[1]}, {\cmp[1]}}\),
\(\mathbf{N}_2 = \tup{N_2, {\op[2]}, {\cmp[2]}}\),
\ldots,
\(\mathbf{N}_i = \tup{N_i, {\op[i]}, {\cmp[i]}}\),
and their vector,
\(\mathbf{M} = \tup{M, {\op[*]}, {\cmp[*]}}\),
\(\mathbf{M}\) is a cost measure.
\end{freetheorem}

\begin{nproof}
All of \(\mathbf{N}_1, \mathbf{N}_2, \ldots, \mathbf{N}_i\) are cost measures.
By the definition of cost measure, they are all abelian monoids, and thus are
all closed, associative, and commutative, and all have identities. Using
\(\mathbf{N}_1\) as an example, this implies:
\begin{enumerate}
\item \(\forall a, b \in N_1, a \op[1] b \in N_1\)
\item \(\forall a, b, c \in N_1, (a \op[1] b) \op[1] c = a \op[1] (b \op[1] c)\)
\item \(\forall a, b \in N_1, a \op[1] b = b \op[1] a\)
\item \(\exists 0_1 \in N_1, \forall a \in N_1, a \op[1] 0_1 = a\)
\end{enumerate}

Let \(A, B, C \in \mathbf{M}\). By the definition of vector of cost measures,
\[A = \tup{a_1, a_2, \ldots, a_i}\]
where
\[a_1 \in \mathbf{N}_1, a_2 \in \mathbf{N}_2, \ldots, a_i \in \mathbf{N}_i\]
and similarly for \(B\) and \(C\).

By the definition of vector,
\[A \op[*] B = \tup{a_1 \op[1] b_1, a_2 \op[2] b_2, \ldots, a_i \op[i] b_i}\]
By the closure of \(\mathbf{N}_1, \mathbf{N}_2, \ldots, \mathbf{N}_i\),
\[a_1 \op[1] b_1 \in \mathbf{N}_1, a_2 \op[2] b_2 \in \mathbf{N}_2, \ldots, a_i \op[i] b_i \in \mathbf{N}_i\]
\[A \op[*] B \in \mathbf{M}\]

Thus, \(\mathbf{M}\) satisfies the property of \emph{closure}.

By the definition of vector,
\begin{align*}
\prm{A \op[*] B}& \op[*] C =\\
&\tup{\prm{a_1 \op[1] b_1} \op[1] c_1, \prm{a_2 \op[2] b_2} \op[2] c_2, \ldots, \prm{a_i \op[i] b_i} \op[i] c_i}
\end{align*}
By the associativity of \(\mathbf{N}_1, \mathbf{N}_2, \ldots, \mathbf{N}_i\),
\begin{align*}
\prm{A \op[*] B}& \op[*] C =\\
&\tup{a_1 \op[1] \prm{b_1 \op[1] c_1}, a_2 \op[2] \prm{b_2 \op[2] c_2}, \ldots, a_i \op[i] \prm{b_i \op[i] c_i}}
\end{align*}
\[\prm{A \op[*] B} \op[*] C = A \op[*] \prm{B \op[*] C}\]

Thus, \(\mathbf{M}\) satisfies the property of \emph{associativity}.

By the definition of vector,
\[A \op[*] B = \tup{a_1 \op[1] b_1, a_2 \op[2] b_2, \ldots, a_i \op[i] b_i}\]
By the commutativity of \(\mathbf{N}_1, \mathbf{N}_2, \ldots, \mathbf{N}_i\),
\[A \op[*] B = \tup{b_1 \op[1] a_1, b_2 \op[2] a_2, \ldots, b_i \op[i] a_i}\]
\[A \op[*] B = B \op[*] A\]

Thus, \(\mathbf{M}\) satisfies the property of \emph{commutativity}.

By the definition of vector,
\[0 \op[*] A = \tup{0_1 \op[1] a_1, 0_2 \op[2] a_2, \ldots, 0_i \op[i] a_i}\]
By the identity of \(\mathbf{N}_1, \mathbf{N}_2, \ldots, \mathbf{N}_i\),
\[0 \op[*] A = \tup{a_1, a_2, \ldots, a_i}\]
\[0 \op[*] A = A\]

Thus, \(\mathbf{M}\) satisfies the property of \emph{identity}.

Since \(\mathbf{M}\) satisfies closure, associativity, commutativity, and
identity, \(\tup{M, {\op[*]}}\) is an abelian monoid.

All of \(\mathbf{N}_1, \mathbf{N}_2, \ldots, \mathbf{N}_i\) are cost measures.
Thus, they are all partially ordered sets, and thus are all reflexive,
antisymmetric, and transitive. Using \(\mathbf{N}_1\) as an example, this
implies:
\begin{enumerate}
\item \(\forall a \in N_1, a \cmp[1] a\)
\item \(\forall a, b \in N_1, a \cmp[1] b \land b \cmp[1] a \Rightarrow a = b\)
\item \(\forall a, b, c \in N_1, a \cmp[1] b \land b \cmp[1] c \Rightarrow a \cmp[1] c\)
\end{enumerate}

Let \(A, B, C \in \mathbf{M}\). By the definition of vector of cost measures,
\[A = \tup{a_1, a_2, \ldots, a_i}\]
where
\[a_1 \in \mathbf{N}_1, a_2 \in \mathbf{N}_2, \ldots, a_i \in \mathbf{N}_i\]
and similarly for \(B\) and \(C\).

By the reflexivity of \(\mathbf{N}_1, \mathbf{N}_2, \ldots, \mathbf{N}_i\),
\[a_1 \cmp[1] a_1, a_2 \cmp[2] a_2, \ldots, a_i \cmp[i] a_i\]
\[A \cmp[*] A\]

Thus, \(\mathbf{M}\) satisfies the property of \emph{reflexivity}.

Assume \(A \cmp[*] B \land B \cmp[*] A\).
By the definition of vector,
\[a_1 \cmp[1] b_1 \land b_1 \cmp[1] a_1 \land a_2 \cmp[2] b_2 \land b_2 \cmp[2] a_2 \land \ldots \land a_i \cmp[i] b_i \land b_i \cmp[i] a_i\]
By the antisymmetry of \(\mathbf{N}_1, \mathbf{N}_2, \ldots, \mathbf{N}_i\),
\[a_1 = b_1 \land a_2 = b_2 \land \ldots \land a_i = b_i\]
\[A = B\]
\[A \cmp[*] B \land B \cmp[*] A \Rightarrow A = B\]

Thus, \(\mathbf{M}\) satisfies the property of \emph{antisymmetry}.

Assume \(A \cmp[*] B \land B \cmp[*] C\).
By the definition of vector,
\[a_1 \cmp[1] b_1 \land b_1 \cmp[1] c_1 \land a_2 \cmp[2] b_2 \land b_2 \cmp[2] c_2 \land \ldots \land a_i \cmp[i] b_i \land b_i \cmp[i] c_i\]
By the transitivity of \(\mathbf{N}_1, \mathbf{N}_2, \ldots, \mathbf{N}_i\),
\[a_1 \cmp[1] c_1 \land a_2 \cmp[2] c_2 \land \ldots \land a_i \cmp[i] c_i\]
\[A \cmp[*] C\]
\[A \cmp[*] B \land B \cmp[*] C \Rightarrow A \cmp[*] C\]

Thus, \(\mathbf{M}\) satisfies the property of \emph{transitivity}.

Since \(\mathbf{M}\) satisfies reflexivity, antisymmetry, and transitivity,
\(\tup{M, {\cmp}}\) is a partially ordered set.

Since \(\tup{M, {\op[*]}}\) is an abelian monoid and \(\tup{M, {\cmp[*]}}\) is
a partially ordered set, \(\mathbf{M} = \tup{M, {\op[*]}, {\cmp[*]}}\) is an
ordered abelian monoid.

All of \(\mathbf{N}_1, \mathbf{N}_2, \ldots, \mathbf{N}_i\) are cost measures.
Thus, they all satisfy \emph{non-negativity}. Using \(\mathbf{N}_1\) as an
example, this means:
\[\forall a, b \in N_1, a \cmp[1] a \op[1] b\]

By the definition of vector,
\[A \op[*] B = \tup{a_1 \op[1] b_1, a_2 \op[2] b_2, \ldots, a_i \op[i] b_i}\]
By the non-negativity of \(\mathbf{N}_1, \mathbf{N}_2, \ldots, \mathbf{N}_i\),
\[a_1 \cmp[1] a_1 \op[1] b_1 \land a_2 \cmp[2] a_2 \op[1] b_2 \land \cdots \land a_i \cmp[i] a_i \op[i] b_i\]
By the definition of vector,
\[A \cmp[*] B\]

Thus, \(\mathbf{M}\) satisfies the property of \emph{non-negativity}.

Since \(\mathbf{M}\) is an ordered abelian monoid and satisfies non-negativity,
\(\mathbf{M}\) is a cost measure.
\end{nproof}

\subsection{Proof of \Cref{thm:complexity}}
\label[appendix]{appx:proofs:complexity}

\renewcommand{\freeheader}{\Cref{thm:complexity}}
\begin{freetheorem}
Assuming that workflow constraints are restricted to the binary operators
\(\set{{=}, {\neq}}\) (i.e., constraints expressing binding of duty and
separation of duty), the simulation procedure described in \cref{alg:simulation}
is pseudo-polynomial in the number of simulated steps and \FPT with parameter
\(\alpha\), the number of actions in the largest task (i.e., the size of the
largest disjoint subgraph of the workflow graph).
\end{freetheorem}

\begin{nproof}
Our proof is by observation of \cref{alg:simulation}. The first loop
(\textbf{for all} \(\scm{S} =\)\ldots) handles assignments and initializations.
The final loop (\textbf{for all} \(\scm{S} \in\)\ldots) outputs results. The
main loop, then, contains all of the computationally intensive code.

The expensive section of the algorithm starts after several nested loops, adding
multiplicative factors for number of time steps (\(T_f / t\)), number of schemes
(\(\card{\Sigma}\)), and number of actors. The steps with computational overhead
are \proc{nextAction}, which polls an actor machine for the next action, and
\proc{WSat}, which calculates whether a particular action can be taken by an
actor without causing any workflow instances to become unsatisfiable. We defer
in-depth discussion of the \WSP problem and its complexity to previous
work~\cite{wang2007esorics,crampton2012ccs}, but it is an \NP-complete problem
with known algorithms that run in fixed parameterized time with parameter
\(\alpha\), the largest number of steps in a workflow task.

By previous work~\cite{wang2007esorics}, \WSP can be solved in \(\bigO{C \cdot
A^\alpha}\), where \(C\) is the number of constraints, \(A\) is the maximum
number of actors, and \(\alpha\) is the number of steps in the largest task
(i.e., the size of the largest disjoint subgraph of the workflow graph). This
greatly exceeds \proc{nextAction}, which executes a single step in a
continuous-time probabilistic machine (polynomial in actor machine size). Thus,
the dominant factor in the complexity of \cref{alg:simulation} is \(\bigO{S
\cdot C \cdot T \cdot A^{\alpha+1}}\), where \(S\) is the number of schemes and
\(T\) is the number of time steps to simulate (\(T_f / t\)). Since \(T\) is an
input, this means the algorithm is pseudo-polynomial in \(T\) and \FPT in
\(\alpha\). Since some consider \FPT to be a generalization of pseudo-polynomial
time~\cite{hromkovic}, we refer to the complexity of \cref{alg:simulation} as
\FPT, thus meeting our definition of tractable.
\end{nproof}

\subsection{Proof of \Cref{thm:auxm}}
\label[appendix]{appx:proofs:auxm}

\renewcommand{\freeheader}{\Cref{thm:auxm}}
\begin{freetheorem}
Given access control scheme \(\scm{S} = \tup{\Gamma^\scm{S}, \Psi^\scm{S},
Q^\scm{S}}\) and access control \aum \(\scm{U} = \tup{\Gamma^\scm{U},
\Psi^\scm{U}, Q^\scm{U}}\), there exists a state-matching implementation of
\(\scm{S}\) in \(\scm{S} \circ \scm{U}\).
\end{freetheorem}

\begin{nproof}
By construction. Presented is a mapping, and proof that the mapping satisfies
the two properties for it to be a state-matching implementation.

Let \(\scm{Q} = \scm{S} \circ \scm{U}\).

The mapping, \(\sigma\), needs to be able to map every \(\gamma \in
\Gamma^\scm{S}\), \(\psi \in \Psi^\scm{S}\), and \(q \in Q^\scm{S}\) in scheme
\(\scm{S}\) to \(\gamma^\scm{Q} \in \Gamma^\scm{Q}\), \(\psi^\scm{Q} \in
\Psi^\scm{Q}\), and \(q^\scm{Q} \in Q^\scm{Q}\) in scheme \(\scm{Q} = \scm{S}
\circ \scm{U}\).

Let \(\sigma\prm{\gamma} = \tup{\gamma, \gamma_{*}}\), where \(\gamma_{*}\) is
an arbitrary \aum-state for \AM \(\scm{U}\). That is, let the \AM component of
the state be arbitrary, but maintain the original scheme component of the state.

Let \(\sigma\prm{\psi} = \psi\) and \(\sigma\prm{q} = q\), since by
\cref{def:augmented} the commands and queries in \(\scm{S}\) exist unaltered in
\(\scm{S} \circ \scm{U}\).

Let \(\gamma_0\) be a start state in \(\scm{S}\). Produce \(\gamma^\scm{Q}_0\)
in \(\scm{S} \circ \scm{U}\) using \(\sigma\). Given \(\gamma_k\) such that
\(\gamma_0 \Mapsto_\psi \gamma_k\), we show that there exists
\(\gamma^\scm{Q}_k\) such that \(\gamma^\scm{Q}_0 \Mapsto_{\psi^\scm{Q}}
\gamma^\scm{Q}_k\) where, for all \(q\) and all parameterizations \(P\),
\(\gamma^\scm{Q}_k \entails_{q^\scm{Q}}\prm{\sigma\prm{P}}\) if and only if
\(\gamma_k \entails_q\prm{P}\).

From \(\gamma^\scm{Q}_0 = \tup{\gamma_0, \gamma_{*}}\), construct
\(\gamma^\scm{Q}_k\) by following the same string of commands that were executed
in transitioning from \(\gamma_0\) to \(\gamma_k\). Since, by
\cref{def:augmented}, commands in \(\scm{S}\) exist unaltered in \(\scm{Q}\),
the resulting state is \(\gamma^\scm{Q}_k = \tup{\gamma_k, \gamma_{*}}\). Thus,
since \(\scm{S}\)'s queries also exist in \(\scm{Q}\), \(\gamma^\scm{Q}_k
\entails_{q^\scm{Q}}\prm{\sigma\prm{P}}\) if and only if \(\gamma_k
\entails_q\prm{P}\).

Therefore, we have proven property~(1) for the state-matching implementation.

We prove that property~(2) for a state-matching implementation is satisfied by
our mapping also by construction. Let \(\gamma^\scm{Q}_0\) be the start-state in
\(\scm{S} \circ \scm{U}\) corresponding to \(\gamma_0\), the start-state in
\(\scm{S}\). Then, if \(\gamma^\scm{Q}_k\) is a state reachable from
\(\gamma^\scm{Q}_0\) and \(q^\scm{Q}\) is a query in \(\scm{S} \circ \scm{U}\)
whose corresponding query in \(\scm{S}\) is \(q\), we construct \(\gamma_k\)
from \(\gamma_0\) by executing each \(\psi_i \in \Psi^\scm{Q} = \tup{\psi_1,
\ldots, \psi_k}\) such that \(\exists \psi\Prime_i \in \Psi^\scm{S} :
\delta\prm{\psi\Prime_i} = \psi_i\). That is, we execute the same string of
commands used in transitioning from \(\gamma^\scm{Q}_0\) to
\(\gamma^\scm{Q}_k\), excluding the commands that are a part of
\(\Psi^\scm{U}\). By \cref{def:augmented}, and by an argument similar to above,
\(\gamma_k \entails q\) if and only if \(\gamma^\scm{Q}_k \entails q^\scm{Q}\).

Therefore, we have proven property~(2) for state-matching implementations, and
proven that our mapping \(\sigma\) is a state-matching implementation.
\end{nproof}

\section{Expressiveness Evaluation Details}
\label[appendix]{appx:expressiveness}

\subsection{GMS}
\label[appendix]{appx:expressiveness:gms}

The GMS scheme is defined as \(\scm{G} = \tup{\Gamma^\scm{G}, \Psi^\scm{G},
Q^\scm{G}}\). Its states, \(\Gamma^\scm{G}\), are defined by the sets \(\tup{U,
G, M, T, T_c, O, A, R, TX}\), where:
\begin{itemize}
\item \(U\) is the set of users
\item \(G\) is the set of groups
\item \(M\) is the set of messages
\item \(T\) is the ordered set of timestamps, including special timestamp \(\infty\)
\item \(T_c\) is the current timestamp
\item \(O \subseteq U \times G\) is the group ownership relation
\item \(A \subseteq U \times G\) is the group administration relation
\item \(R \subseteq U \times G \times T \times T\) is the group membership record
\item \(TX \subseteq G \times M \times T\) is the messaging transcript
\end{itemize}

GMS's commands, \(\Psi^\scm{G}\), include the following.
\begin{alltt}\smaller
CreateGroup(\(u\),\(g\))
\+\(G\is{}G\cup{}\set{g}\)
\+\(O\is{}O\cup{}\set{\tup{u,g}}\)
\+\(A\is{}A\cup{}\set{\tup{u,g}}\)
\+\(R\is{}R\cup{}\set{\tup{u,g,0,\infty}}\)
\+
GrantAdmin(\(o\),\(u\),\(g\))
\+if \(\tup{o,g}\in{}O\)
\+\+\(A\is{}A\cup{}\set{\tup{u,g}}\)
\+
RevokeAdmin(\(o\),\(u\),\(g\))
\+if \(\tup{o,g}\in{}O \lor o=u\)
\+\+\(A\is{}A-\set{\tup{u,g}}\)
\+
SAddMember(\(a\),\(u\),\(g\))
\+if \(\tup{a,g}\in{}A\)
\+\+\(R\is{}R\cup{}\set{\tup{u,g,T\sb{c},\infty}}\)
\+\+\(T\sb{c}\is{}T\sb{c}+1\)
\+
LAddMember(\(a\),\(u\),\(g\))
\+if \(\tup{a,g}\in{}A\)
\+\+\(R\is{}R\cup{}\set{\tup{u,g,0,\infty}}\)
\+
SRemoveMember(\(a\),\(u\),\(g\))
\+if \(\tup{a,g}\in{}A \lor a=u\)
\+\+\(R\is{}R-\set{\tup{u,g,t,t\sp\prime}:\tup{u,g,t,t\Prime}\in{}R}\)
\+
LRemoveMember(\(a\),\(u\),\(g\))
\+if \(\tup{a,g}\in{}A\)
\+\+\(R\is{}R\cup{}\set{\tup{u,g,t,T\sb{c}}:\tup{u,g,t,\infty}\in{}R}\)
\+\+\(R\is{}R-\set{\tup{u,g,t,\infty}:\tup{u,g,t,\infty}\in{}R}\)
\+
Post(\(u\),\(g\),\(m\))
\+if \(\exists{}t\in{}T:\tup{u,g,t,\infty}\in{}R\)
\+\+\(TX\is{}TX\cup{}\set{\tup{g,m,T\sb{c}}}\)
\+\+\(T\sb{c}\is{}T\sb{c}+1\)
\end{alltt}

Finally, GMS's queries, \(Q^\scm{G}\), include the following.
\begin{alltt}\smaller
Access(\(u\),\(m\))
\+\(\exists{}g\in{}G,t\sb{l},t,t\sb{u}\in{}T:\)
\+\+\+\(\tup{u,g,t\sb{l},t\sb{u}}\in{}R \land \tup{g,m,t}\in{}TX \land t\sb{l}\leq{}t\leq{}t\sb{u}\)
\end{alltt}

\subsection{RBAC}
\label[appendix]{appx:expressiveness:rbac}

{\newcommand{\sigmaR}{\sigma^\scm{R}}

RBAC is a role-based access control scheme,\footnote{There are many competing
definitions for role-based access control schemes in the literature. We derive
our definition of RBAC from NIST RBAC~\cite{nist-rbac}. We exclude from the
state elements to maintain sessions as well as several derived relations,
changes which have also been suggested by others~\cite{LiByunBertino2007}.}
\(\scm{R} = \tup{\Gamma^\scm{R}, \Psi^\scm{R}, Q^\scm{R}}\). Its states,
\(\Gamma^\scm{R}\), are defined by the sets \(\tup{U, R, P, UA, PA}\), where:
\begin{itemize}
\item \(U\) is the set of users
\item \(R\) is the set of roles
\item \(P\) is the set of permissions
\item \(UA \subseteq U \times R\) is the user-assignment relation
\item \(PA \subseteq P \times R\) is the permission-assignment relation
\end{itemize}

RBAC's commands, \(\Psi^\scm{R}\), include the following.
\begin{alltt}\smaller
AddRole(\(a\), \(r\))
\+if \(\tup{a,\text{admin}}\in{}UA\)
\+\+\(R\is{}R\cup{}\set{r}\)
\+
DeleteRole(\(a\), \(r\))
\+if \(\tup{a,\text{admin}}\in{}UA\)
\+\+\(R\is{}R-\set{r}\)
\+
AssignUser(\(a\), \(u\), \(r\))
\+if \(\tup{a,\text{admin}}\in{}UA\)
\+\+\(UA\is{}UA\cup{}\set{\tup{u,r}}\)
\+
DeassignUser(\(a\), \(u\), \(r\))
\+if \(\tup{a,\text{admin}}\in{}UA\)
\+\+\(UA\is{}UA-\set{\tup{u,r}}\)
\+
GrantPermission(\(a\), \(p\), \(r\))
\+if \(\tup{a,\text{admin}}\in{}UA\)
\+\+\(PA\is{}PA\cup{}\set{\tup{p,r}}\)
\+
RevokePermission(\(a\), \(p\), \(r\))
\+if \(\tup{a,\text{admin}}\in{}UA\)
\+\+\(PA\is{}PA-\set{\tup{p,r}}\)
\end{alltt}

Finally, RBAC's queries, \(Q^\scm{R}\), include the following.
\begin{alltt}\smaller
Access(\(u\), \(p\))
\+\(\exists{}r\in{}R:\tup{u,r}\in{}UR \land \tup{r,p}\in{}PA\)
\+
Assigned(\(u\), \(r\))
\+\(\tup{u,r}\in{}UR\)
\end{alltt}

We extend RBAC with \AM \(\scm{U} = \tup{\Gamma^\scm{U}, \Psi^\scm{U},
Q^\scm{U}}\). The \AM's states, \(\Gamma^\scm{U}\), are defined by the sets
\(\tup{G, GM}\), where:
\begin{itemize}
\item \(G\) is the set of groups
\item \(GM \subseteq G \times P\) is the group-message relation
\end{itemize}

The extension's commands, \(\Psi^\scm{U}\), include the following.
\begin{alltt}\smaller
CreateGroup(\(u\), \(g\))
\+if \(\tup{u,\text{admin}}\in{}UA\)
\+\+\(G\is{}G\cup{}\set{g}\)
\+
AssociateWithGroup(\(u\), \(g\), \(p\))
\+if \(\tup{u,\text{admin}}\in{}UA\)
\+\+\(GM\is{}GM\cup{}\set{\tup{g,p}}\)
\end{alltt}

Finally, \(Q^\scm{U} = \emptyset\), and thus the extension does not add any
queries to the scheme.

We can now demonstrate the implementation of GMS using \(\text{RBAC} \circ
\scm{U}\). To describe an implementation \(\sigmaR\) of GMS in \(\text{RBAC}
\circ \scm{U}\), we must describe the state-to-state mapping
(\(\sigmaR_\Gamma\)), the command-to-command mapping (\(\sigmaR_\Psi\)), and the
query-to-query mapping (\(\sigmaR_Q\)).

First, we describe \(\sigmaR_\Gamma\), which maps a state in GMS,
\(\gamma^\scm{G} \in \Gamma^\scm{G}\), to a state in \(\text{RBAC} \circ
\scm{U}\), \(\sigmaR_\Gamma\prm{\gamma^\scm{G}} = \gamma^\scm{R} \in
\Gamma^\scm{R}\), as follows. Users are mapped in the obvious way. Each message
\(m\) posted to any group is mapped to a permission \(m \in P\), which grants
read access to the message. Each such permission is then assigned to role \(r^m
\in R\). Each user is assigned to role \(r^m\) for each message \(m\) she has
access to. We also store and assign roles \(m^g\), \(o^g\), and \(a^g\) for
current membership, ownership, and administration of group \(g \in G\),
respectively. These roles allow certain commands to be executed, but do not
correspond to a permission in \(P\).

Now, we describe \(\sigmaR_\Psi\), which maps commands in GMS to strings of
commands in \(\text{RBAC} \circ \scm{U}\).
\begin{itemize}
\item \textstt{CreateGroup(\(u\), \(g\))} in GMS is mapped to the sequence
\textstt{CreateGroup(\(u\), \(g\))}, \textstt{AddRole(\(u\), \(m^g\))},
\textstt{AssignUser(\(u\), \(u\), \(m^g\))}, \textstt{AddRole(\(u\), \(o^g\))},
\textstt{AssignUser(\(u\), \(u\), \(o^g\))}, \textstt{AddRole(\(u\), \(a^g\))},
\textstt{AssignUser(\(u\), \(u\), \(a^g\))} in \(\text{RBAC} \circ \scm{U}\).
\item \textstt{GrantAdmin(\(u\), \(u_2\), \(g\))} in GMS is mapped to
\textstt{AssignUser(\(u\), \(u_2\), \(a^g\))} in \(\text{RBAC} \circ \scm{U}\).
\item \textstt{RevokeAdmin(\(u\), \(u_2\), \(g\))} in GMS is mapped to
\textstt{DeassignUser(\(u\), \(u_2\), \(a^g\))} in \(\text{RBAC} \circ
\scm{U}\).
\item \textstt{SAddMember(\(u\), \(u_2\), \(g\))} in GMS is mapped to
\textstt{AssignUser(\(u\), \(u_2\), \(m^g\))} in \(\text{RBAC} \circ \scm{U}\).
\item \textstt{LAddMember(\(u\), \(u_2\), \(g\))} in GMS is mapped to
\textstt{AssignUser(\(u\), \(u_2\), \(m^g\))} in \(\text{RBAC} \circ \scm{U}\),
followed by \textstt{AssignUser(\(u\), \(u_2\), \(r^m\))} for each \(m\) such
that \(\tup{g, m} \in GM\).
\item \textstt{SRemoveMember(\(u\), \(u_2\), \(g\))} in GMS is mapped to
\textstt{DeassignUser(\(u\), \(u_2\), \(m^g\))} in \(\text{RBAC} \circ
\scm{U}\), followed by \textstt{DeassignUser(\(u\), \(u_2\), \(r^m\))} for each
\(m\) such that \(\tup{g, m} \in GM\).
\item \textstt{LRemoveMember(\(u\), \(u_2\), \(g\))} in GMS is mapped to
\textstt{DeassignUser(\(u\), \(u_2\), \(m^g\))} in \(\text{RBAC} \circ
\scm{U}\).
\item \textstt{Post(\(u\), \(g\), \(m\))} in GMS is mapped to
\textstt{AssociateWithGroup(\(u\), \(g\), \(m\))} in \(\text{RBAC} \circ
\scm{U}\), followed by \textstt{AssignUser(\(u\), \(u_2\), \(r^m\))} for each
\(u_2\) such that \(\tup{u_2, m^g} \in UA\).
\end{itemize}

Finally, \(\sigmaR_Q\) maps \textstt{Access(\(u\), \(m\))} in GMS to
\textstt{Access(\(u\), \(p^m\))} in \(\text{RBAC} \circ \scm{U}\).

\begin{theorem}
\label{thm:implementation:rbac}
\(\sigmaR\) is a state-matching implementation of GMS in \(\text{RBAC} \circ
\scm{U}\).
\end{theorem}

\begin{nproof}
First, we prove property~(1) for state-matching implementations.

Let \(\gamma_0\) be a start state in GMS. Produce \(\gamma^\scm{R}_0\) in
\(\text{RBAC} \circ \scm{U}\) using \(\sigmaR_\Gamma\). Given \(\gamma_k\) such
that \(\gamma_0 \Mapsto \gamma_k\), we show that there exists
\(\gamma^\scm{R}_k\) such that \(\gamma^\scm{R}_0 \Mapsto \gamma^\scm{R}_k\)
where, for all queries \(q = \tup{n, P, {\entails}} \in Q^\scm{G}\) and
parameterizations \(p \in P^*\), \(\gamma^\scm{R}_k \entails q^\scm{R}\prm{p}\)
if and only if \(\gamma_k \entails q\prm{p}\).

Consider the case where \(\gamma_k = \gamma_0\), then let \(\gamma^\scm{R}_k =
\gamma^\scm{R}_0\). By inspection of the procedure for \(\sigmaR_\Gamma\),
\(\gamma_k \entails q\prm{p}\) if and only if \(\gamma^\scm{R}_k \entails
q^\scm{R}\prm{p}\).

Next, consider some arbitrary \(\gamma_k\) reachable from \(\gamma_0\). We
construct \(\gamma^\scm{R}_k\) that is reachable from \(\gamma^\scm{R}_0\) and
that answers every \(q^\scm{R}\prm{p}\) in the same way that \(\gamma_k\)
answers \(q\prm{p}\), as per \(\sigmaR_\Psi\). Since \(\gamma_0 \Mapsto
\gamma_k\), there exists a sequence of commands \(\tup{\psi_1 = \tup{n_1, P_1,
e_1}, \ldots, \psi_k = \tup{n_k, P_k, e_k}}\) and a sequence of
parameterizations \(\tup{p_1 \in P_1^*, \ldots, p_k \in P_k^*}\) of these
commands such that \(\gamma_k = e_k\prm{\ldots e_1\prm{\gamma_0, p_1}, \ldots,
p_k}\). For each command\slash{}parameterization pair \(\tup{\psi_i, p_i}\), we
show that the same queries change value between \(\gamma_{i-1}\) and \(\gamma_i
= e_i\prm{\gamma_{i-1}, p_i}\) and between \(\gamma^\scm{R}_{i-1} =
\sigmaR_\Psi\prm{\gamma_{i-1}}\) and \(\gamma^\scm{R}_i =
\sigmaR_\Psi\prm{\gamma_i}\). Thus, by induction it will be clear that
\(\gamma_k \entails q\prm{p}\) if and only if \(\gamma^\scm{R}_k \entails
q^\scm{R}\prm{p}\).

\begin{itemize}
\item If \(\tup{\psi_i, p_i}\) is an instance of \textstt{CreateGroup},
\textstt{GrantAdmin}, \textstt{RevokeAdmin}, \textstt{SAddMember}, or
\textstt{LRemoveMember}, no queries are changed between \(\gamma_{i-1}\) and
\(\gamma_i\). Since the corresponding operations in \(\text{RBAC} \circ
\scm{U}\) alter only the role relation for roles with no permissions, similarly
no queries are changed between \(\gamma^\scm{R}_{i-1}\) and
\(\gamma^\scm{R}_i\).
\item If \(\psi_i\) is \textstt{LAddMember}, let \(p_i = \tup{u, u_2, g}\), then
\textstt{Access} queries are changed to \true for user \(u_2\) and all messages
in group \(g\). These same \textstt{Access} queries are explicitly made \true by
\(\sigmaR_\Psi\) by adding \(u_2\) to roles that grant precisely these
permissions.
\item If \(\psi_i\) is \textstt{SRemoveMember}, let \(p_i = \tup{u, u_2, g}\),
then \textstt{Access} queries are made \false for user \(u_2\) and all messages
in group \(g\). These same \textstt{Access} queries are explicitly made \false
by \(\sigmaR_\Psi\) by removing \(u_2\) from the only roles with these
permissions.
\item If \(\psi_i\) is \textstt{Post}, let \(p_i = \tup{u, g, m}\), then
\textstt{Access} queries are changed to \true for all users in group \(g\) and
message \(m\). These same \textstt{Access} queries are explicitly made \true by
\(\sigmaR_\Psi\) by adding all users in group \(g\) to the role with the
permission corresponding to \(m\).
\end{itemize}

Thus, we have proven property~(1) for state-matching implementations, and we
proceed to prove property~(2).

Let \(\gamma^\scm{R}_0\) be the start-state in \(\text{RBAC} \circ \scm{U}\)
corresponding to \(\gamma_0\), the start-state in GMS. Then, if
\(\gamma^\scm{R}_k\) is a state reachable from \(\gamma^\scm{R}_0\), we
construct \(\gamma_k\), a state in GMS reachable from \(\gamma_0\), as follows.
\begin{enumerate}
\item Consider each \textstt{Access} query changed to \true (i.e., each
permission granted) between \(\gamma^\scm{R}_0\) and \(\gamma^\scm{R}_k\). Let
\(p = \tup{u, m}\) be the parameterization of the \textstt{Access} query in
question. If permission \(m\) corresponds to a message in GMS, execute
\textstt{CreateGroup} to create a new group, and use \textstt{SAddMember} to add
\(u\) to this group (note that no queries have changed yet, since the new group
has no messages). Finally, \textstt{Post} message \(m\) in the new group,
granting only the access in question.
\item Consider each \textstt{Access} query changed to \false (i.e., each
permission revoked) between \(\gamma^\scm{R}_0\) and \(\gamma^\scm{R}_k\). Let
\(p = \tup{u, m}\) be the parameterization of the \textstt{Access} query in
question. If permission \(m\) corresponds to a message in GMS, then since \(u\)
can access \(m\) in \(\gamma_0\), there exists group \(g\) that \(u\) access to
\(m\) through (i.e., \(\exists t_l, t, t_u \in T: \tup{u, g, t_l, t_u} \in R
\land \tup{g, m, t} \in TX \land t_l \leq t \leq t_u\)). Execute
\textstt{CreateGroup} to create a new group, and use \textstt{SAddMember} to add
\(u\) to this group. Next, \textstt{Post} all messages that \(u\) has access to
through \(g\) to this new group, with the exception of \(m\) (note that no
queries have changed yet; user \(u\) has not gained or lost any accesses).
Finally, use \textstt{SRemoveMember} to remove \(u\) from \(g\), revoking only
the access in question.
\end{enumerate}

These changes to transition between \(\gamma_0\) and \(\gamma_k\) in GMS allow
\(\gamma_k\) to answer each query in the same way as \(\gamma^\scm{R}_k\). Thus,
\(\gamma_k \entails q\prm{p}\) if and only if \(\gamma^\scm{R}_k \entails
q^\scm{R}\prm{p}\). Therefore, we have proven property~(2) for state-matching
implementations, and proven that the implementation \(\sigmaR\) is a
state-matching implementation.
\end{nproof}

} 

\subsection{DAC}
\label[appendix]{appx:expressiveness:dac}

{\newcommand{\sigmaD}{\sigma^\scm{D}}

DAC is a discretionary access control scheme based on the Graham-Denning
scheme~\cite{grahamdenning}, \(\scm{D} = \tup{\Gamma^\scm{D}, \Psi^\scm{D},
Q^\scm{D}}\). Its states, \(\Gamma^\scm{D}\), are defined by the sets \(\tup{S,
O, I, M}\), where:
\begin{itemize}
\item \(S\) is the set of subjects
\item \(O\) is the set of objects
\item \(I\) is the set of access rights
\item \(M : S \times O \to 2^I\) is the access matrix
\end{itemize}

DAC's commands, \(\Psi^\scm{D}\), include the following.
\begin{alltt}\smaller
Grant(\(s\), \(t\), \(o\), \(i\))
\+if \(\text{own}\in{}M\prm{s,o} \land i\neq\text{own}\)
\+\+\(M\prm{t,o}\is{}M\prm{t,o}\cup{}\set{i}\)
\+
Revoke(\(s\), \(t\), \(o\), \(i\))
\+if \(\text{own}\in{}M\prm{s,o} \land i\neq\text{own}\)
\+\+\(M\prm{t,o}\is{}M\prm{t,o}-\set{i}\)
\end{alltt}

Finally, DAC's queries, \(Q^\scm{D}\), include the following.
\begin{alltt}\smaller
Access(\(s\), \(o\), \(i\))
\+\(i\in{}M\prm{s,o}\)
\end{alltt}

We extend DAC with \AM \(\scm{V} = \tup{\Gamma^\scm{V}, \Psi^\scm{V},
Q^\scm{V}}\). The \AM's states, \(\Gamma^\scm{V}\), are defined by the sets
\(\tup{G, GM, W, A, B}\), where:
\begin{itemize}
\item \(G\) is the set of groups
\item \(GM \subseteq G \times O\) is the group-message relation
\item \(W \subseteq S \times G\) is the group ownership relation
\item \(A \subseteq S \times G\) is the group administration relation
\item \(B \subseteq S \times G\) is the group membership relation
\end{itemize}

The extension's commands, \(\Psi^\scm{V}\), include the following.
\begin{alltt}\smaller
CreateGroup(\(s\), \(g\))
\+\(G\is{}G\cup{}\set{g}\)
\+\(W\is{}W\cup{}\set{\tup{s,g}}\)
\+\(A\is{}A\cup{}\set{\tup{s,g}}\)
\+\(B\is{}B\cup{}\set{\tup{s,g}}\)
\+
AssociateWithGroup(\(s\), \(g\), \(o\))
\+if \(\tup{s,g}\in{}B\)
\+\+\(GM\is{}GM\cup{}\set{\tup{g,o}}\)
\+
GrantAdmin(\(s\), \(t\), \(g\))
\+if \(\tup{s,g}\in{}W\)
\+\+\(A\is{}A\cup{}\set{\tup{t,g}}\)
\+
RevokeAdmin(\(s\), \(t\), \(g\))
\+if \(\tup{s,g}\in{}W \lor s=t\)
\+\+\(A\is{}A-\set{\tup{t,g}}\)
\+
GrantMember(\(s\), \(t\), \(g\))
\+if \(\tup{s,g}\in{}A\)
\+\+\(B\is{}B\cup{}\set{\tup{u,g}}\)
\+
RevokeMember(\(s\), \(t\), \(g\))
\+if \(\tup{s,g}\in{}A\)
\+\+\(B\is{}B-\set{\tup{u,g}}\)
\end{alltt}

Finally, \(Q^\scm{V} = \emptyset\), and thus the extension does not add any
queries to the scheme.

We can now demonstrate the implementation of GMS using \(\text{DAC} \circ
\scm{V}\). To describe an implementation \(\sigmaD\) of GMS in \(\text{DAC}
\circ \scm{V}\), we must describe the state-to-state mapping
(\(\sigmaD_\Gamma\)), the command-to-command mapping (\(\sigmaD_\Psi\)), and the
query-to-query mapping (\(\sigmaD_Q\)).

First, we describe \(\sigmaD_\Gamma\), which maps a state in GMS,
\(\gamma^\scm{G} \in \Gamma^\scm{G}\), to a state in \(\text{DAC} \circ
\scm{V}\), \(\sigmaD_\Gamma\prm{\gamma^\scm{G}} = \gamma^\scm{D} \in
\Gamma^\scm{D}\), as follows. Users in GMS are mapped to subjects in
\(\text{DAC} \circ \scm{V}\). Each message \(m\) posted to any group is mapped
to an object \(m \in O\). Since GMS considers only read access, DAC's \(I\) is
statically set to \(\set{\textstt{r}}\). The group-message relation is stored in
\(\scm{V}\) along with relations for group ownership, administration, and
membership. DAC's \(M\) maintains a ``flattened'' view of the current accesses,
and thus \(M\prm{s,o} = \set{\textstt{r}}\) if the GMS user corresponding to
\(s\) has access to the GMS message corresponding to object \(o\). The
projection of the accesses maintained in \(M\) will be updated by
\(\sigmaD_\Psi\) whenever the more semantically meaningful structures in
\(\scm{V}\)'s state are changed.

Next, we describe \(\sigmaD_\Psi\), which maps commands in GMS to strings of
commands in \(\text{DAC} \circ \scm{V}\).
\begin{itemize}
\item \textstt{CreateGroup(\(u\), \(g\))} in GMS is mapped to
\textstt{CreateGroup(\(u\), \(g\))} in \(\text{DAC} \circ \scm{V}\).
\item \textstt{GrantAdmin(\(u\), \(u_2\), \(g\))} in GMS is mapped to
\textstt{GrantAdmin(\(u\), \(u_2\), \(g\))} in \(\text{DAC} \circ \scm{V}\).
\item \textstt{RevokeAdmin(\(u\), \(u_2\), \(g\))} in GMS is mapped to
\textstt{RevokeAdmin(\(u\), \(u_2\), \(g\))} in \(\text{DAC} \circ \scm{V}\).
\item \textstt{SAddMember(\(u\), \(u_2\), \(g\))} in GMS is mapped to
\textstt{GrantMember(\(u\), \(u_2\), \(g\))} in \(\text{DAC} \circ \scm{V}\).
\item \textstt{LAddMember(\(u\), \(u_2\), \(g\))} in GMS is mapped to
\textstt{GrantMember(\(u\), \(u_2\), \(g\))} in \(\text{DAC} \circ \scm{V}\),
followed by \textstt{Grant(\(u\), \(u_2\), \(m\), \(\text{r}\))} for each \(m\)
such that \(\tup{g, m} \in GM\).
\item \textstt{SRemoveMember(\(u\), \(u_2\), \(g\))} in GMS is mapped to
\textstt{RevokeMember(\(u\), \(u_2\), \(g\))} in \(\text{DAC} \circ \scm{V}\),
followed by \textstt{Revoke(\(u\), \(u_2\), \(m\), \(\text{r}\))} for each \(m\)
such that \(\tup{g, m} \in GM\).
\item \textstt{LRemoveMember(\(u\), \(u_2\), \(g\))} in GMS is mapped to
\textstt{RevokeMember(\(u\), \(u_2\), \(g\))} in \(\text{DAC} \circ \scm{V}\).
\item \textstt{Post(\(u\), \(g\), \(m\))} in GMS is mapped to
\textstt{AssociateWithGroup(\(u\), \(g\), \(m\))} in \(\text{DAC} \circ
\scm{V}\), followed by \textstt{Grant(\(u\), \(u_2\), \(m\), \(\text{r}\))} for
each \(u_2\) such that \(\tup{u_2, g} \in B\).
\end{itemize}

Finally, \(\sigmaD_Q\) maps \textstt{Access(\(u\), \(m\))} in GMS to
\textstt{Access(\(u\), \(m\), \(\text{r}\))} in \(\text{DAC} \circ \scm{V}\).

\begin{theorem}
\label{thm:implementation:dac}
\(\sigmaD\) is a state-matching implementation of GMS in \(\text{DAC} \circ \scm{V}\).
\end{theorem}

\begin{nproof}
First, we prove property~(1) for state-matching implementations.

Let \(\gamma_0\) be a start state in GMS. Produce \(\gamma^\scm{D}_0\) in
\(\text{DAC} \circ \scm{V}\) using \(\sigmaD_\Gamma\). Given \(\gamma_k\) such
that \(\gamma_0 \Mapsto \gamma_k\), we show that there exists
\(\gamma^\scm{D}_k\) such that \(\gamma^\scm{D}_0 \Mapsto \gamma^\scm{D}_k\)
where, for all queries \(q = \tup{n, P, {\entails}} \in Q^\scm{G}\) and
parameterizations \(p \in P^*\), \(\gamma^\scm{D}_k \entails q^\scm{D}\prm{p}\)
if and only if \(\gamma_k \entails q\prm{p}\).

Consider the case where \(\gamma_k = \gamma_0\), then let \(\gamma^\scm{D}_k =
\gamma^\scm{D}_0\). By inspection of the procedure for \(\sigmaD_\Gamma\),
\(\gamma_k \entails q\prm{p}\) if and only if \(\gamma^\scm{D}_k \entails
q^\scm{D}\prm{p}\).

Next, consider some arbitrary \(\gamma_k\) reachable from \(\gamma_0\). We
construct \(\gamma^\scm{D}_k\) that is reachable from \(\gamma^\scm{D}_0\) and
that answers every \(q^\scm{D}\prm{p}\) in the same way that \(\gamma_k\)
answers \(q\prm{p}\), as per \(\sigmaD_\Psi\). Since \(\gamma_0 \Mapsto
\gamma_k\), there exists a sequence of commands \(\tup{\psi_1 = \tup{n_1, P_1,
e_1}, \ldots, \psi_k = \tup{n_k, P_k, e_k}}\) and a sequence of
parameterizations \(\tup{p_1 \in P_1^*, \ldots, p_k \in P_k^*}\) of these
commands such that \(\gamma_k = e_k\prm{\ldots e_1\prm{\gamma_0, p_1}, \ldots,
p_k}\). For each command\slash{}parameterization pair \(\tup{\psi_i, p_i}\), we
show that the same queries change value between \(\gamma_{i-1}\) and \(\gamma_i
= e_i\prm{\gamma_{i-1}, p_i}\) and between \(\gamma^\scm{D}_{i-1} =
\sigmaD_\Psi\prm{\gamma_{i-1}}\) and \(\gamma^\scm{D}_i =
\sigmaD_\Psi\prm{\gamma_i}\). Thus, by induction it will be clear that
\(\gamma_k \entails q\prm{p}\) if and only if \(\gamma^\scm{D}_k \entails
q^\scm{D}\prm{p}\).

\begin{itemize}
\item If \(\tup{\psi_i, p_i}\) is an instance of \textstt{CreateGroup},
\textstt{GrantAdmin}, \textstt{RevokeAdmin}, \textstt{SAddMember}, or
\textstt{LRemoveMember}, no queries are changed between \(\gamma_{i-1}\) and
\(\gamma_i\). Since the corresponding operations in \(\text{DAC} \circ \scm{V}\)
alter only the extension state (not granting any new accesses), similarly no
queries are changed between \(\gamma^\scm{D}_{i-1}\) and \(\gamma^\scm{D}_i\).
\item If \(\psi_i\) is \textstt{LAddMember}, let \(p_i = \tup{u, u_2, g}\), then
\textstt{Access} queries are changed to \true for user \(u_2\) and all messages
in group \(g\). These same \textstt{Access} queries are explicitly made \true by
\(\sigmaD_\Psi\) through executions of the \textstt{Grant} command.
\item If \(\psi_i\) is \textstt{SRemoveMember}, let \(p_i = \tup{u, u_2, g}\),
then \textstt{Access} queries are made \false for user \(u_2\) and all messages
in group \(g\). These same \textstt{Access} queries are explicitly made \false by
\(\sigmaD_\Psi\) through executions of the \textstt{Revoke} command.
\item If \(\psi_i\) is \textstt{Post}, let \(p_i = \tup{u, g, m}\), then
\textstt{Access} queries are changed to \true for all users in group \(g\) and
message \(m\). These same \textstt{Access} queries are explicitly made \true by
\(\sigmaD_\Psi\) through executions of the \textstt{Grant} command.
\end{itemize}

Thus, we have proven property~(1) for state-matching implementations, and we
proceed to prove property~(2).

Let \(\gamma^\scm{D}_0\) be the start-state in \(\text{DAC} \circ \scm{V}\)
corresponding to \(\gamma_0\), the start-state in GMS. Then, if
\(\gamma^\scm{D}_k\) is a state reachable from \(\gamma^\scm{D}_0\), we
construct \(\gamma_k\), a state in GMS reachable from \(\gamma_0\), as follows.
\begin{enumerate}
\item Consider each \textstt{Access} query changed to \true (i.e., each access
granted) between \(\gamma^\scm{D}_0\) and \(\gamma^\scm{D}_k\). Let \(p =
\tup{u, m, \textstt{r}}\) be the parameterization of the \textstt{Access} query
in question (if the access is any right but \textstt{r}, it will not affect the
GMS state). If object \(m\) corresponds to a message in GMS, execute
\textstt{CreateGroup} to create a new group, and use \textstt{SAddMember} to add
\(u\) to this group (note that no queries have changed yet, since the new group
has no messages). Finally, \textstt{Post} message \(m\) in the new group,
granting only the access in question.
\item Consider each \textstt{Access} query changed to \false (i.e., each access
revoked) between \(\gamma^\scm{D}_0\) and \(\gamma^\scm{D}_k\). Let \(p =
\tup{u, m, \textstt{r}}\) be the parameterization of the \textstt{Access} query
in question. If object \(m\) corresponds to a message in GMS, then since \(u\)
can access \(m\) in \(\gamma_0\), there exists group \(g\) that \(u\) has access
to \(m\) through (i.e., \(\exists t_l, t, t_u \in T: \tup{u, g, t_l, t_u} \in R
\land \tup{g, m, t} \in TX \land t_l \leq t \leq t_u\)). Execute
\textstt{CreateGroup} to create a new group, and use \textstt{SAddMember} to add
\(u\) to this group. Next, \textstt{Post} all messages that \(u\) has access to
through \(g\) to this new group, with the exception of \(m\) (note that no
queries have changed yet; user \(u\) has not gained or lost any accesses).
Finally, use \textstt{SRemoveMember} to remove \(u\) from \(g\), revoking only
the access in question.
\end{enumerate}

These changes to transition between \(\gamma_0\) and \(\gamma_k\) in GMS allow
\(\gamma_k\) to answer each query in the same way as \(\gamma^\scm{D}_k\). Thus,
\(\gamma_k \entails q\prm{p}\) if and only if \(\gamma^\scm{D}_k \entails
q^\scm{D}\prm{p}\). Therefore, we have proven property~(2) for state-matching
implementations, and proven that the implementation \(\sigmaD\) is a
state-matching implementation.
\end{nproof}

} 

\subsection{SD3-GM}
\label[appendix]{appx:expressiveness:sd3gm}

{\newcommand{\sigmaS}{\sigma^\scm{S}}

SD3-GM is the group-messaging instantiation of the SD3 trust management scheme,
\(\scm{S} = \tup{\Gamma^\scm{S}, \Psi^\scm{S}, Q^\scm{S}}\). Its states,
\(\Gamma^\scm{S}\), are defined by the set \(P\), the set of policy sentences
written in the SD3 policy language. The following static policy sentences
enforce the access semantics and the current membership semantics.
{\smaller
\begin{align*}
\texttt{ACCESS(U, M) \sdif{} {}}
& \texttt{MEMBER(U, G, T_1, T_2),}\\
& \texttt{POST(G, M, T),}\\
& \texttt{LESSEQ(T_1, T),}\\
& \texttt{LESSEQ(T, T_2)}\\
\texttt{CURRMEMBER(U, G) \sdif{} {}}
& \texttt{MEMBER(U, G, T, \(\infty\))}
\end{align*}
}
Here, \textstt{LESSEQ} is the inherent ``less-than-or-equal'' predicate for
timestamps.

SD3-GM's commands, \(\Psi^\scm{S}\), include the following.
\begin{alltt}\smaller
CreateGroup(\(u\),\(g\))
\+\(P\is{}P\cup{}\set{\sent{\texttt{OWN(\(u\),\(g\))}},\sent{\texttt{ADMIN(\(u\),\(g\))}},\sent{\texttt{MEMBER(\(u\),\(g\),\(0\),\(\infty\))}}}\)
\+
GrantAdmin(\(o\),\(u\),\(g\))
\+if \(eval(\sent{\texttt{OWN(\(o\),\(g\))}})\)
\+\+\(P\is{}P\cup{}\set{\sent{\texttt{ADMIN(\(u\),\(g\))}}}\)
\+
RevokeAdmin(\(o\),\(u\),\(g\))
\+if \(eval(\sent{\texttt{OWN(\(o\),\(g\))}}) \lor o = u\)
\+\+\(P\is{}P-\set{\sent{\texttt{ADMIN(\(u\),\(g\))}}}\)
\+
SAddMember(\(a\),\(u\),\(g\))
\+if \(eval(\sent{\texttt{ADMIN(\(a\),\(g\))}})\)
\+\+\(P\is{}P\cup{}\set{\sent{\texttt{MEMBER(\(u\),\(g\),\(T\sb{c}\),\(\infty\))}}}\)
\+\+\(P\is{}P\cup{}\set{\sent{\texttt{TIME(\(T\sb{c}+1\))}}}-\set{\sent{\texttt{TIME(\(T\sb{c}\))}}}\)
\+
LAddMember(\(a\),\(u\),\(g\))
\+if \(eval(\sent{\texttt{ADMIN(\(a\),\(g\))}})\)
\+\+\(P\is{}P\cup{}\set{\sent{\texttt{MEMBER(\(u\),\(g\),\(0\),\(\infty\))}}}\)
\+
SRemoveMember(\(a\),\(u\),\(g\))
\+if \(eval(\sent{\texttt{ADMIN(\(o\),\(g\))}}) \lor o=u\)
\+\+\(P\is{}P-\set{\sent{\texttt{MEMBER(\(u\),\(g\),\(*\),\(*\))}}}\)
\+
LRemoveMember(\(a\),\(u\),\(g\))
\+if \(eval(\sent{\texttt{ADMIN(\(a\),\(g\))}})\)
\+\+\(P\is{}P\cup{}\set{\sent{\texttt{MEMBER(\(u\),\(g\),\(t\),\(T\sb{c}\))}}}-\set{\sent{\texttt{MEMBER(\(u\),\(g\),\(t\),\(\infty\))}}}\)
\+\+(where \(\sent{\texttt{MEMBER(\(u\),\(g\),\(t\),\(\infty\))}}\in{}P\))
\+
Post(\(u\),\(g\),\(m\))
\+if \(eval(\sent{\texttt{CURRMEMBER(\(u\),\(g\))}})\)
\+\+\(P\is{}P\cup{}\set{\sent{\texttt{POST(\(g\),\(m\),\(T\sb{c}\))}}}\)
\+\+\(P\is{}P\cup{}\set{\sent{\texttt{TIME(\(T\sb{c}+1\))}}}-\set{\sent{\texttt{TIME(\(T\sb{c}\))}}}\)
\end{alltt}

Finally, SD3-GM's queries, \(Q^\scm{S}\), include the following.
\begin{alltt}\smaller
Access(\(u\),\(m\))
\+\(eval(\sent{\texttt{ACCESS(\(u\),\(m\))}})\)
\end{alltt}

To describe an implementation \(\sigmaS\) of GMS in SD3-GM, we must describe the
state-to-state mapping (\(\sigmaS_\Gamma\)), the command-to-command mapping
(\(\sigmaS_\Psi\)), and the query-to-query mapping (\(\sigmaS_Q\)).

First, we describe \(\sigmaS_\Gamma\), which maps a state in GMS,
\(\gamma^\scm{G} \in \Gamma^\scm{G}\), to a state in SD3-GM,
\(\sigmaS_\Gamma\prm{\gamma^\scm{G}} = \gamma^\scm{S} \in \Gamma^\scm{S}\), as
follows.
\begin{itemize}
\item For \(T_c\) in GMS, \sent{\textstt{TIME(\(T_c\))}} is added to \(P\) in
SD3-GM.
\item For each \(\tup{u, g} \in O\) in GMS, \sent{\textstt{OWN(\(u\),\(g\))}} is
added to \(P\) in SD3-GM.
\item For each \(\tup{u, g} \in A\) in GMS, \sent{\textstt{ADMIN(\(u\),\(g\))}}
is added to \(P\) in SD3-GM.
\item For each \(\tup{u, g, t_1, t_2} \in R\) in GMS,
\sent{\textstt{MEMBER(\(u\),\(g\),\(t_1\),\(t_2\))}} is added to \(P\) in
SD3-GM.
\item For each \(\tup{g, m, t} \in TX\) in GMS,
\sent{\textstt{POST(\(g\),\(m\),\(t\))}} is added to \(P\) in SD3-GM.
\end{itemize}

Then, \(\sigmaS_\Psi\) and \(\sigmaS_Q\) are both identity mappings. That is,
commands and queries are both mapped to their identically-named versions in
SD3-GM.

\begin{theorem}
\label{thm:implementation:sd3}
\(\sigmaS\) is a state-matching implementation of GMS in SD3-GM.
\end{theorem}

\begin{nproof}
First, we prove property~(1) for state-matching implementations.

Let \(\gamma_0\) be a start state in GMS. Produce \(\gamma^\scm{S}_0\) in SD3-GM
using \(\sigmaS_\Gamma\). Given \(\gamma_k\) such that \(\gamma_0 \Mapsto
\gamma_k\), we show that there exists \(\gamma^\scm{S}_k\) such that
\(\gamma^\scm{S}_0 \Mapsto \gamma^\scm{S}_k\) where, for all queries \(q =
\tup{n, P, {\entails}} \in Q^\scm{G}\) and parameterizations \(p \in P^*\),
\(\gamma^\scm{S}_k \entails q^\scm{S}\prm{p}\) if and only if \(\gamma_k
\entails q\prm{p}\).

Consider the case where \(\gamma_k = \gamma_0\), then let \(\gamma^\scm{S}_k =
\gamma^\scm{S}_0\). By inspection of the procedure for \(\sigmaS_\Gamma\),
\(\gamma_k \entails q\prm{p}\) if and only if \(\gamma^\scm{S}_k \entails
q^\scm{S}\prm{p}\).

Next, consider some arbitrary \(\gamma_k\) reachable from \(\gamma_0\). We
construct \(\gamma^\scm{S}_k\) this is reachable from \(\gamma^\scm{S}_0\) and
that answers every \(q^\scm{S}\prm{p}\) in the same way that \(\gamma_k\)
answers \(q\prm{p}\), as per \(\sigmaS_\Psi\). Since \(\gamma_0 \Mapsto
\gamma_k\), there exists a sequence of commands \(\tup{\psi_1 = \tup{n_1, P_1,
e_1}, \ldots, \psi_k = \tup{n_k, P_k, e_k}}\) and a sequence of
parameterizations \(\tup{p_1 \in P_1^*, \ldots, p_k \in P_k^*}\) of these
commands such that \(\gamma_k = e_k\prm{\ldots e_1\prm{\gamma_0, p_1}, \ldots,
p_k}\). For each command\slash{}parameterization pair \(\tup{\psi_i, p_i}\), we
show that the same queries change value between \(\gamma_{i-1}\) and \(\gamma_i
= e_i\prm{\gamma_{i-1}, p_i}\) and between \(\gamma^\scm{S}_{i-1} =
\sigmaS_\Psi\prm{\gamma_{i-1}}\) and \(\gamma^\scm{S}_i =
\sigmaS_\Psi\prm{\gamma_i}\). Thus, by induction it will be clear that
\(\gamma_k \entails q\prm{p}\) if and only if \(\gamma^\scm{S}_k \entails
q^\scm{S}\prm{p}\). In the case of \(\sigmaS\), the implementation is a strict
bisimulation, and GMS and SD3-GM move in strict lock-step.

\begin{itemize}
\item If \(\tup{\psi_i, p_i}\) is an instance of \textstt{CreateGroup},
\textstt{GrantAdmin}, \textstt{RevokeAdmin}, \textstt{SAddMember}, or
\textstt{LRemoveMember}, no queries are changed between \(\gamma_{i-1}\) and
\(\gamma_i\). Since the corresponding operations in SD3-GM behave identically,
no queries are changed between \(\gamma^\scm{R}_{i-1}\) and
\(\gamma^\scm{R}_i\).
\item If \(\psi_i\) is \textstt{LAddMember}, let \(p_i = \tup{u, u_2, g}\), then
\textstt{Access} queries are changed to \true for user \(u_2\) and all messages
in group \(g\). These same \textstt{Access} queries are also made \true by
\(\sigmaS_\Psi\).
\item If \(\psi_i\) is \textstt{SRemoveMember}, let \(p_i = \tup{u, u_2, g}\),
then \textstt{Access} queries are made \false for user \(u_2\) and all messages
in group \(g\). These same \textstt{Access} queries are also made \false by
\(\sigmaS_\Psi\).
\item If \(\psi_i\) is \textstt{Post}, let \(p_i = \tup{u, g, m}\), then
\textstt{Access} queries are changed to \true for all users in group \(g\) and
message \(m\). These same \textstt{Access} queries are also made \true by
\(\sigmaS_\Psi\).
\end{itemize}

Thus, we have proven property~(1) for state-matching implementations, and we
proceed to prove property~(2).

Let \(\gamma^\scm{S}_0\) be the start-state in SD3-GM corresponding to
\(\gamma_0\), the start-state in GMS. Then, if \(\gamma^\scm{S}_k\) is a state
reachable from \(\gamma^\scm{S}_0\), we construct \(\gamma_k\), a state in GMS
reachable from \(\gamma_0\), as follows. Since both \(\sigmaS_\Gamma\) and
\(\sigmaS_Q\) are the identity mapping, for each command and parameterization
executed between \(\gamma^\scm{S}_0\) and \(\gamma^\scm{S}_k\), we can execute
the identically-named command with the same parameterization in GMS, leading to
a state in which all queries are answered in the same way.

Thus, \(\gamma_k \entails q\prm{p}\) if and only if \(\gamma^\scm{S}_k \entails
q^\scm{S}\prm{p}\). Therefore, we have proven property~(2) for state-matching
implementations, and proven that the implementation \(\sigmaS\) is a
state-matching implementation.
\end{nproof}

} 

\end{document}